\documentclass[letter]{aa}  
\usepackage[colorlinks=true, linkcolor=blue, citecolor=blue, urlcolor=blue]{hyperref}
\usepackage{graphicx}
\usepackage{txfonts}
\usepackage{graphicx}
\usepackage{subcaption}

\usepackage{amsmath}
\graphicspath{{fig/}}
\usepackage[figure,figure*]{hypcap}
\usepackage{lscape}

\newcommand{\TB}{\textcolor{black}}

\def\nh2d{$\rm{NH_2D}$}

\def\nh3{$\rm{NH_3}$}
\def\NH3{$\rm{NH_3}$}

\def\n2hp{$\rm{N_2H^+}$}

\def\h2o{$\rm{H_2O}$}
\def\h2{$\rm{H_2}$}

\def\msun{\,$M_\odot$}

\def\lsun{\,$L_\odot$}

\def\um{\,$\mu\mathrm{m}$}

\def\kms{\,km~s$^{-1}$}

\def\cm2{\,$\rm{cm^{-2}}$}
\def\cm3{\,$\rm{cm^{-3}}$}

\def\11{(1,1)}
\def\22{(2,2)}
\def\33{(3,3)}
\def\44{(4,4)}
\def\55{(5,5)}
\def\t21{$T_{21}$}
\def\r31{$R_{31}$}

\def\her{\emph{Herschel}}

\let \amp = \&

\defcitealias{me16}{I}
\defcitealias{Ge22}{II}
\defcitealias{Ge23}{III}

\begin{document}

\title{The Milky Way Atlas for Linear Filaments}

\authorrunning{Wang et al. }
\titlerunning{Milky Way Atlas for Linear Filaments}

\author{Ke Wang \inst{1}
          \and Yifei Ge \inst{2}
          \and Tapas Baug \inst{3}
          }

\institute{
Kavli Institute for Astronomy and Astrophysics, Peking University, 5 Yiheyuan Road, Haidian District, Beijing 100871, China \\ 
\email{kwang.astro@pku.edu.cn}  
\and 
Department of Astronomy, School of Physics, Peking University, 5 Yiheyuan Road, Haidian District, Beijing 100871, China 
\and
S. N. Bose National Centre for Basic Sciences, Block JD, Sector III, Salt Lake, Kolkata 700106, India
}

\date{Received ---; accepted ---}

 
  \abstract
   {Filamentary structure is important for the ISM and star formation. Galactic distribution of filaments may regulate the star formation rate in the Milky Way. 
However, interstellar filaments are intrinsically complex, making it difficult to study quantitatively.}
   {Here, we focus on linear filaments, the simplest morphology that can be treated as building blocks of any filamentary structure.}
   {We present the first catalog of 42 ``straight-line'' filaments across the full Galactic plane, identified by clustering \TB{of} far-IR \textit{Herschel} HiGAL clumps in position-position-velocity space. We use molecular line cubes to investigate the dynamics along the filaments; compare the filaments with Galactic spiral arms; and compare ambient magnetic fields with the filaments' orientation.
   }
   {The {selected} filaments show extreme linearity ($>$10), aspect ratio (7-48), and velocity coherence over a length of 3-40 pc (mostly $>$10 pc). \TB{About} 1/3 of them are associated with spiral arms, but only one is located in arm center, a.k.a. ``bones'' of the Milky Way.
A few of them extend perpendicular to the Galactic plane, and none is located in the Central Molecular Zone (CMZ) near the Galactic center.
Along the filaments, prevalent periodic oscillation (both in velocity and density) is consistent with gas flows channeled by the filaments and feeding the clumps which harbor diverse star formation activities.
No correlation is found between the filament orientations with \textit{Planck} measured global magnetic field lines.}
{This work highlights some of the fundamental properties of molecular filaments and provides a golden sample for follow-up studies on star formation, ISM structure, and Milky Way structure.
}



\maketitle




\section{Introduction} \label{sec:intro}

Filamentary structure is the dominant morphology of the interstellar medium (ISM), and molecular filaments can play an important role in star formation \citep{Hacar23-PP7}. The Galactic distribution of filaments may regulate the global star formation rate in the Milky Way. 
The largest filaments can map out the skeleton (``bones'') of the Milky Way in spiral arms \citep{me15,Zucker2015,Vallee2016}.
However, the inherent complexity and hierarchy of molecular filaments make it challenging to characterize the structure and dynamics important for star formation.
Traditionally, to study star formation, maps of molecular clouds are often decomposed into cores \citep[e.g.][]{soft:CLUMPFIND,soft:FellWalker}, and in recent years, also to filaments \citep[e.g.][]{soft:getfilaments-Menshchikov2013,soft:FilFinder-Koch2015}.
While spherical cores, by definition, are simple in morphology and thus are relatively easy to treat theoretically, filaments are not. More critically, the community has not yet reached a consensus on the definition of what is a filament, resulting a wide range of ``filaments'' reported in the literature, from simple filaments of linear L-shape, C-shape, S-shape filaments, to a network of filamentary structure of X-shape \citep{me15}, and Hub-Filament Systems \citep{Myers2010}. A meaningful comparison amongst those filaments are thus difficult, because they are intrinsically different entities \citep[c.f.][]{Zucker2018}.

We proposed a physically driven definition of filaments \citep{me16} inspired by the ``sausage instability'' of a gaseous cylinder \citep{Chandra1953}. Under self-gravity, a supercritical filament radially collapse and fragments into a string of equally-spaced clumps. Built on this picture, we developed a customized minimum spanning tree (MST) algorithm to identify filaments by clustering clumps in position-position-velocity (PPV) space. We applied this to PPV clump catalogs decomposed from three Galactic plane surveys:
BGPS, ATLASGAL, and SEDIGISM 
\citep[][hereafter Paper \citetalias{me16}, \citetalias{Ge22}, \citetalias{Ge23}]{me16,Ge22,Ge23}. 
However, those surveys cover only a portion of the Galactic plane. More importantly, those filaments and others reported in the literature so far have a large variety of linearity and morphology classes. Such limitations prevent a further characterization of the fundamental common properties of filaments. A full Galactic plane census of the ``same entities'' is highly demanded.

Here, we identify and characterize the most collimated, ``straight-line'' like filaments across the full Galactic plane, using data from the 
\textit{Herschel} HiGAL survey.
Among all the morphology types \citep{me15}, linear filaments are the simplest, and can be treated as ``unit'' filaments, or building blocks of more complicate filamentary structures. Physical characterization of these units can pave the way towards a unified understanding of filaments \citep[][]{LiuHL18-straightFL,Zucker2018}.
For example, how long can a straight-line filament extend in our Galaxy?
How do they distribute in the Galaxy and map out the spiral arms?
What role do they play in star formation?
Answer to these questions are crucial for a quantitative understanding of star formation in our and other galaxies, as interferometric observations are routinely resolving the largest filaments in nearby galaxies \citep[][]{me15}.
This work provides a key step forward.

\begin{figure*}[!ht]
\centering
(a)
\includegraphics[width=.23\textwidth]{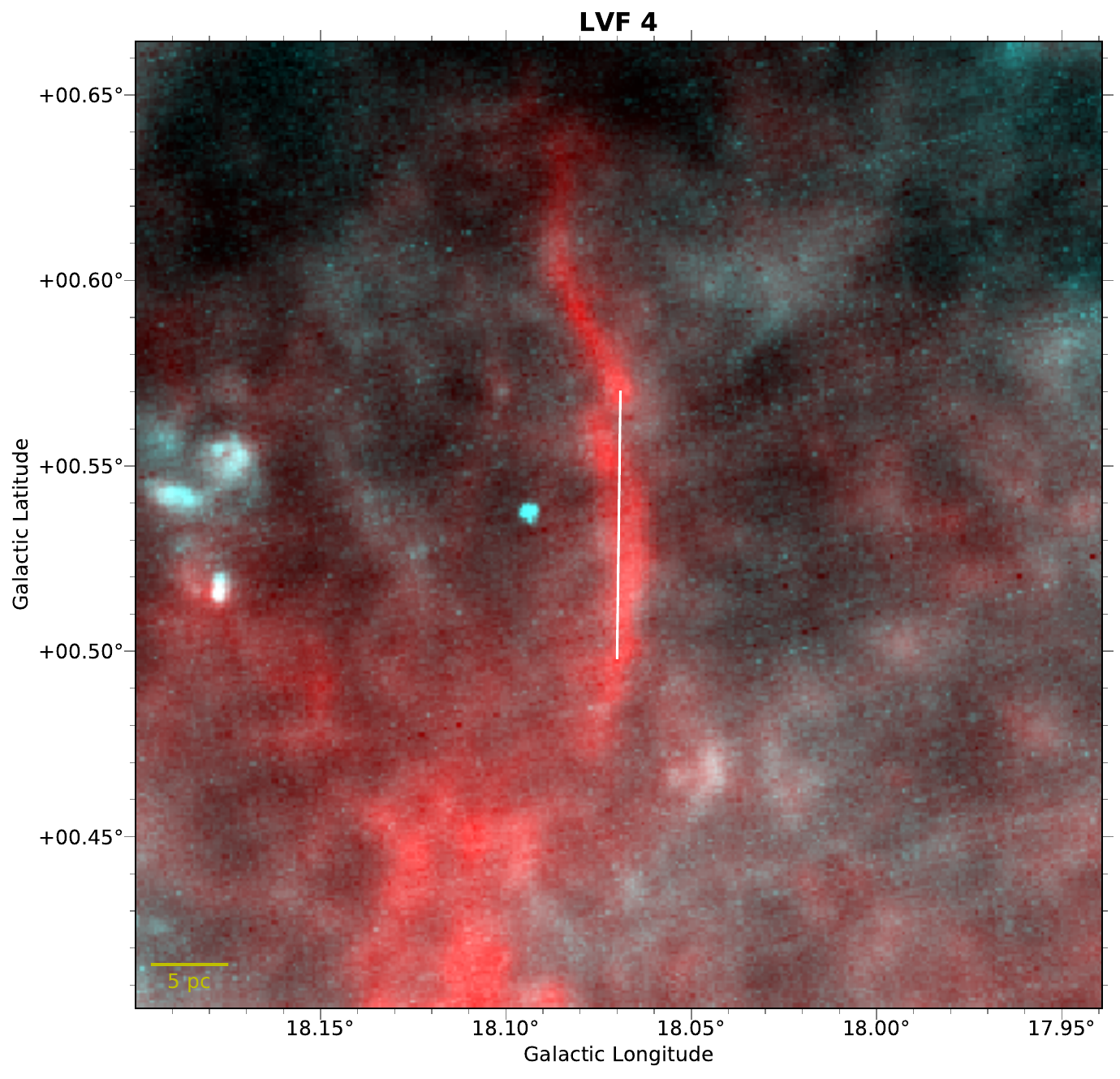}
\includegraphics[width=.23\textwidth]{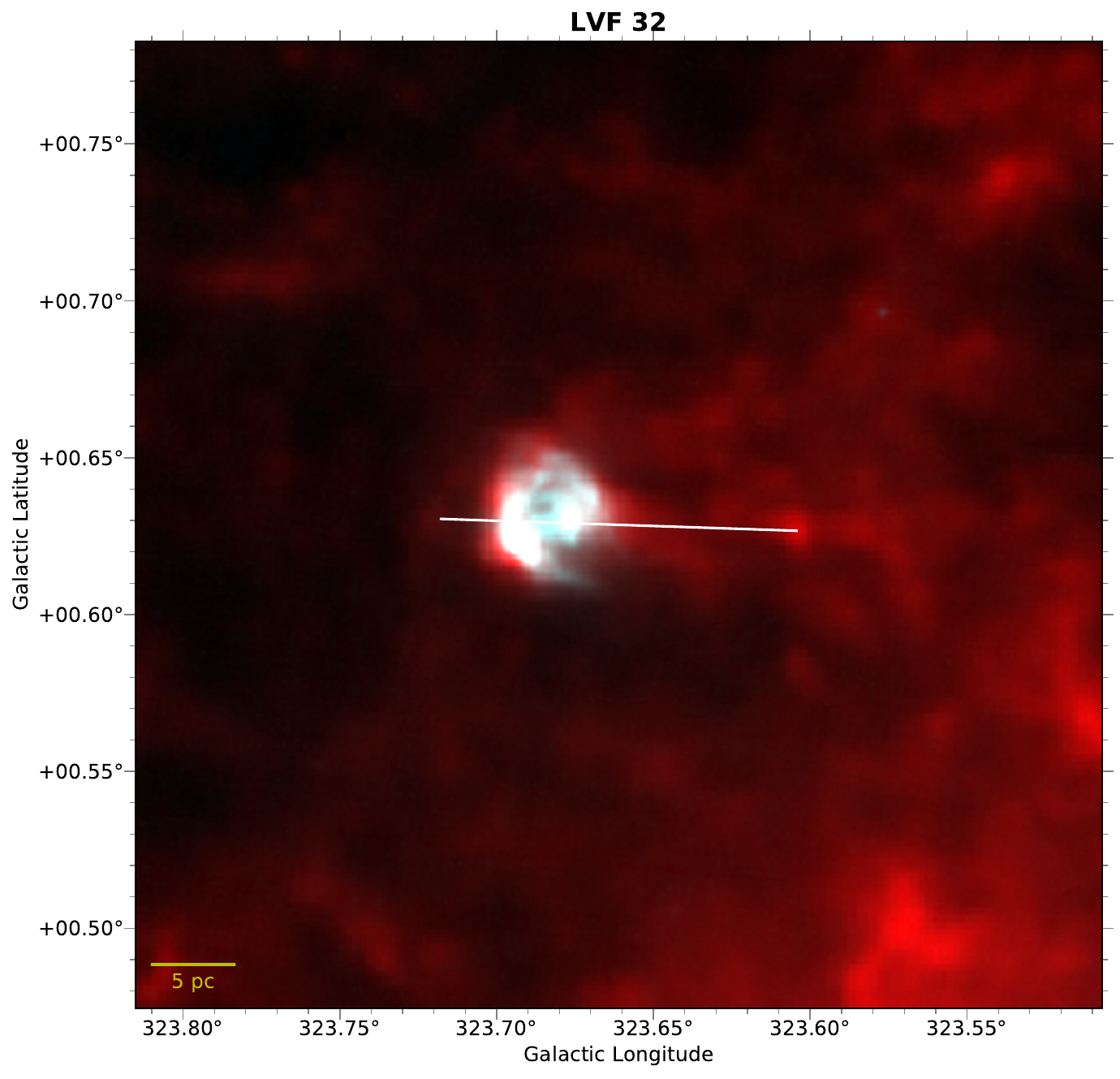}
\includegraphics[width=.23\textwidth]{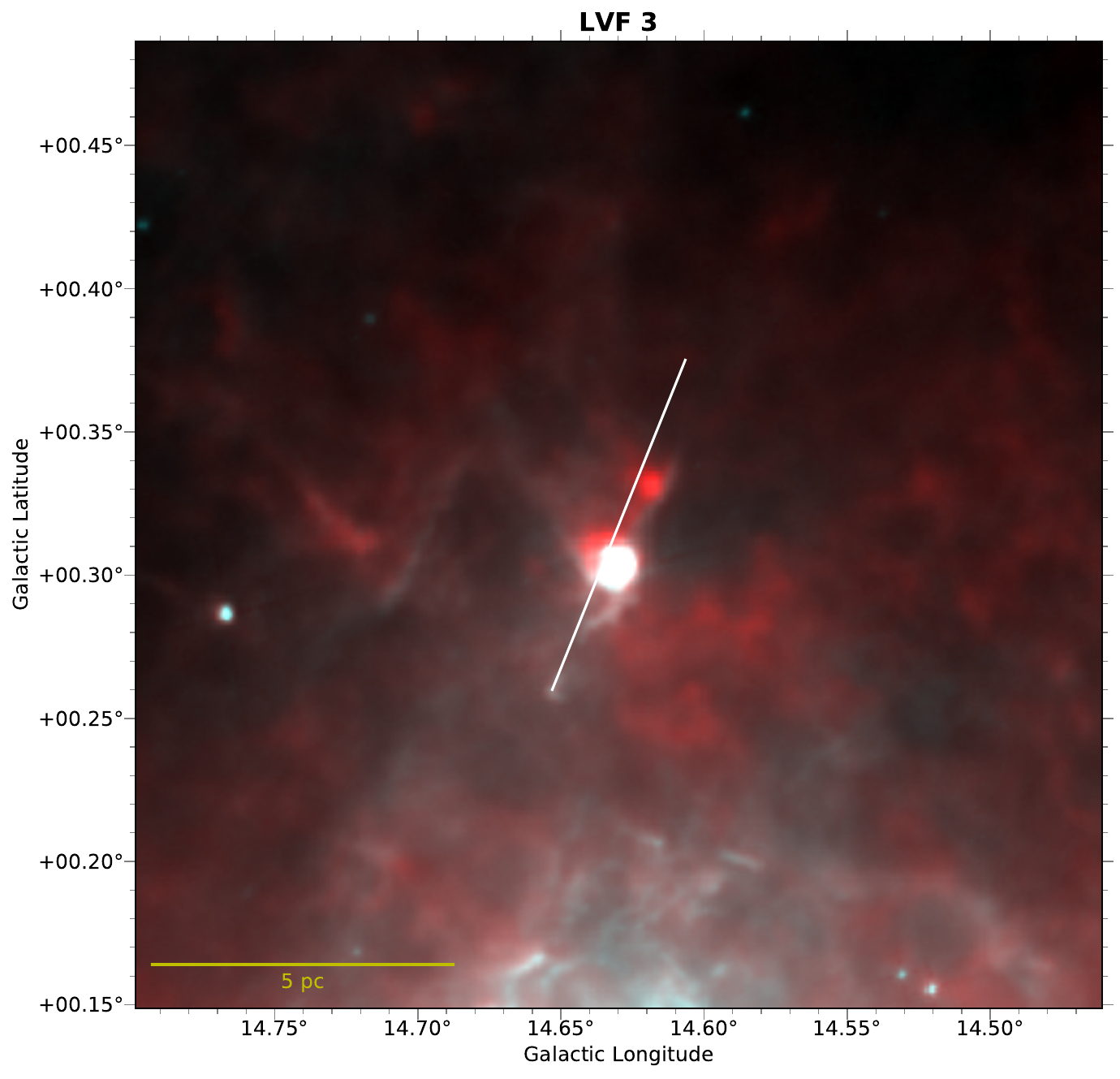}
\includegraphics[width=.23\textwidth]{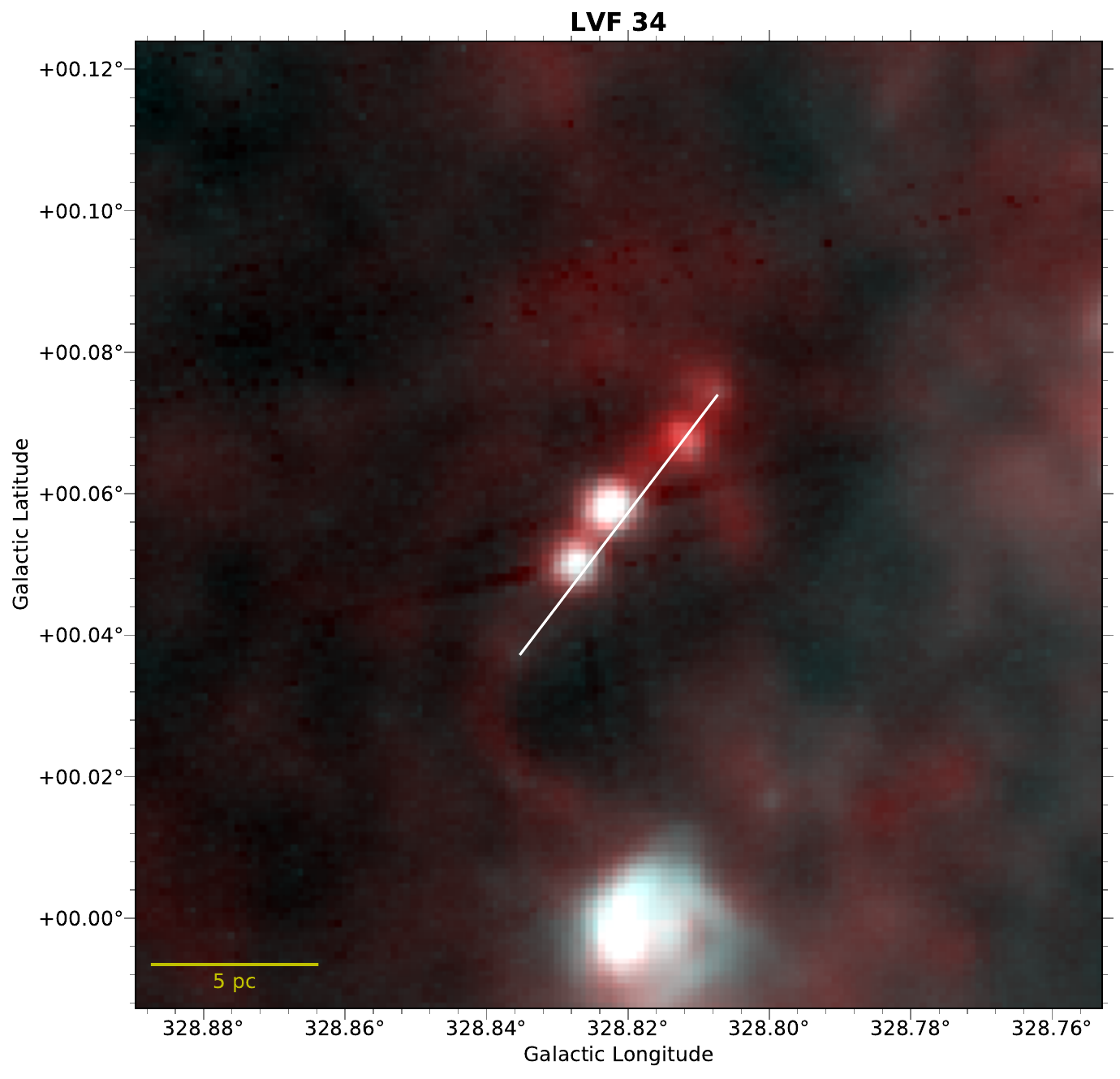}\\
(b)
\includegraphics[width=.7\textwidth, trim={0 5cm 0 5cm},clip]{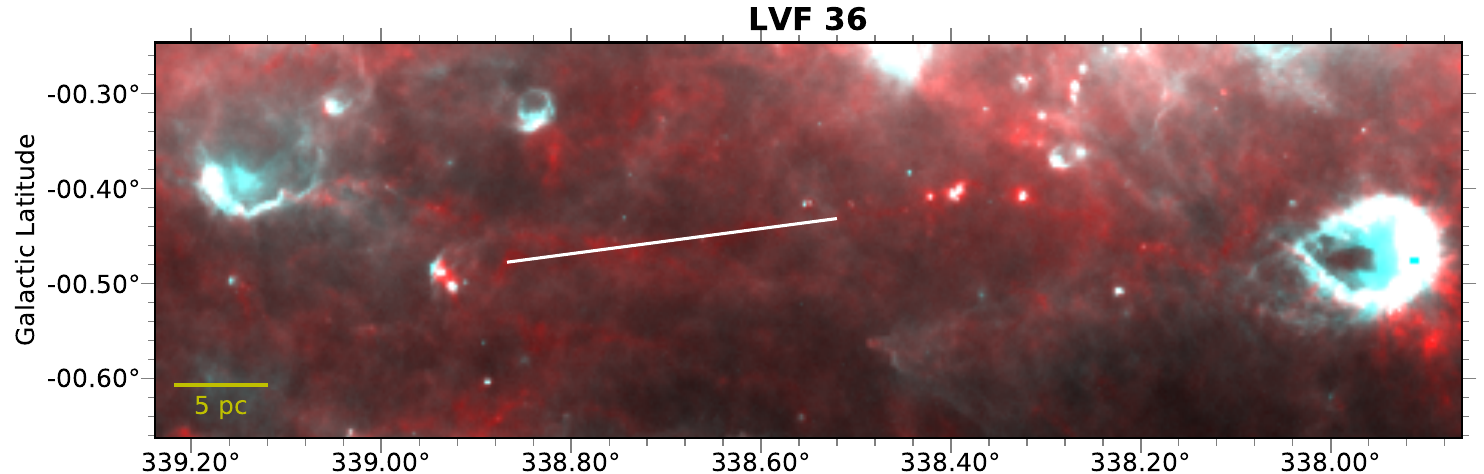}\\
(c)
\includegraphics[width=.72\textwidth]{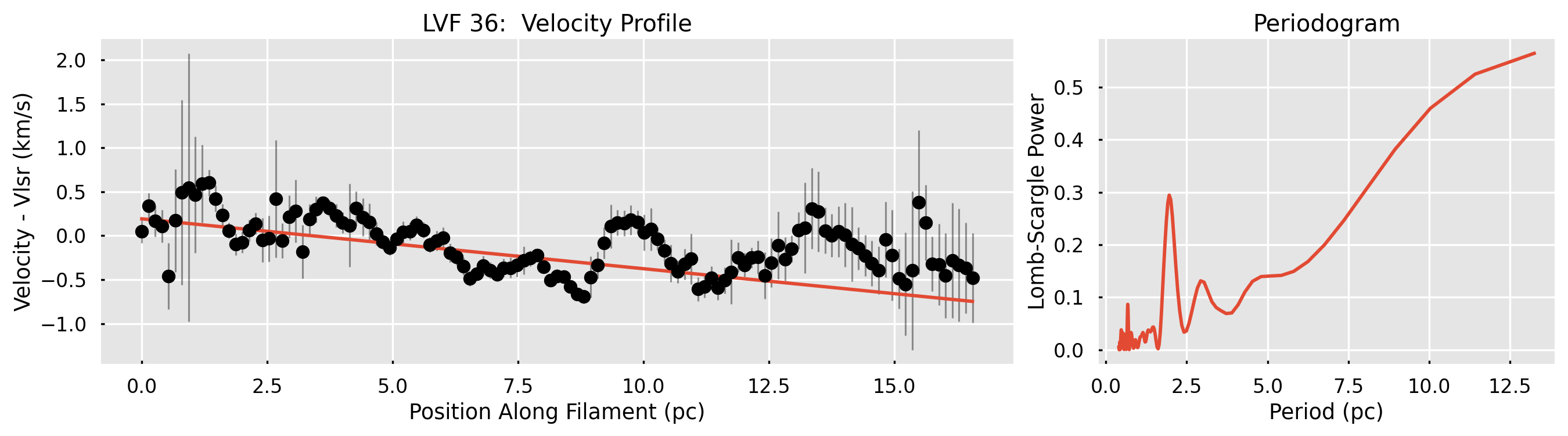}
\caption{
\textbf{(a)}
Representative filaments in a two-color view, where \her\ 250\um\ and 70\um\ emission are shown in red and cyan, respectively.
A filament is marked by end-to-end line connecting two clumps at the filament tips.
From left to right are four representative categorizes: ``Quiescent'',
``Flower'', 
``Central'',
and ``Necklace''.
\autoref{fig:rgb_a} show all filaments.
\textbf{(b)}
same two-color view but for F36, part of ``Nessie''.
\textbf{(c)}
$^{13}$CO(2--1)
PV plot of F36 along curved filament length, and fitting to the global velocity gradient (red line), where the error bars correspond to $1\sigma$ dispersion. The periodogram of the PV curve after removing the global velocity gradient shows a period at 2 pc, consistent to clump separation (\S \ref{sec:VeloOscillation}).
}
\label{fig1:fl5PV}
\end{figure*}

\section{Data and Method} \label{sec:DataMethod}

The \textit{Herschel} infrared Galactic Plane Survey (HiGAL) mapped the entire Galactic plane at five bands: 70, 160, 250, 350, and 500\um\ \citep{Hi-GAL,survey:HiGAL-DR1-Molinari2016}. A band-merged catalog has been released, containing a total of $\sim$150k sources with physical properties determined for most sources \citep{Elia17-HiGALcatInner,Elia21-HiGAL360cat}.
Of the $\sim$150k sources, $\sim$126k sources have radial velocity resolved using literature data. These sources form by far the deepest, largest, and uniform PPV catalog of dust clumps across the entire Galactic plane,
excellent for finding filaments using our MST method. Note that \cite{Elia21-HiGAL360cat} assigned HiGAL clumps into ``high-relaible'' and ``low-reliable'' catalogs based on the quality of SED fitting. Here we merged both catalogs as we are most interested in PPV information to begin with.
The HiGAL sources have an average diameter of $25''$ and 0.6 pc.
For simplicity, we refer them as ``clumps'' in this paper, although their physical scales can be up to few parsec for sources at large distances.


We search for filaments using the MST method following 
Paper \citetalias{me16}, \citetalias{Ge22}, \citetalias{Ge23}.
In brief, any two clumps in the \cite{Elia21-HiGAL360cat} catalog are connected only if they are close enough both in position {(spatial separation $\Delta s <0.06^{\circ}$)} and velocity ($\Delta v <2$\kms). These criteria are resulted from robust tests (details are given in Paper \citetalias{me16}, \citetalias{Ge22}, \citetalias{Ge23}).
This results to $\sim$3.4k MST ``trees'' containing at least five clumps. A strict linearity of $>$10 is applied to select the most prominent straight-line filaments. Linearity is defined as the ratio between the spread of clumps along the long and short axes (Paper \citetalias{me16}, \citetalias{Ge22}, \autoref{fig:mst_a}). 
The selection results in 42 linear and velocity-coherent (LV) filaments.
Most of them are new identification, and only six of them are associated with previously known filaments (\autoref{tab:fl42}).
For example, F36 and F7 are the central and linear part of the well kown S-shaped ``Nessie'' filament \citep{Jackson2010,me15}, and the X-shaped \her\ cold filament CFG26 \citep{me15}.

Thanks to HiGAL's full coverage of the Galactic plane, and the $>$10 times larger number of PPV clumps compared to previous PPV catalogs used in Paper \citetalias{me16}, \citetalias{Ge22}, \citetalias{Ge23}, we are able to extract an atlas of linear filaments in the Milky Way's disk. Unlike in Paper \citetalias{me16}, \citetalias{Ge22}, \citetalias{Ge23}, we do not put a length cut ($>$10 pc for large-scale filaments) but strictly limit the linearity to select straight-line filaments. They are the building blocks of more complex filamentary structure, and it is of great interest to investigate their properties.

After finding the filaments (as strings of clumps), we use properties of the HiGAL clumps \citep{Elia21-HiGAL360cat} as basis to characterize the filaments (\S \ref{sec:GlobalProperties}, \ref{sec:DiverseSF}). These include clump position, velocity, distance, diameter, mass, luminosity, temperature, association with 70\um\ point source, and evolutionary stages (starless, prestellar, and protostellar).
A Galactic spiral arm model is compared to the filaments (\S \ref{sec:GalDistrib}).
Molecular line maps are retrieved from public surveys to analyze the dynamics (\S \ref{sec:VeloOscillation}, \ref{sec:Vcoherence}). 
Dust polarization maps from the \textit{Planck} 
are used to derive global magnetic fields of the filaments (\S \ref{sec:B-fields}).

Note that our MST approach is different than other automated identification that directly decompose continuum images \citep[e.g.][]{Schisano20-HiGAL-FLcat,LiGX16-FLcat}, which resulted orders of magnitude more filaments, and did not use velocity information in identification. Our approach uses a uniform PPV catalog as input, and can have a physically driven criteria for filaments. 
In comparison, \cite{LiGX16-FLcat} identified 517 ATLASGAL filaments merely in the inner Galactic plane; only three have overlaps with our filaments (\autoref{tab:fl42}).
\cite{Schisano20-HiGAL-FLcat} identified 32059 candidate filaments using HiGAL images. Generally, those are not the same kind of features as our filaments.

\section{Results and Discussion} \label{sec:ResultsDisc}
\subsection{Global Properties} \label{sec:GlobalProperties}

\autoref{fig1:fl5PV}(a,b) present example filaments in far-IR. \her\ 250\um\ emission in red provides a rough proxy for column density, and the 70\um\ emission in cyan highlights point sources of embedded young stellar objects, indicative of star formation. The white lines outline the filaments end to end.
The 42 filaments can be divided into four categories of star formation activities:
``Quiet'' (no 70\um\ point source), ``Flower'' (star formation developed at one tip, or head-tail), ``Central'' (star formation developed in central clumps), and ``Necklace'' (star formation developed in almost all clumps).
Evolution and dynamic effects, including edge collapse \citep{Yuan20-EdgeCollapse,Bhadari22-EdgeCollapse} likely play a role in shaping the appearance of the filaments, which deserve further study.

\autoref{tab:fl42}
lists physical properties of the filaments, and \autoref{fig2:hist} illustrates the distribution of some parameters.

Distance is determined from the distances of the clumps in a filament. For most filaments, the clumps have consistent distances reported in \cite{Elia21-HiGAL360cat}. Only in few cases (F1, 9, 39, 41, 42), the clump distances are spread and containing near and far distances. In those cases, additional judgements are made based on the infrared extinction/emission to adopt either near or far distance for the filaments. Clumps in three filaments (F13, 22, 30) have no available distance; we compute distances using the parallax-based Bayesian distance estimator provided by \citep{Reid19-distance,Reid16-distance}.

No obvious selection effect is seen in the distance histogram (\autoref{fig2:hist}), which shows a broad range from 1.2-16.3 kpc, and a weak peak at 4-5 kpc, corresponding to the ``Galactic molecular ring'' \citep{Jackson2006-GRS}.
After determining the filament distance, related clump parameters in \cite{Elia21-HiGAL360cat} are scaled accordingly to computer filament properties.

{Originated from selection, the}
filaments show remarkable linearity and aspect ratios. 
Filament length is measured by the sum of the MST edges (thus following the curvature), and adding sizes of the two edge clumps to account for the actual length of the filament. This ``edge length'' is on average only 3\% (and up to 12\%) larger than the ``end-to-end length'' by simply connecting the two clumps at the tips with a straight line (shown in \autoref{fig1:fl5PV}). {The small curvature provides another measure of the extreme linearity resulted from the selection criteria.} 
Of the 42 filaments, 28 are longer than 10 pc, satisfying the ``large-scale'' filaments in Paper \citetalias{me16}, \citetalias{Ge22}, \citetalias{Ge23}.
The aspect ratios (7-48) are among the highest reported in the literature. 

A lower limit of filament mass ($5\times 10^1$ to $5\times 10^4$\msun) is given by the sum of clump masses. According to Paper \citetalias{Ge22}, this lower limit is typically 36\% of the total filament mass.

A global velocity gradient is determined by a linear fit to the ($l_i, v_i$) plot, where $l_i$ is the length along filament, and $v_i$ the LSR velocity of the $i$th clump in a filament. Note that the slope (velocity gradient) of the fitted lines can be positive or negative, depending on the orientation.

Mass, luminosity, and luminosity-to-mass ratio are among the sharpest distributions (Kurtosis $K = 12, 22.5, 8$, \autoref{tab:fl42}). But those are dominated by a long tail at the right side made of only three large values. If those outliers are removed, the distributions would be quite flat.
Aspect ratio, number of clumps, and angular length also show sharp distributions ($K=7.5, 10, 10$) with no obvious outliers.

\begin{figure}
\centering
\includegraphics[width=.45\textwidth, trim={0 0 1cm 1.5cm},clip]{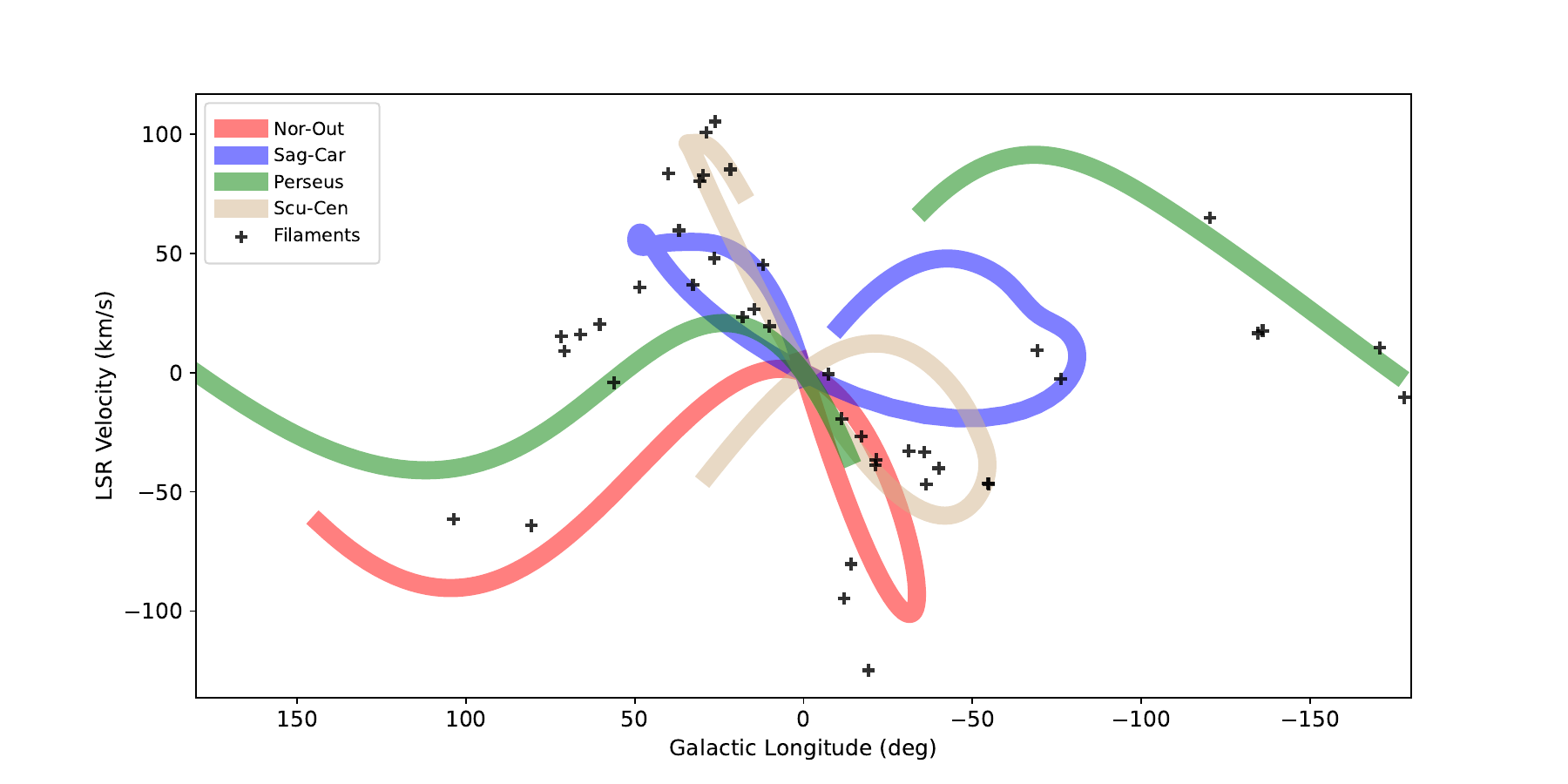}
\includegraphics[width=.35\textwidth, trim={0 4cm 0 4cm},clip]{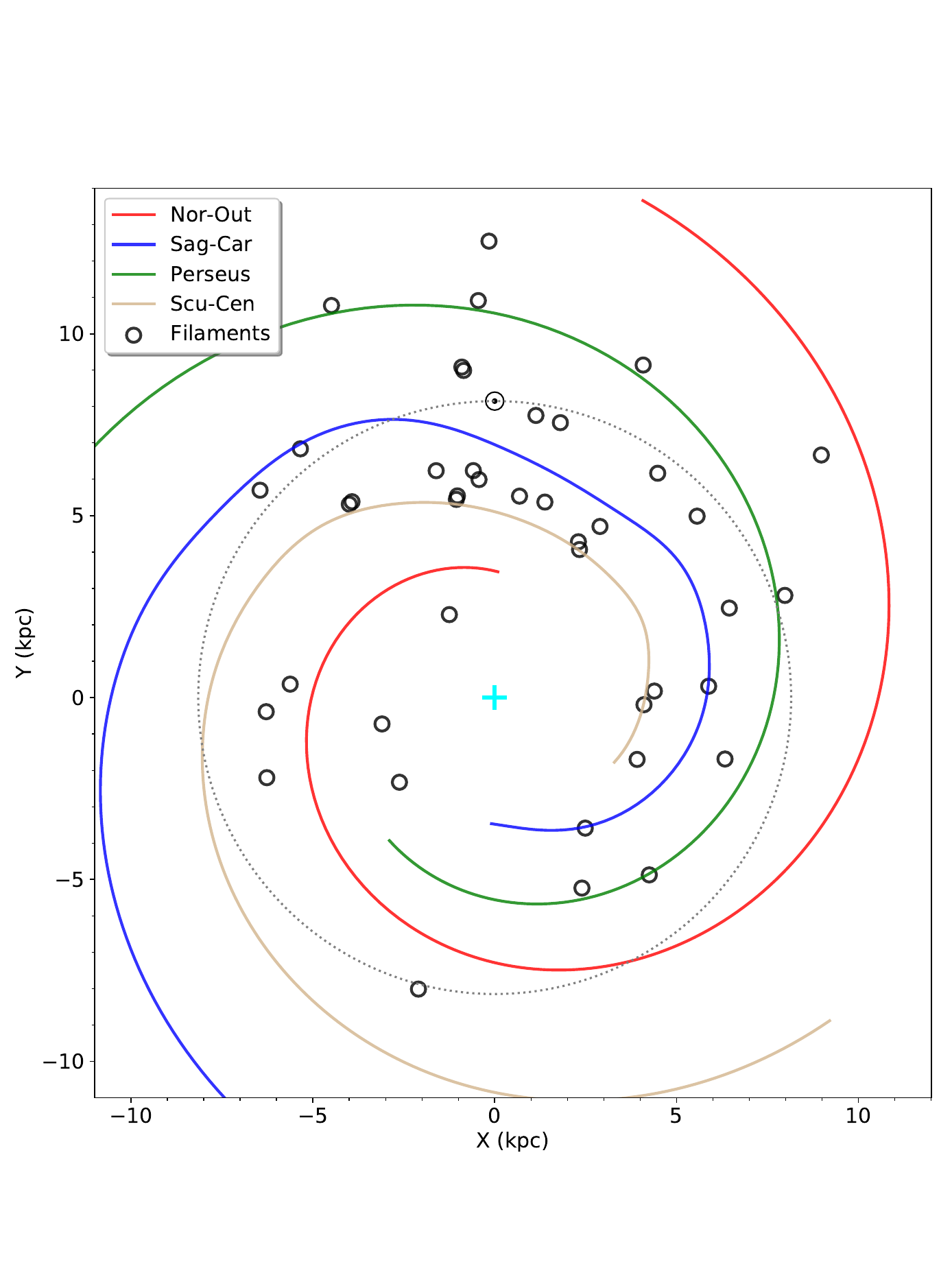}
\caption{Galactic distribution of the 42 linear filaments compared with spiral arms of the \cite{Taylor1993} model, with a width of 10\kms. Upper panel shows the Galactic longitude vs. LSR velocity, and the bottom panel shows a ``face-on'' view of the Milky Way, where the Sun ($\odot$) and the Galactic center ($+$) are marked. The dashed circle marks the solar circle that separates inner and outer Galaxy.
}
\label{fig3:gal}
\end{figure}

\subsection{Galactic Distribution and Association with Spiral Arms} \label{sec:GalDistrib}

The 42 linear filaments are located in all the four Galactic quadrants, concentrating in the I (20 filaments) and IV (16 filaments) quadrants, while the II and III quadrants have only 2 and 5 filaments, respectively (\autoref{fig2:hist}, \autoref{fig3:gal}). 
Interestingly, no linear filaments are found near the Galactic center, in the Central Molecular Zone \citep[CMZ,][]{LuX19}, and at low longitudes of $-8^{\circ} <l< 10^{\circ}$.
The averaged Galactic longitude and latitude of the filaments are
$(l,b)=(-8.92, -0.05)$ deg. Distributions of both $l$ and $b$ slightly skew towards negative side (Skewness $S_l=-0.9, S_b=-0.2$). The $l$ distribution is slightly more centrally peaked than normal distribution, and the $b$ is flatter ($K_l=0.5, K_b=-0.1$).
After taking distance into account, most filaments are found within the solar circle, with 31 in the inner Galaxy and 11 in the outer Galaxy (\autoref{fig3:gal}).

\autoref{fig2:hist} shows that
the filaments are concentrated on the Galactic mid-plane ($K_z=4.1$), most of which are within height $|z|<100$ pc; only 7 of them are located at a larger height up to 320 pc.
In contrast, the orientation of the filaments with respect to Galactic mid-plane ($\theta$) do not show obvious preference ($K_\theta=-1.5, S_\theta=0$), i.e., these well-defined filaments appear to be randomly orientated
\citep[c.f. bimodal distribution in][]{Ge24-BimodalFL}. Notably, a few filaments extend perpendicular to the Galactic plane (e.g., F4 in \autoref{fig1:fl5PV}).
These properties are important to further understanding the formation of these filaments \citep[e.g.][]{FengJC24-FLsim,Zucker19}.

We compare the filaments to spiral arms following the criteria of Paper \citetalias{me16}, \citetalias{Ge22}, \citetalias{Ge23}. We adopt the \cite{Taylor1993} spiral arm model, because it is derived based on measurements in all the four Galactic quadrants. More recent models do not have sufficient observations in the southern sky (Paper \citetalias{Ge23}).
A filament is considered to be associated with a spiral arm when its LSR velocity is within $\pm$5\kms\ from the spiral arms in the Galactic longitude-velocity tracks (\autoref{fig3:gal}).
This results 14 filaments associated with arms: 5 on Perseus, 4 on Sagittarius-Carina, 3 on Scutum-Centaurus, and 2 on Norma-Outer arm.
We note that these 14 on-arm filaments have rather spread orientation angles (similar to the entire sample), with only 5 filaments having $\theta \le 30^{\circ}$, i.e., running almost parallel to the mid-plane (after projection).
Surprisingly, among these, only one (F36, central part of Nessie) fulfills additional criterion for a Galactic ``bone'', i.e., lying in the center of mid-plane ($|z| \le 20$ pc). Another 4 ``parallel filaments'' are offset higher to the mid-plane ($|z| = 27-50$\,pc).

\subsection{Velocity Oscillation along Filaments} \label{sec:VeloOscillation}
Our sample of filaments represent a string of clumps semi equally spaced along the filaments. This ``necklace'' configuration is a natural result of the ``sausage instability'' \citep{Chandra1953,Ostriker1964_FL,Fiege2000}. Once this configuration is formed, materials can flow along the filaments to feed the clumps, enhancing star formation activities therein. Indeed, fluctuation of velocity along filaments have been observed, consistent with mass flows \citep{Hacar2011,qz15,me18,Henshaw20NatAs}.

Our unbiased sample provides an ideal test bed to investigate whether mass flows are common in filaments, and further revealing their properties.
Of the 42 filaments, we are able to retrieve  molecular line cubes for 20 filaments from public surveys
\citep{Jackson2006-GRS,Barnes2015-ThrUMMS,Umemoto17-FUGIN,Schuller-SEDIGISM,Schuller21-SEDIGISM-DR1,Cubuk23MopraCOsurveyDR4final}. These provide $^{13}$CO(1-0), $^{12}$CO(1-0), or $^{13}$CO(2-1) cubes that well resolve the filaments. Few other filaments are also covered by the surveys, but the data quality are insufficient for our following analysis.
A position-velocity (PV) cut is extracted along the filaments by connecting the clumps. At a given position along the filament, the intensity weighted centroid velocity (moment 1) and dispersion (moment 2) are plotted in \autoref{fig1:fl5PV}(c).

All the 20 PV curves show some variation in velocity along the filaments, most of which appear to be semi-periodic. We first fit a global slope in the PV curve, and then perform periodicity analysis on the residual PV curve, following \cite{me18}. For 19/20 filaments, the periodicity analysis reveals one or two dominant periods ($P_{\rm v}$). 
(The only exception is F12 with a weak peak of $\sim$4 pc period.)
In 11 filaments, the primary period match well with $P_{\rm cl}$, defined as the mean separation between the clumps.
In the rest of 8 filaments, the secondary period match well with $P_{\rm cl}$.
So, in all but one cases, the PV curves are oscillating with either a primary or secondary period that match well with $P_{\rm cl}$.
The averaged ratio $P_{\rm v}/P_{\rm cl}$ is $1.0\pm 0.2$.
This is in striking consistent with the model proposed by \cite{Hacar2011}, and strongly suggest that gas flows are channelled by these linear filaments, and are feeding the clumps. Each of the clumps are potential or ongoing sites for star or star cluster formation (\S \ref{sec:DiverseSF}).
The global PV slope may represent net bulk gas motions from end to end of the filament, at a fitted rate of 0.0-0.35\kms/pc (average 0.09\kms/pc). Localized to the clumps, the velocity gradient reaches typically 10 times higher, due to velocity oscillation (\autoref{fig1:fl5PV}(c)).

\subsection{High Velocity-Coherence} \label{sec:Vcoherence}
With no exception, all of the filaments show remarkable coherence in velocity. Along the full lengths up to nearly 40 pc, the velocity difference between the maximum and minimum velocities of all the clumps in a given filament is in the range of 0.0-4.3 \kms\ (average 1.5 \kms). Normalized by length, this corresponds to only 0.0-0.5 \kms\ change of velocity per pc. 
Extreme cases are F27 (length 7.4 pc) and F24 (length 2.6 pc), having the same velocity for all clumps. Those are resulted from a low-resolution of the literature molecular line data used in \cite{Elia21-HiGAL360cat}, and warrant observations with higher resolution.

Note that we have limited $\Delta v <2$\kms\ for adjacent clumps in selection (\S \ref{sec:DataMethod}).
In principle these filaments could have much higher velocity differences of $\Delta v \times (N_{\rm clump}-1)$, which amounts to 8-20\kms\ for these filaments harbouring 5-11 clumps (\autoref{tab:fl42}). 
Thus, the extreme velocity-coherence observed here is not a direct outcome of the selection. Rather, it {may reflect} intrinsic dynamics of these extremely linear filaments. Numerical simulations \citep[e.g.][]{FengJC24-FLsim} can be employed to explore the link between linearity, coherence, {projection effect,} and general structure of filaments.

\subsection{Diverse Star Formation Activities} \label{sec:DiverseSF}

After the initial fragmentation, clumps continue to accumulate gas while stars are to form, leading to far-IR point sources along the filaments.
We use the $L/M$ ratio and 70\um\ point sources to characterize the star formation activities. 
Of the 42 filaments, 23 have a global luminosity-to-mass ratio $L/M<1$\,\lsun/\msun, 
excellent for early evolution \citep{me11,me15,survey:HiGAL-DR1-Molinari2016,YuanJH17-HMSC,LuX19}. These filaments have a mean dust temperature of 8.6-14.5 K (\autoref{fig2:hist}).
16 of the filaments have intermediate $L/M$ in range 1.3-7.7\,\lsun/\msun,
and the remaining 3 of them have developed to 21.6-27.9\,\lsun/\msun.
We note that filaments without 70\um\ point sources are young ($L/M<1$\,\lsun/\msun, with one exception of $L/M=1.7$\,\lsun/\msun); all the three evolved filaments ($L/M>20$\,\lsun/\msun) have multiple 70\um\ point sources. On the other hand, filaments with multiple 70\um\ point sources can also be young in terms of global $L/M$. For example, F36 (central part of Nessie) has 7 point sources detected, but has a global $L/M<1$\,\lsun/\msun.
These suggest that far-IR point sources are efficient to reveal deeply embedded star formation activities. In particular, HiGAL is sensitive enough to detect 70\um\ sources, indicator of internal heating from accreting young protostars, at all distances of the filaments in this study, up to 16 kpc (e.g., F42).

\begin{figure}
\centering
\includegraphics[width=0.45\textwidth]{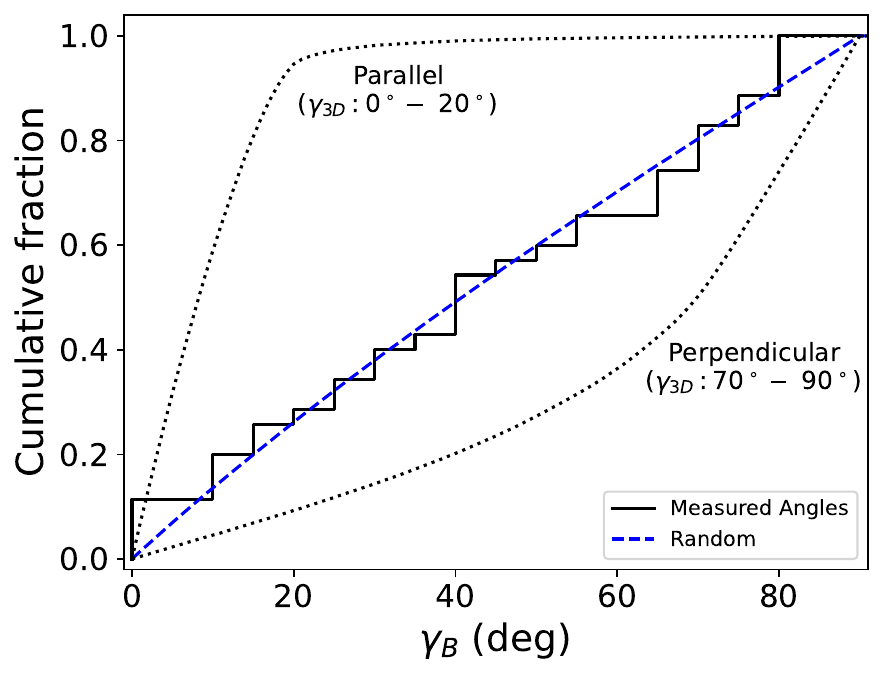}
\caption{Cumulative distribution function (CDF) of plane of sky projected angles between the filaments and the ambient B-fields ($\gamma_B$). The blue dashed line shows the CDF of randomly distributed $\gamma_{3D}$ values. The two black dotted lines show expected CDF for parallel ($\gamma_{3D} \sim$0$^\circ$--20$^\circ$) and perpendicular distributions ($\gamma_{3D} \sim$70$^\circ$--90$^\circ$).} 
\label{fig4:Bfield}
\end{figure}

\subsection{Filament Orientation Compared to Magnetic Fields}
\label{sec:B-fields}
Geometrical configuration and alignment of the magnetic fields (hereafter B-fields) with filamentary structures could hint about
the formation mechanisms of the filaments. Both numerical and observational studies have shown that B-fields are preferably oriented
either perpendicular or parallel to the filaments depending on the column density ($N_{\rm{H_2}}$) of the filaments and strength of
B-field \citep[e.g.,][]{liHB13, soler19,Pillai20NatAs-SOFIA,Jiao24}. The switch over from parallel to perpendicular orientation takes
place at an $N_{\mathrm{H_2}}$ of 10$^{21}-$10$^{22}$ cm$^{-2}$. Simulations of \citet{soler17} showed that the transition from a
parallel to a perpendicular field orientation is introduced by anisotropic converging flows regulated by a dynamically important magnetic
field on larger scales. It is thus important to explore the relative orientation of filaments in reference to the ambient B-fields.

Plane-of-sky orientation of B-fields toward our target filaments were estimated using the {\sl Planck}
 353 GHz (870$\mu$m) dust continuum polarization observations. Stokes $I, Q, U$ maps were extracted from the
 {\sl Planck} Public Data Release 2 \citep{planck16-PolarizationData,planck16b}. The maps have a pixel scale of $\sim1\arcmin$ and beam size of $\sim5\arcmin$.
As our sole purpose of using {\sl Planck} polarization data is obtaining the mean B-field orientation toward the filaments, the estimation of the position angles of the B-field was performed over \TB{a box with length equal to the length of the filament listed in \autoref{tab:fl42} and a width of 6$'$} around each filament. More technical details 
can be found in \citet{baug20}. 

{A cumulative histogram of the relative 2-dimensional position angles between the filaments and the corresponding B-fields ($\gamma_{B}$) is shown in \autoref{fig4:Bfield}. Note that the calculated position angles are 2-dimensional projections of actual 3-dimensional angles. To examine the effect of projection on observed distribution of $\gamma_{B}$, we simulated about two million radially outward pairs of
unit vectors randomly distributed on the surface of a sphere \citep[see][for more details]{stephens17, baug20}. Subsequently, we calculated their actual 3-dimensional angles ($\gamma_{3D}$) and also their 2-dimensional projection ($\gamma_{2D}$), assuming they are projected onto the y–z plane. These $\gamma_{2D}$ values could be considered as equivalent to the observed $\gamma_{B}$. Finally, we generated three different cumulative distribution functions (CDFs) of $\gamma_{2D}$, based on their actual 3-dimensional position angles -- (1) parallel, with $\gamma_{3D}$ ranging between 0$^\circ -$20$^\circ$; (2) perpendicular, with $\gamma_{3D}$ ranging between 70$^\circ -$90$^\circ$; and (3) random, with $\gamma_{3D}$ for all possible angels ranging between 0$^\circ -$90$^\circ$. Along with the cumulative histogram of $\gamma_{B}$, three CDFs are also shown in \autoref{fig4:Bfield} for comparison.}

No specific trend is noted in the \TB{cumulative} histogram indicating toward a non-correlation of filaments with the large-scale B-field orientation. This is broadly consistent with the SOFIA (Stratospheric Observatory for Infrared Astronomy) observations of filament G47 \citep{Stephens22-SOFIA_G47Pol}.
{However, we caution that the polarization data could largely be affected by line-of-sight depolarization \citep{planck16b} due to unknown B-field structure within the volume sampled by the beam. Also, averaging of multiple emitting layers with different position angles along the line-of-sight may lead to depolarization \citep{Jones2015-Bfield-GrainAlignment}. Note that the angular resolution of the \textit{Planck} data is comparable to the size of most of the filaments in our sample.} The orientation of B-fields could vary substantially at the scale (and sub-scale) of filaments than they are in the larger scales, and the \textit{Planck} data is incapable of resolving any such existing trends for the filaments in our sample. Dust polarization data at higher resolution are needed for a better comparison \citep[e.g.][]{Stephens22-SOFIA_G47Pol}.

\section{Summary}

We present the first catalog of 42 large-scale linear filaments across the full Galactic plane, and highlight some fundamental properties of them, including extreme linearity and velocity-coherence, velocity oscillation and filamentary gas flows, and diverse star formation activities. 1/3 of them are associated with the spiral arms, and 2/3 of them are located inter-arm, but only one is located at the arm center (a.k.a. Milky Way ``bone''); none found near the Galactic center and in the Central Molecular Zone. A few filaments extend perpendicular to the Galactic plane while no obvious trend is found in orientation of the full sample. The filaments also appear to be randomly orientated compared to global magnetic fields.

As linear filaments are building blocks for more complex filamentary structure, these properties are important for further studies towards a unified understanding of more general filamentary structure, their role in star formation, and the structure of the Milky Way ISM.

\begin{acknowledgements}
This work has been supported by the National Science Foundation of China (12041305),
the National Key R\&D Program of China (2022YFA1603100),
the China Manned Space Project (CMS-CSST-2021-A09, CMS-CSST-2021-B06), the Tianchi Talent Program of Xinjiang Uygur Autonomous Region, and the China-Chile Joint Research Fund (CCJRF No. 2211). CCJRF is provided by Chinese Academy of Sciences South America Center for Astronomy (CASSACA) and established by National Astronomical Observatories, Chinese Academy of Sciences (NAOC) and Chilean Astronomy Society (SOCHIAS) to support China-Chile collaborations in astronomy.
\end{acknowledgements}

\bibliography{my.astro}{}

\begin{thebibliography}{}
\expandafter\ifx\csname natexlab\endcsname\relax\def\natexlab#1{#1}\fi
\providecommand{\url}[1]{\href{#1}{#1}}
\providecommand{\dodoi}[1]{doi:~\href{http://doi.org/#1}{\nolinkurl{#1}}}
\providecommand{\doeprint}[1]{\href{http://ascl.net/#1}{\nolinkurl{http://ascl.net/#1}}}
\providecommand{\doarXiv}[1]{\href{https://arxiv.org/abs/#1}{\nolinkurl{https://arxiv.org/abs/#1}}}

\bibitem[{{Barnes} {et~al.}(2015){Barnes}, {Muller}, {Indermuehle}, {O'Dougherty}, {Lowe}, {Cunningham}, {Hernandez}, \& {Fuller}}]{Barnes2015-ThrUMMS}
{Barnes}, P.~J., {Muller}, E., {Indermuehle}, B., {et~al.} 2015, \apj, 812, 6, \dodoi{10.1088/0004-637X/812/1/6}

\bibitem[{{Baug} {et~al.}(2020){Baug}, {Wang}, {Liu}, {Tang}, {Zhang}, {Li}, {Eswaraiah}, {Liu}, {Tej}, {Goldsmith}, {Bronfman}, {Qin}, {T{\'o}th}, {Li}, \& {Kim}}]{baug20}
{Baug}, T., {Wang}, K., {Liu}, T., {et~al.} 2020, \apj, 890, 44, \dodoi{10.3847/1538-4357/ab66b6}

\bibitem[{{Berry}(2015)}]{soft:FellWalker}
{Berry}, D.~S. 2015, Astronomy and Computing, 10, 22, \dodoi{10.1016/j.ascom.2014.11.004}

\bibitem[{{Bhadari} {et~al.}(2022){Bhadari}, {Dewangan}, {Ojha}, {Pirogov}, \& {Maity}}]{Bhadari22-EdgeCollapse}
{Bhadari}, N.~K., {Dewangan}, L.~K., {Ojha}, D.~K., {Pirogov}, L.~E., \& {Maity}, A.~K. 2022, \apj, 930, 169, \dodoi{10.3847/1538-4357/ac65e9}

\bibitem[{{Chandrasekhar} \& {Fermi}(1953)}]{Chandra1953}
{Chandrasekhar}, S., \& {Fermi}, E. 1953, \apj, 118, 116, \dodoi{10.1086/145732}

\bibitem[{{Cubuk} {et~al.}(2023){Cubuk}, {Burton}, {Braiding}, {Wong}, {Rowell}, {Maxted}, {Eden}, {Alsaberi}, {Blackwell}, {Enokiya}, {Feijen}, {Filipovi{\'c}}, {Freeman}, {Fujita}, {Ghavam}, {Gunay}, {Indermuehle}, {Hayashi}, {Kohno}, {Nagaya}, {Nishimura}, {Okawa}, {Rebolledo}, {Romano}, {Sano}, {Snoswell}, {Tothill}, {Tsuge}, {Voisin}, {Yamane}, \& {Yoshiike}}]{Cubuk23MopraCOsurveyDR4final}
{Cubuk}, K.~O., {Burton}, M.~G., {Braiding}, C., {et~al.} 2023, \pasa, 40, e047, \dodoi{10.1017/pasa.2023.44}

\bibitem[{{Elia} {et~al.}(2017){Elia}, {Molinari}, {Schisano}, {Pestalozzi}, {Pezzuto}, {Merello}, {Noriega-Crespo}, {Moore}, {Russeil}, {Mottram}, {Paladini}, {Strafella}, {Benedettini}, {Bernard}, {Di Giorgio}, {Eden}, {Fukui}, {Plume}, {Bally}, {Martin}, {Ragan}, {Jaffa}, {Motte}, {Olmi}, {Schneider}, {Testi}, {Wyrowski}, {Zavagno}, {Calzoletti}, {Faustini}, {Natoli}, {Palmeirim}, {Piacentini}, {Piazzo}, {Pilbratt}, {Polychroni}, {Baldeschi}, {Beltr{\'a}n}, {Billot}, {Cambr{\'e}sy}, {Cesaroni}, {Garc{\'\i}a-Lario}, {Hoare}, {Huang}, {Joncas}, {Liu}, {Maiolo}, {Marsh}, {Maruccia}, {M{\`e}ge}, {Peretto}, {Rygl}, {Schilke}, {Thompson}, {Traficante}, {Umana}, {Veneziani}, {Ward-Thompson}, {Whitworth}, {Arab}, {Bandieramonte}, {Becciani}, {Brescia}, {Buemi}, {Bufano}, {Butora}, {Cavuoti}, {Costa}, {Fiorellino}, {Hajnal}, {Hayakawa}, {Kacsuk}, {Leto}, {Li Causi}, {Marchili}, {Martinavarro-Armengol}, {Mercurio}, {Molinaro}, {Riccio}, {Sano}, {Sciacca}, {Tachihara}, {Torii}, {Trigilio}, {Vitello}, \&
  {Yamamoto}}]{Elia17-HiGALcatInner}
{Elia}, D., {Molinari}, S., {Schisano}, E., {et~al.} 2017, \mnras, 471, 100, \dodoi{10.1093/mnras/stx1357}

\bibitem[{{Elia} {et~al.}(2021){Elia}, {Merello}, {Molinari}, {Schisano}, {Zavagno}, {Russeil}, {M{\`e}ge}, {Martin}, {Olmi}, {Pestalozzi}, {Plume}, {Ragan}, {Benedettini}, {Eden}, {Moore}, {Noriega-Crespo}, {Paladini}, {Palmeirim}, {Pezzuto}, {Pilbratt}, {Rygl}, {Schilke}, {Strafella}, {Tan}, {Traficante}, {Baldeschi}, {Bally}, {di Giorgio}, {Fiorellino}, {Liu}, {Piazzo}, \& {Polychroni}}]{Elia21-HiGAL360cat}
{Elia}, D., {Merello}, M., {Molinari}, S., {et~al.} 2021, \mnras, 504, 2742, \dodoi{10.1093/mnras/stab1038}

\bibitem[{{Feng} {et~al.}(2024){Feng}, {Smith}, {Hacar}, {Clark}, \& {Seifried}}]{FengJC24-FLsim}
{Feng}, J., {Smith}, R.~J., {Hacar}, A., {Clark}, S.~E., \& {Seifried}, D. 2024, \mnras, 528, 6370, \dodoi{10.1093/mnras/stae407}

\bibitem[{{Fiege} \& {Pudritz}(2000)}]{Fiege2000}
{Fiege}, J.~D., \& {Pudritz}, R.~E. 2000, \apj, 544, 830, \dodoi{10.1086/317228}

\bibitem[{{Ge} {et~al.}(2024){Ge}, {Du}, \& {Yuan}}]{Ge24-BimodalFL}
{Ge}, W., {Du}, F., \& {Yuan}, L. 2024, \mnras, 529, 3060, \dodoi{10.1093/mnras/stae680}

\bibitem[{{Ge} \& {Wang}(2022)}]{Ge22}
{Ge}, Y., \& {Wang}, K. 2022, \apjs, 259, 36, \dodoi{10.3847/1538-4365/ac4a76}

\bibitem[{{Ge} {et~al.}(2023){Ge}, {Wang}, {Duarte-Cabral}, {Pettitt}, {Dobbs}, {S{\'a}nchez-Monge}, {Neralwar}, {Urquhart}, {Colombo}, {Dur{\'a}n-Camacho}, {Beuther}, {Bronfman}, {Rigby}, {Eden}, {Neupane}, {Barnes}, {Henning}, \& {Yang}}]{Ge23}
{Ge}, Y., {Wang}, K., {Duarte-Cabral}, A., {et~al.} 2023, \aap, 675, A119, \dodoi{10.1051/0004-6361/202245784}

\bibitem[{{Hacar} {et~al.}(2023){Hacar}, {Clark}, {Heitsch}, {Kainulainen}, {Panopoulou}, {Seifried}, \& {Smith}}]{Hacar23-PP7}
{Hacar}, A., {Clark}, S.~E., {Heitsch}, F., {et~al.} 2023, in Astronomical Society of the Pacific Conference Series, Vol. 534, Protostars and Planets VII, ed. S.~{Inutsuka}, Y.~{Aikawa}, T.~{Muto}, K.~{Tomida}, \& M.~{Tamura}, 153, \dodoi{10.48550/arXiv.2203.09562}

\bibitem[{{Hacar} \& {Tafalla}(2011)}]{Hacar2011}
{Hacar}, A., \& {Tafalla}, M. 2011, \aap, 533, A34, \dodoi{10.1051/0004-6361/201117039}

\bibitem[{{Henshaw} {et~al.}(2020){Henshaw}, {Kruijssen}, {Longmore}, {Riener}, {Leroy}, {Rosolowsky}, {Ginsburg}, {Battersby}, {Chevance}, {Meidt}, {Glover}, {Hughes}, {Kainulainen}, {Klessen}, {Schinnerer}, {Schruba}, {Beuther}, {Bigiel}, {Blanc}, {Emsellem}, {Henning}, {Herrera}, {Koch}, {Pety}, {Ragan}, \& {Sun}}]{Henshaw20NatAs}
{Henshaw}, J.~D., {Kruijssen}, J.~M.~D., {Longmore}, S.~N., {et~al.} 2020, Nature Astronomy, 4, 1064, \dodoi{10.1038/s41550-020-1126-z}

\bibitem[{{Jackson} {et~al.}(2010){Jackson}, {Finn}, {Chambers}, {Rathborne}, \& {Simon}}]{Jackson2010}
{Jackson}, J.~M., {Finn}, S.~C., {Chambers}, E.~T., {Rathborne}, J.~M., \& {Simon}, R. 2010, \apjl, 719, L185, \dodoi{10.1088/2041-8205/719/2/L185}

\bibitem[{{Jackson} {et~al.}(2006){Jackson}, {Rathborne}, {Shah}, {Simon}, {Bania}, {Clemens}, {Chambers}, {Johnson}, {Dormody}, {Lavoie}, \& {Heyer}}]{Jackson2006-GRS}
{Jackson}, J.~M., {Rathborne}, J.~M., {Shah}, R.~Y., {et~al.} 2006, \apjs, 163, 145, \dodoi{10.1086/500091}

\bibitem[{{Jiao} {et~al.}(2024){Jiao}, {Wang}, {Xu}, {Wang}, \& {Beuther}}]{Jiao24}
{Jiao}, W., {Wang}, K., {Xu}, F., {Wang}, C., \& {Beuther}, H. 2024, arXiv e-prints, arXiv:2403.04274, \dodoi{10.48550/arXiv.2403.04274}

\bibitem[{{Jones} {et~al.}(2015){Jones}, {Bagley}, {Krejny}, {Andersson}, \& {Bastien}}]{Jones2015-Bfield-GrainAlignment}
{Jones}, T.~J., {Bagley}, M., {Krejny}, M., {Andersson}, B.~G., \& {Bastien}, P. 2015, \aj, 149, 31, \dodoi{10.1088/0004-6256/149/1/31}

\bibitem[{{Koch} \& {Rosolowsky}(2015)}]{soft:FilFinder-Koch2015}
{Koch}, E.~W., \& {Rosolowsky}, E.~W. 2015, \mnras, 452, 3435, \dodoi{10.1093/mnras/stv1521}

\bibitem[{{Li} {et~al.}(2016){Li}, {Urquhart}, {Leurini}, {Csengeri}, {Wyrowski}, {Menten}, \& {Schuller}}]{LiGX16-FLcat}
{Li}, G.-X., {Urquhart}, J.~S., {Leurini}, S., {et~al.} 2016, \aap, 591, A5, \dodoi{10.1051/0004-6361/201527468}

\bibitem[{{Li} {et~al.}(2013){Li}, {Fang}, {Henning}, \& {Kainulainen}}]{liHB13}
{Li}, H.-b., {Fang}, M., {Henning}, T., \& {Kainulainen}, J. 2013, \mnras, 436, 3707, \dodoi{10.1093/mnras/stt1849}

\bibitem[{{Liu} {et~al.}(2018){Liu}, {Stutz}, \& {Yuan}}]{LiuHL18-straightFL}
{Liu}, H.-L., {Stutz}, A., \& {Yuan}, J.-H. 2018, \mnras, 478, 2119, \dodoi{10.1093/mnras/sty1270}

\bibitem[{{Lu} {et~al.}(2019){Lu}, {Mills}, {Ginsburg}, {Walker}, {Barnes}, {Butterfield}, {Henshaw}, {Battersby}, {Kruijssen}, {Longmore}, {Zhang}, {Bally}, {Kauffmann}, {Ott}, {Rickert}, \& {Wang}}]{LuX19}
{Lu}, X., {Mills}, E. A.~C., {Ginsburg}, A., {et~al.} 2019, \apjs, 244, 35, \dodoi{10.3847/1538-4365/ab4258}

\bibitem[{{Men'shchikov}(2013)}]{soft:getfilaments-Menshchikov2013}
{Men'shchikov}, A. 2013, \aap, 560, A63, \dodoi{10.1051/0004-6361/201321885}

\bibitem[{{Molinari} {et~al.}(2010){Molinari}, {Swinyard}, {Bally}, {Barlow}, {Bernard}, {Martin}, {Moore}, {Noriega-Crespo}, {Plume}, {Testi}, {Zavagno}, {Abergel}, {Ali}, {Andr{\'e}}, {Baluteau}, {Benedettini}, {Bern{\'e}}, {Billot}, {Blommaert}, {Bontemps}, {Boulanger}, {Brand}, {Brunt}, {Burton}, {Campeggio}, {Carey}, {Caselli}, {Cesaroni}, {Cernicharo}, {Chakrabarti}, {Chrysostomou}, {Codella}, {Cohen}, {Compiegne}, {Davis}, {de Bernardis}, {de Gasperis}, {Di Francesco}, {di Giorgio}, {Elia}, {Faustini}, {Fischera}, {Fukui}, {Fuller}, {Ganga}, {Garcia-Lario}, {Giard}, {Giardino}, {Glenn}, {Goldsmith}, {Griffin}, {Hoare}, {Huang}, {Jiang}, {Joblin}, {Joncas}, {Juvela}, {Kirk}, {Lagache}, {Li}, {Lim}, {Lord}, {Lucas}, {Maiolo}, {Marengo}, {Marshall}, {Masi}, {Massi}, {Matsuura}, {Meny}, {Minier}, {Miville-Desch{\^e}nes}, {Montier}, {Motte}, {M{\"u}ller}, {Natoli}, {Neves}, {Olmi}, {Paladini}, {Paradis}, {Pestalozzi}, {Pezzuto}, {Piacentini}, {Pomar{\`e}s}, {Popescu}, {Reach}, {Richer}, {Ristorcelli},
  {Roy}, {Royer}, {Russeil}, {Saraceno}, {Sauvage}, {Schilke}, {Schneider-Bontemps}, {Schuller}, {Schultz}, {Shepherd}, {Sibthorpe}, {Smith}, {Smith}, {Spinoglio}, {Stamatellos}, {Strafella}, {Stringfellow}, {Sturm}, {Taylor}, {Thompson}, {Tuffs}, {Umana}, {Valenziano}, {Vavrek}, {Viti}, {Waelkens}, {Ward-Thompson}, {White}, {Wyrowski}, {Yorke}, \& {Zhang}}]{Hi-GAL}
{Molinari}, S., {Swinyard}, B., {Bally}, J., {et~al.} 2010, \pasp, 122, 314, \dodoi{10.1086/651314}

\bibitem[{{Molinari} {et~al.}(2016){Molinari}, {Schisano}, {Elia}, {Pestalozzi}, {Traficante}, {Pezzuto}, {Swinyard}, {Noriega-Crespo}, {Bally}, {Moore}, {Plume}, {Zavagno}, {di Giorgio}, {Liu}, {Pilbratt}, {Mottram}, {Russeil}, {Piazzo}, {Veneziani}, {Benedettini}, {Calzoletti}, {Faustini}, {Natoli}, {Piacentini}, {Merello}, {Palmese}, {Del Grande}, {Polychroni}, {Rygl}, {Polenta}, {Barlow}, {Bernard}, {Martin}, {Testi}, {Ali}, {Andr{\'e}}, {Beltr{\'a}n}, {Billot}, {Carey}, {Cesaroni}, {Compi{\`e}gne}, {Eden}, {Fukui}, {Garcia-Lario}, {Hoare}, {Huang}, {Joncas}, {Lim}, {Lord}, {Martinavarro-Armengol}, {Motte}, {Paladini}, {Paradis}, {Peretto}, {Robitaille}, {Schilke}, {Schneider}, {Schulz}, {Sibthorpe}, {Strafella}, {Thompson}, {Umana}, {Ward-Thompson}, \& {Wyrowski}}]{survey:HiGAL-DR1-Molinari2016}
{Molinari}, S., {Schisano}, E., {Elia}, D., {et~al.} 2016, \aap, 591, A149, \dodoi{10.1051/0004-6361/201526380}

\bibitem[{{Myers}(2010)}]{Myers2010}
{Myers}, P.~C. 2010, \apj, 714, 1280, \dodoi{10.1088/0004-637X/714/2/1280}

\bibitem[{{Ostriker}(1964)}]{Ostriker1964_FL}
{Ostriker}, J. 1964, \apj, 140, 1056, \dodoi{10.1086/148005}

\bibitem[{{Pillai} {et~al.}(2020){Pillai}, {Clemens}, {Reissl}, {Myers}, {Kauffmann}, {Lopez-Rodriguez}, {Alves}, {Franco}, {Henshaw}, {Menten}, {Nakamura}, {Seifried}, {Sugitani}, \& {Wiesemeyer}}]{Pillai20NatAs-SOFIA}
{Pillai}, T. G.~S., {Clemens}, D.~P., {Reissl}, S., {et~al.} 2020, Nature Astronomy, 4, 1195, \dodoi{10.1038/s41550-020-1172-6}

\bibitem[{{Planck Collaboration} {et~al.}(2016{\natexlab{a}}){Planck Collaboration}, {Adam}, {Ade}, {Aghanim}, {Akrami}, {Alves}, {Arg{\"u}eso}, {Arnaud}, {Arroja}, {Ashdown}, {Aumont}, {Baccigalupi}, {Ballardini}, {Banday}, {Barreiro}, {Bartlett}, {Bartolo}, {Basak}, {Battaglia}, {Battaner}, {Battye}, {Benabed}, {Beno{\^\i}t}, {Benoit-L{\'e}vy}, {Bernard}, {Bersanelli}, {Bertincourt}, {Bielewicz}, {Bikmaev}, {Bock}, {B{\"o}hringer}, {Bonaldi}, {Bonavera}, {Bond}, {Borrill}, {Bouchet}, {Boulanger}, {Bucher}, {Burenin}, {Burigana}, {Butler}, {Calabrese}, {Cardoso}, {Carvalho}, {Casaponsa}, {Castex}, {Catalano}, {Challinor}, {Chamballu}, {Chary}, {Chiang}, {Chluba}, {Chon}, {Christensen}, {Church}, {Clemens}, {Clements}, {Colombi}, {Colombo}, {Combet}, {Comis}, {Contreras}, {Couchot}, {Coulais}, {Crill}, {Cruz}, {Curto}, {Cuttaia}, {Danese}, {Davies}, {Davis}, {de Bernardis}, {de Rosa}, {de Zotti}, {Delabrouille}, {Delouis}, {D{\'e}sert}, {Di Valentino}, {Dickinson}, {Diego}, {Dolag}, {Dole}, {Donzelli},
  {Dor{\'e}}, {Douspis}, {Ducout}, {Dunkley}, {Dupac}, {Efstathiou}, {Eisenhardt}, {Elsner}, {En{\ss}lin}, {Eriksen}, {Falgarone}, {Fantaye}, {Farhang}, {Feeney}, {Fergusson}, {Fernandez-Cobos}, {Feroz}, {Finelli}, {Florido}, {Forni}, {Frailis}, {Fraisse}, {Franceschet}, {Franceschi}, {Frejsel}, {Frolov}, {Galeotta}, {Galli}, {Ganga}, {Gauthier}, {G{\'e}nova-Santos}, {Gerbino}, {Ghosh}, {Giard}, {Giraud-H{\'e}raud}, {Giusarma}, {Gjerl{\o}w}, {Gonz{\'a}lez-Nuevo}, {G{\'o}rski}, {Grainge}, {Gratton}, {Gregorio}, {Gruppuso}, {Gudmundsson}, {Hamann}, {Handley}, {Hansen}, {Hanson}, {Harrison}, {Heavens}, {Helou}, {Henrot-Versill{\'e}}, {Hern{\'a}ndez-Monteagudo}, {Herranz}, {Hildebrandt}, {Hivon}, {Hobson}, {Holmes}, {Hornstrup}, {Hovest}, {Huang}, {Huffenberger}, {Hurier}, {Ili{\'c}}, {Jaffe}, {Jaffe}, {Jin}, {Jones}, {Juvela}, {Karakci}, {Keih{\"a}nen}, {Keskitalo}, {Khamitov}, {Kiiveri}, {Kim}, {Kisner}, {Kneissl}, {Knoche}, {Knox}, {Krachmalnicoff}, {Kunz}, {Kurki-Suonio}, {Lacasa}, {Lagache},
  {L{\"a}hteenm{\"a}ki}, {Lamarre}, {Langer}, {Lasenby}, {Lattanzi}, {Lawrence}, {Le Jeune}, {Leahy}, {Lellouch}, {Leonardi}, {Le{\'o}n-Tavares}, {Lesgourgues}, {Levrier}, {Lewis}, {Liguori}, {Lilje}, {Lilley}, {Linden-V{\o}rnle}, {Lindholm}, {Liu}, {L{\'o}pez-Caniego}, {Lubin}, {Ma}, {Mac{\'\i}as-P{\'e}rez}, {Maggio}, {Maino}, {Mak}, {Mandolesi}, {Mangilli}, {Marchini}, {Marcos-Caballero}, {Marinucci}, {Maris}, {Marshall}, {Martin}, {Martinelli}, {Mart{\'\i}nez-Gonz{\'a}lez}, {Masi}, {Matarrese}, {Mazzotta}, {McEwen}, {McGehee}, {Mei}, {Meinhold}, {Melchiorri}, {Melin}, {Mendes}, {Mennella}, {Migliaccio}, {Mikkelsen}, {Millea}, {Mitra}, {Miville-Desch{\^e}nes}, {Molinari}, {Moneti}, {Montier}, {Moreno}, {Morgante}, {Mortlock}, {Moss}, {Mottet}, {M{\"u}nchmeyer}, {Munshi}, {Murphy}, {Narimani}, {Naselsky}, {Nastasi}, {Nati}, {Natoli}, {Negrello}, {Netterfield}, {N{\o}rgaard-Nielsen}, {Noviello}, {Novikov}, {Novikov}, {Olamaie}, {Oppermann}, {Orlando}, {Oxborrow}, {Paci}, {Pagano}, {Pajot}, {Paladini},
  {Pandolfi}, {Paoletti}, {Partridge}, {Pasian}, {Patanchon}, {Pearson}, {Peel}, {Peiris}, {Pelkonen}, {Perdereau}, {Perotto}, {Perrott}, {Perrotta}, {Pettorino}, {Piacentini}, {Piat}, {Pierpaoli}, {Pietrobon}, {Plaszczynski}, {Pogosyan}, {Pointecouteau}, {Polenta}, {Popa}, {Pratt}, {Pr{\'e}zeau}, {Prunet}, {Puget}, {Rachen}, {Racine}, {Reach}, {Rebolo}, {Reinecke}, {Remazeilles}, {Renault}, {Renzi}, {Ristorcelli}, {Rocha}, {Roman}, {Romelli}, {Rosset}, {Rossetti}, {Rotti}, {Roudier}, {Rouill{\'e} d'Orfeuil}, {Rowan-Robinson}, {Rubi{\~n}o-Mart{\'\i}n}, {Ruiz-Granados}, {Rumsey}, {Rusholme}, {Said}, {Salvatelli}, {Salvati}, {Sandri}, {Sanghera}, {Santos}, {Saunders}, {Sauv{\'e}}, {Savelainen}, {Savini}, {Schaefer}, {Schammel}, {Scott}, {Seiffert}, {Serra}, {Shellard}, {Shimwell}, {Shiraishi}, {Smith}, {Souradeep}, {Spencer}, {Spinelli}, {Stanford}, {Stern}, {Stolyarov}, {Stompor}, {Strong}, {Sudiwala}, {Sunyaev}, {Sutter}, {Sutton}, {Suur-Uski}, {Sygnet}, {Tauber}, {Tavagnacco}, {Terenzi}, {Texier},
  {Toffolatti}, {Tomasi}, {Tornikoski}, {Tramonte}, {Tristram}, {Troja}, {Trombetti}, {Tucci}, {Tuovinen}, {T{\"u}rler}, {Umana}, {Valenziano}, {Valiviita}, {Van Tent}, {Vassallo}, {Vibert}, {Vidal}, {Viel}, {Vielva}, {Villa}, {Wade}, {Walter}, {Wandelt}, {Watson}, {Wehus}, {Welikala}, {Weller}, {White}, {White}, {Wilkinson}, {Yvon}, {Zacchei}, {Zibin}, \& {Zonca}}]{planck16-PolarizationData}
{Planck Collaboration}, {Adam}, R., {Ade}, P.~A.~R., {et~al.} 2016{\natexlab{a}}, \aap, 594, A1, \dodoi{10.1051/0004-6361/201527101}

\bibitem[{{Planck Collaboration} {et~al.}(2016{\natexlab{b}}){Planck Collaboration}, {Ade}, {Aghanim}, {Alves}, {Arnaud}, {Arzoumanian}, {Aumont}, {Baccigalupi}, {Banday}, {Barreiro}, {Bartolo}, {Battaner}, {Benabed}, {Benoit-L{\'e}vy}, {Bernard}, {Bern{\'e}}, {Bersanelli}, {Bielewicz}, {Bonaldi}, {Bonavera}, {Bond}, {Borrill}, {Bouchet}, {Boulanger}, {Bracco}, {Burigana}, {Calabrese}, {Cardoso}, {Catalano}, {Chamballu}, {Chiang}, {Christensen}, {Clements}, {Colombi}, {Colombo}, {Combet}, {Couchot}, {Crill}, {Curto}, {Cuttaia}, {Danese}, {Davies}, {Davis}, {de Bernardis}, {de Rosa}, {de Zotti}, {Delabrouille}, {Dickinson}, {Diego}, {Donzelli}, {Dor{\'e}}, {Douspis}, {Ducout}, {Dupac}, {Elsner}, {En{\ss}lin}, {Eriksen}, {Falgarone}, {Ferri{\`e}re}, {Finelli}, {Forni}, {Frailis}, {Fraisse}, {Franceschi}, {Frejsel}, {Galeotta}, {Galli}, {Ganga}, {Ghosh}, {Giard}, {Giraud-H{\'e}raud}, {Gjerl{\o}w}, {Gonz{\'a}lez-Nuevo}, {G{\'o}rski}, {Gregorio}, {Gruppuso}, {Guillet}, {Hansen}, {Hanson}, {Harrison},
  {Hern{\'a}ndez-Monteagudo}, {Herranz}, {Hildebrandt}, {Hivon}, {Hobson}, {Holmes}, {Huffenberger}, {Hurier}, {Jaffe}, {Jaffe}, {Jones}, {Juvela}, {Keskitalo}, {Kisner}, {Knoche}, {Kunz}, {Kurki-Suonio}, {Lagache}, {Lamarre}, {Lasenby}, {Lawrence}, {Leonardi}, {Levrier}, {Liguori}, {Lilje}, {Linden-V{\o}rnle}, {L{\'o}pez-Caniego}, {Lubin}, {Mac{\'\i}as-P{\'e}rez}, {Maffei}, {Mandolesi}, {Mangilli}, {Maris}, {Martin}, {Mart{\'\i}nez-Gonz{\'a}lez}, {Masi}, {Matarrese}, {Mazzotta}, {Melchiorri}, {Mendes}, {Mennella}, {Migliaccio}, {Mitra}, {Miville-Desch{\^e}nes}, {Moneti}, {Montier}, {Morgante}, {Mortlock}, {Munshi}, {Murphy}, {Naselsky}, {Nati}, {Natoli}, {N{\o}rgaard-Nielsen}, {Noviello}, {Novikov}, {Novikov}, {Oppermann}, {Pagano}, {Pajot}, {Paladini}, {Paoletti}, {Pasian}, {Perrotta}, {Pettorino}, {Piacentini}, {Piat}, {Pierpaoli}, {Pietrobon}, {Plaszczynski}, {Pointecouteau}, {Polenta}, {Pratt}, {Puget}, {Rachen}, {Rebolo}, {Reinecke}, {Remazeilles}, {Renault}, {Renzi}, {Ricciardi}, {Ristorcelli},
  {Rocha}, {Rosset}, {Rossetti}, {Roudier}, {Rubi{\~n}o-Mart{\'\i}n}, {Rusholme}, {Sandri}, {Savelainen}, {Savini}, {Scott}, {Soler}, {Stolyarov}, {Sutton}, {Suur-Uski}, {Sygnet}, {Tauber}, {Terenzi}, {Toffolatti}, {Tomasi}, {Tristram}, {Tucci}, {Valenziano}, {Valiviita}, {Van Tent}, {Vielva}, {Villa}, {Wade}, {Wandelt}, {Yvon}, {Zacchei}, \& {Zonca}}]{planck16b}
{Planck Collaboration}, {Ade}, P.~A.~R., {Aghanim}, N., {et~al.} 2016{\natexlab{b}}, \aap, 586, A136, \dodoi{10.1051/0004-6361/201425305}

\bibitem[{{Reid} {et~al.}(2016){Reid}, {Dame}, {Menten}, \& {Brunthaler}}]{Reid16-distance}
{Reid}, M.~J., {Dame}, T.~M., {Menten}, K.~M., \& {Brunthaler}, A. 2016, \apj, 823, 77, \dodoi{10.3847/0004-637X/823/2/77}

\bibitem[{{Reid} {et~al.}(2019){Reid}, {Menten}, {Brunthaler}, {Zheng}, {Dame}, {Xu}, {Li}, {Sakai}, {Wu}, {Immer}, {Zhang}, {Sanna}, {Moscadelli}, {Rygl}, {Bartkiewicz}, {Hu}, {Quiroga-Nu{\~n}ez}, \& {van Langevelde}}]{Reid19-distance}
{Reid}, M.~J., {Menten}, K.~M., {Brunthaler}, A., {et~al.} 2019, \apj, 885, 131, \dodoi{10.3847/1538-4357/ab4a11}

\bibitem[{{Schisano} {et~al.}(2020){Schisano}, {Molinari}, {Elia}, {Benedettini}, {Olmi}, {Pezzuto}, {Traficante}, {Brescia}, {Cavuoti}, {di Giorgio}, {Liu}, {Moore}, {Noriega-Crespo}, {Riccio}, {Baldeschi}, {Becciani}, {Peretto}, {Merello}, {Vitello}, {Zavagno}, {Beltr{\'a}n}, {Cambr{\'e}sy}, {Eden}, {Li Causi}, {Molinaro}, {Palmeirim}, {Sciacca}, {Testi}, {Umana}, \& {Whitworth}}]{Schisano20-HiGAL-FLcat}
{Schisano}, E., {Molinari}, S., {Elia}, D., {et~al.} 2020, \mnras, 492, 5420, \dodoi{10.1093/mnras/stz3466}

\bibitem[{{Schuller} {et~al.}(2017){Schuller}, {Csengeri}, {Urquhart}, {Duarte-Cabral}, {Barnes}, {Giannetti}, {Hernandez}, {Leurini}, {Mattern}, {Medina}, {Agurto}, {Azagra}, {Anderson}, {Beltr{\'a}n}, {Beuther}, {Bontemps}, {Bronfman}, {Dobbs}, {Dumke}, {Finger}, {Ginsburg}, {Gonzalez}, {Henning}, {Kauffmann}, {Mac-Auliffe}, {Menten}, {Montenegro-Montes}, {Moore}, {Muller}, {Parra}, {Perez-Beaupuits}, {Pettitt}, {Russeil}, {S{\'a}nchez-Monge}, {Schilke}, {Schisano}, {Suri}, {Testi}, {Torstensson}, {Venegas}, {Wang}, {Wienen}, {Wyrowski}, \& {Zavagno}}]{Schuller-SEDIGISM}
{Schuller}, F., {Csengeri}, T., {Urquhart}, J.~S., {et~al.} 2017, \aap, 601, A124, \dodoi{10.1051/0004-6361/201628933}

\bibitem[{{Schuller} {et~al.}(2021){Schuller}, {Urquhart}, {Csengeri}, {Colombo}, {Duarte-Cabral}, {Mattern}, {Ginsburg}, {Pettitt}, {Wyrowski}, {Anderson}, {Azagra}, {Barnes}, {Beltran}, {Beuther}, {Billington}, {Bronfman}, {Cesaroni}, {Dobbs}, {Eden}, {Lee}, {Medina}, {Menten}, {Moore}, {Montenegro-Montes}, {Ragan}, {Rigby}, {Riener}, {Russeil}, {Schisano}, {Sanchez-Monge}, {Traficante}, {Zavagno}, {Agurto}, {Bontemps}, {Finger}, {Giannetti}, {Gonzalez}, {Hernandez}, {Henning}, {Kainulainen}, {Kauffmann}, {Leurini}, {Lopez}, {Mac-Auliffe}, {Mazumdar}, {Molinari}, {Motte}, {Muller}, {Nguyen-Luong}, {Parra}, {Perez-Beaupuits}, {Schilke}, {Schneider}, {Suri}, {Testi}, {Torstensson}, {Veena}, {Venegas}, {Wang}, \& {Wienen}}]{Schuller21-SEDIGISM-DR1}
{Schuller}, F., {Urquhart}, J.~S., {Csengeri}, T., {et~al.} 2021, \mnras, 500, 3064, \dodoi{10.1093/mnras/staa2369}

\bibitem[{{Soler}(2019)}]{soler19}
{Soler}, J.~D. 2019, \aap, 629, A96, \dodoi{10.1051/0004-6361/201935779}

\bibitem[{{Soler} \& {Hennebelle}(2017)}]{soler17}
{Soler}, J.~D., \& {Hennebelle}, P. 2017, \aap, 607, A2, \dodoi{10.1051/0004-6361/201731049}

\bibitem[{{Stephens} {et~al.}(2017){Stephens}, {Dunham}, {Myers}, {Pokhrel}, {Sadavoy}, {Vorobyov}, {Tobin}, {Pineda}, {Offner}, {Lee}, {Kristensen}, {J{\o}rgensen}, {Goodman}, {Bourke}, {Arce}, \& {Plunkett}}]{stephens17}
{Stephens}, I.~W., {Dunham}, M.~M., {Myers}, P.~C., {et~al.} 2017, \apj, 846, 16, \dodoi{10.3847/1538-4357/aa8262}

\bibitem[{{Stephens} {et~al.}(2022){Stephens}, {Myers}, {Zucker}, {Jackson}, {Andersson}, {Smith}, {Soam}, {Battersby}, {Sanhueza}, {Hogge}, {Smith}, {Novak}, {Sadavoy}, {Pillai}, {Li}, {Looney}, {Sugitani}, {Coud{\'e}}, {Guzm{\'a}n}, {Goodman}, {Kusune}, {Santos}, {Zuckerman}, \& {Encalada}}]{Stephens22-SOFIA_G47Pol}
{Stephens}, I.~W., {Myers}, P.~C., {Zucker}, C., {et~al.} 2022, \apjl, 926, L6, \dodoi{10.3847/2041-8213/ac4d8f}

\bibitem[{{Taylor} \& {Cordes}(1993)}]{Taylor1993}
{Taylor}, J.~H., \& {Cordes}, J.~M. 1993, \apj, 411, 674, \dodoi{10.1086/172870}

\bibitem[{{Umemoto} {et~al.}(2017){Umemoto}, {Minamidani}, {Kuno}, {Fujita}, {Matsuo}, {Nishimura}, {Torii}, {Tosaki}, {Kohno}, {Kuriki}, {Tsuda}, {Hirota}, {Ohashi}, {Yamagishi}, {Handa}, {Nakanishi}, {Omodaka}, {Koide}, {Matsumoto}, {Onishi}, {Tokuda}, {Seta}, {Kobayashi}, {Tachihara}, {Sano}, {Hattori}, {Onodera}, {Oasa}, {Kamegai}, {Tsuboi}, {Sofue}, {Higuchi}, {Chibueze}, {Mizuno}, {Honma}, {Muller}, {Inoue}, {Morokuma-Matsui}, {Shinnaga}, {Ozawa}, {Takahashi}, {Yoshiike}, {Costes}, \& {Kuwahara}}]{Umemoto17-FUGIN}
{Umemoto}, T., {Minamidani}, T., {Kuno}, N., {et~al.} 2017, \pasj, 69, 78, \dodoi{10.1093/pasj/psx061}

\bibitem[{{Vall{\'e}e}(2016)}]{Vallee2016}
{Vall{\'e}e}, J.~P. 2016, \aj, 151, 55, \dodoi{10.3847/0004-6256/151/3/55}

\bibitem[{{Wang}(2018)}]{me18}
{Wang}, K. 2018, Research Notes of the American Astronomical Society, 2, 52, \dodoi{10.3847/2515-5172/aacb29}

\bibitem[{{Wang} {et~al.}(2016){Wang}, {Testi}, {Burkert}, {Walmsley}, {Beuther}, \& {Henning}}]{me16}
{Wang}, K., {Testi}, L., {Burkert}, A., {et~al.} 2016, \apjs, 226, 9, \dodoi{10.3847/0067-0049/226/1/9}

\bibitem[{{Wang} {et~al.}(2015){Wang}, {Testi}, {Ginsburg}, {Walmsley}, {Molinari}, \& {Schisano}}]{me15}
{Wang}, K., {Testi}, L., {Ginsburg}, A., {et~al.} 2015, \mnras, 450, 4043, \dodoi{10.1093/mnras/stv735}

\bibitem[{{Wang} {et~al.}(2011){Wang}, {Zhang}, {Wu}, \& {Zhang}}]{me11}
{Wang}, K., {Zhang}, Q., {Wu}, Y., \& {Zhang}, H. 2011, \apj, 735, 64, \dodoi{10.1088/0004-637X/735/1/64}

\bibitem[{{Williams} {et~al.}(1994){Williams}, {de Geus}, \& {Blitz}}]{soft:CLUMPFIND}
{Williams}, J.~P., {de Geus}, E.~J., \& {Blitz}, L. 1994, \apj, 428, 693, \dodoi{10.1086/174279}

\bibitem[{{Yuan} {et~al.}(2017){Yuan}, {Wu}, {Ellingsen}, {Evans}, {Henkel}, {Wang}, {Liu}, {Liu}, {Li}, \& {Zavagno}}]{YuanJH17-HMSC}
{Yuan}, J., {Wu}, Y., {Ellingsen}, S.~P., {et~al.} 2017, \apjs, 231, 11, \dodoi{10.3847/1538-4365/aa7204}

\bibitem[{{Yuan} {et~al.}(2020){Yuan}, {Li}, {Zhu}, {Liu}, {Wang}, {Liu}, {Kim}, {Tatematsu}, {Yuan}, \& {Wu}}]{Yuan20-EdgeCollapse}
{Yuan}, L., {Li}, G.-X., {Zhu}, M., {et~al.} 2020, \aap, 637, A67, \dodoi{10.1051/0004-6361/201936625}

\bibitem[{{Zhang} {et~al.}(2015){Zhang}, {Wang}, {Lu}, \& {Jim{\'e}nez-Serra}}]{qz15}
{Zhang}, Q., {Wang}, K., {Lu}, X., \& {Jim{\'e}nez-Serra}, I. 2015, \apj, 804, 141, \dodoi{10.1088/0004-637X/804/2/141}

\bibitem[{{Zucker} {et~al.}(2015){Zucker}, {Battersby}, \& {Goodman}}]{Zucker2015}
{Zucker}, C., {Battersby}, C., \& {Goodman}, A. 2015, \apj, 815, 23, \dodoi{10.1088/0004-637X/815/1/23}

\bibitem[{{Zucker} {et~al.}(2018){Zucker}, {Battersby}, \& {Goodman}}]{Zucker2018}
---. 2018, \apj, 864, 153, \dodoi{10.3847/1538-4357/aacc66}

\bibitem[{{Zucker} {et~al.}(2019){Zucker}, {Smith}, \& {Goodman}}]{Zucker19}
{Zucker}, C., {Smith}, R., \& {Goodman}, A. 2019, \apj, 887, 186, \dodoi{10.3847/1538-4357/ab517d}

\end{thebibliography}
\bibliographystyle{aasjournal}

\begin{appendix} 

\section{Supporting Materials}

\begin{figure*}
\centering
\includegraphics[width=.7\textwidth]{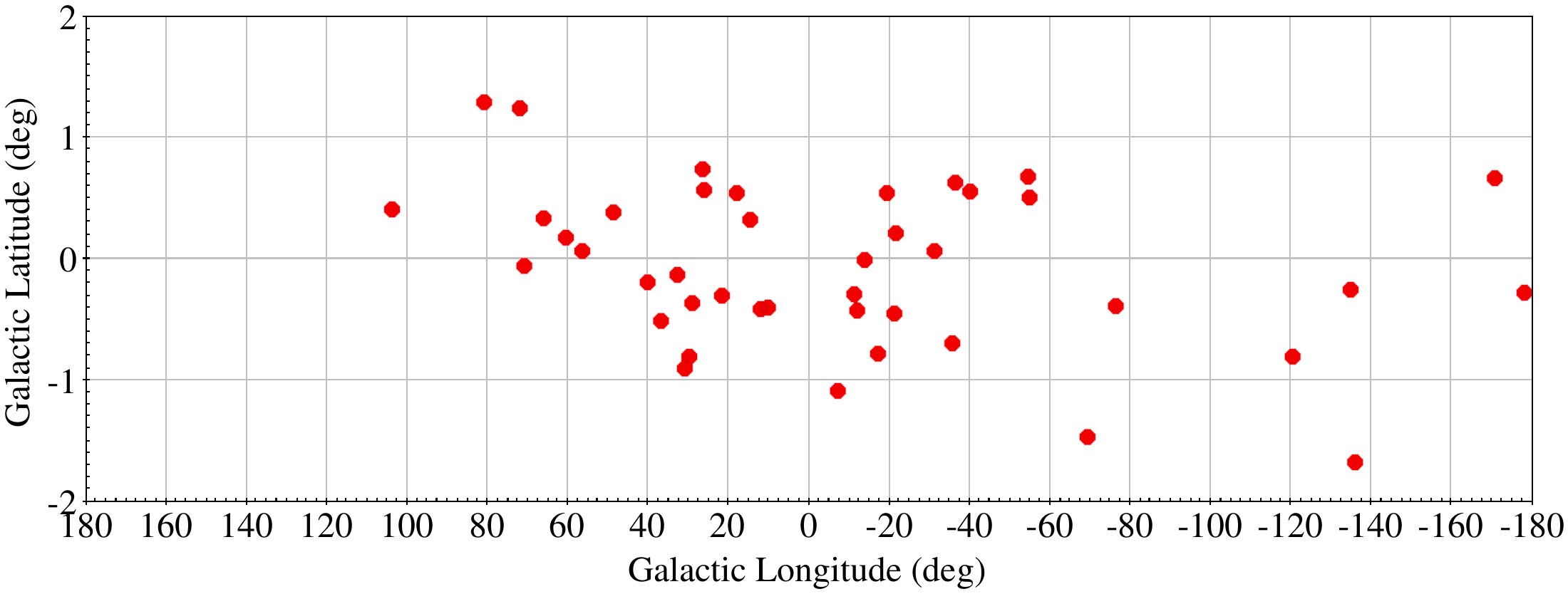}\\
\includegraphics[width=.33\textwidth]{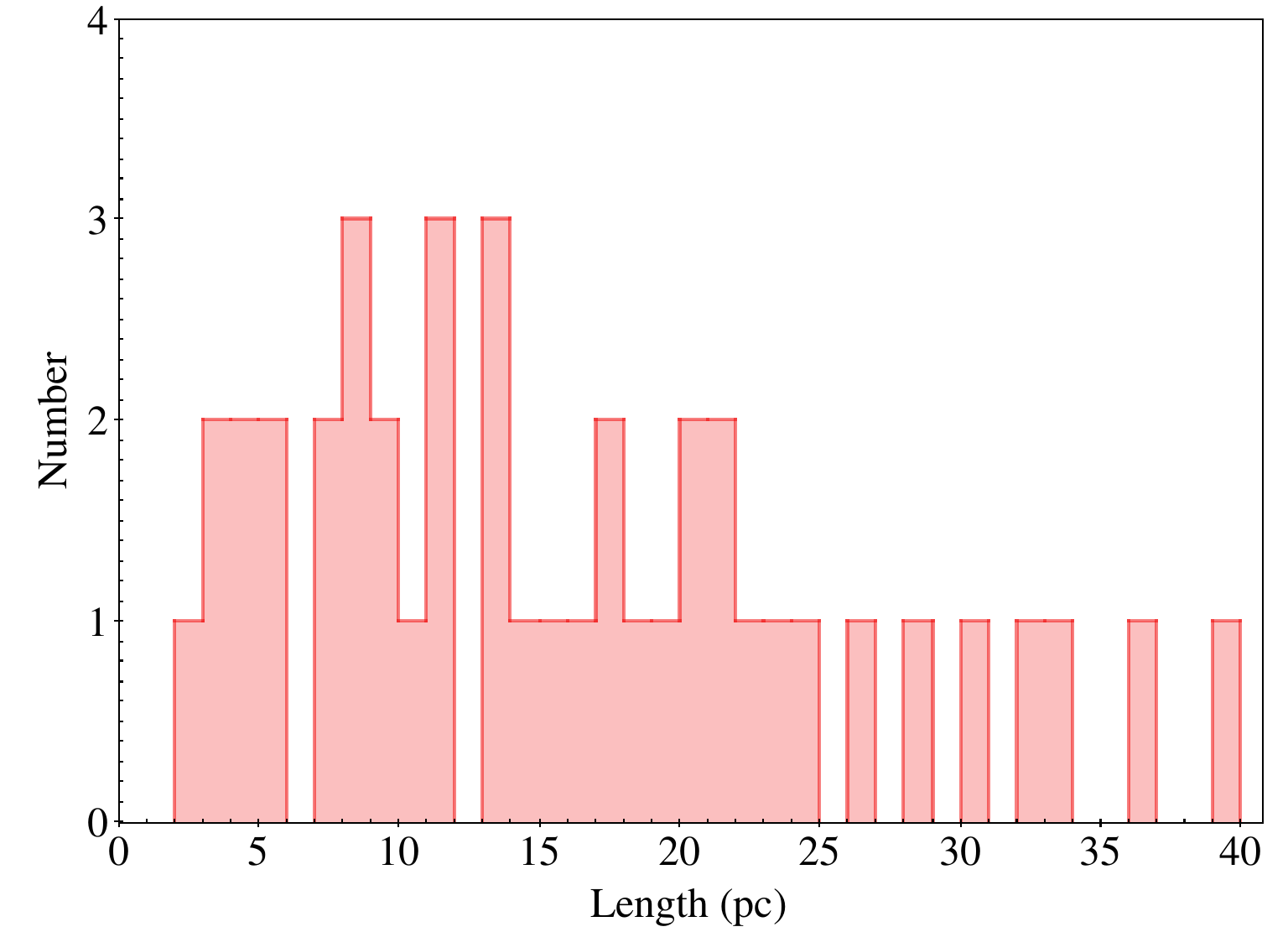}
\includegraphics[width=.33\textwidth]{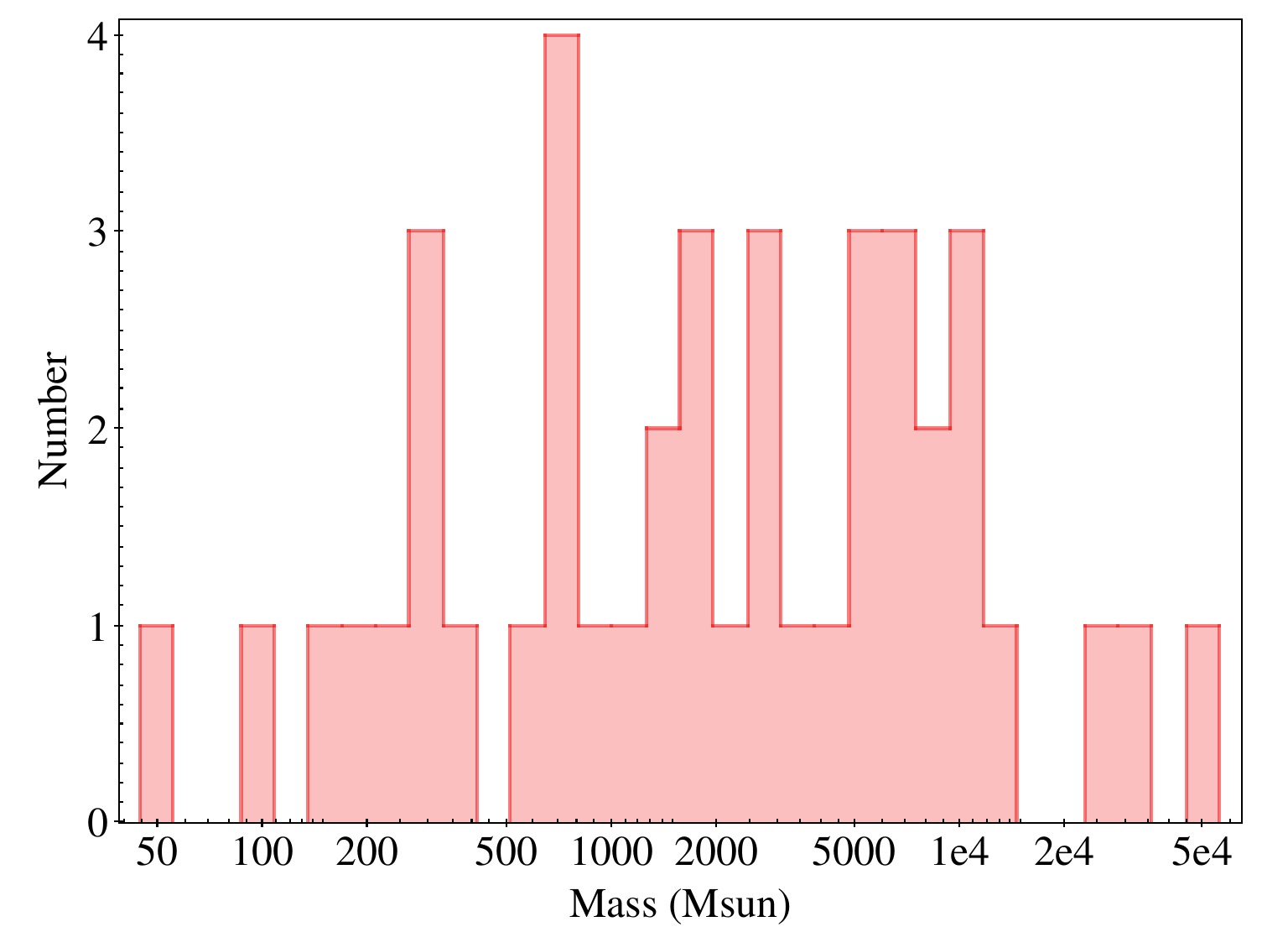}\\
\includegraphics[width=.33\textwidth]{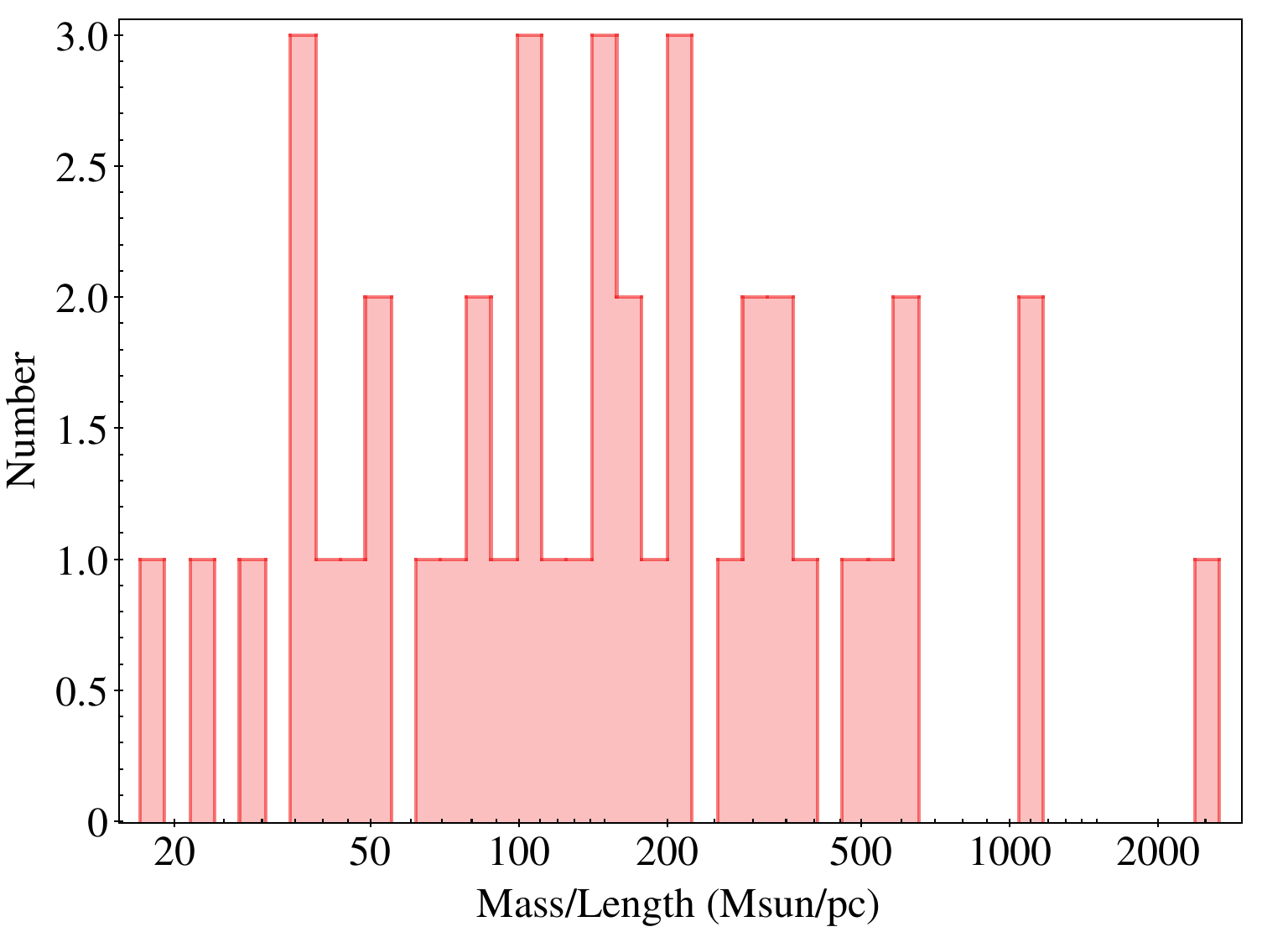}
\includegraphics[width=.33\textwidth]{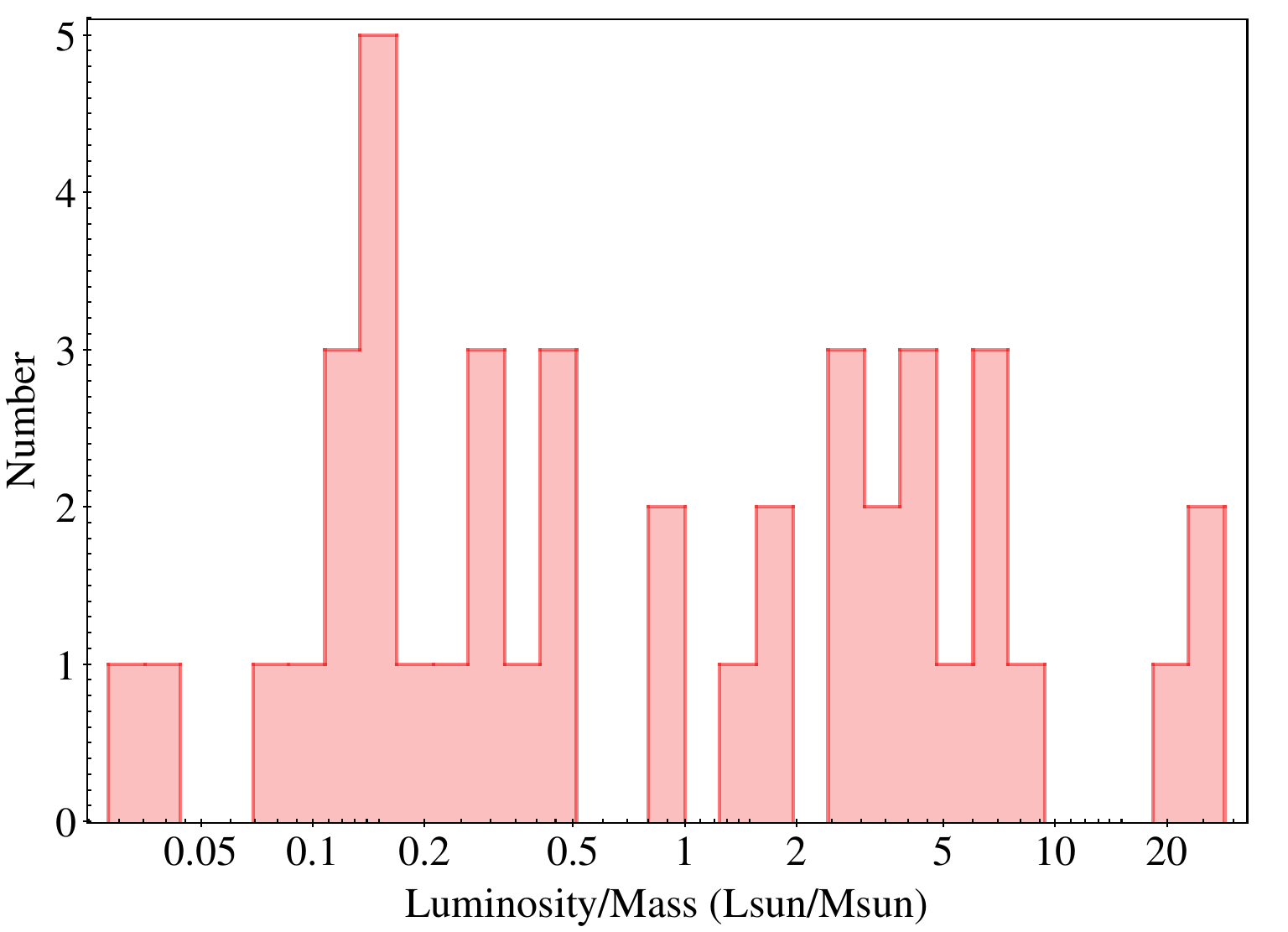}\\
\includegraphics[width=.33\textwidth]{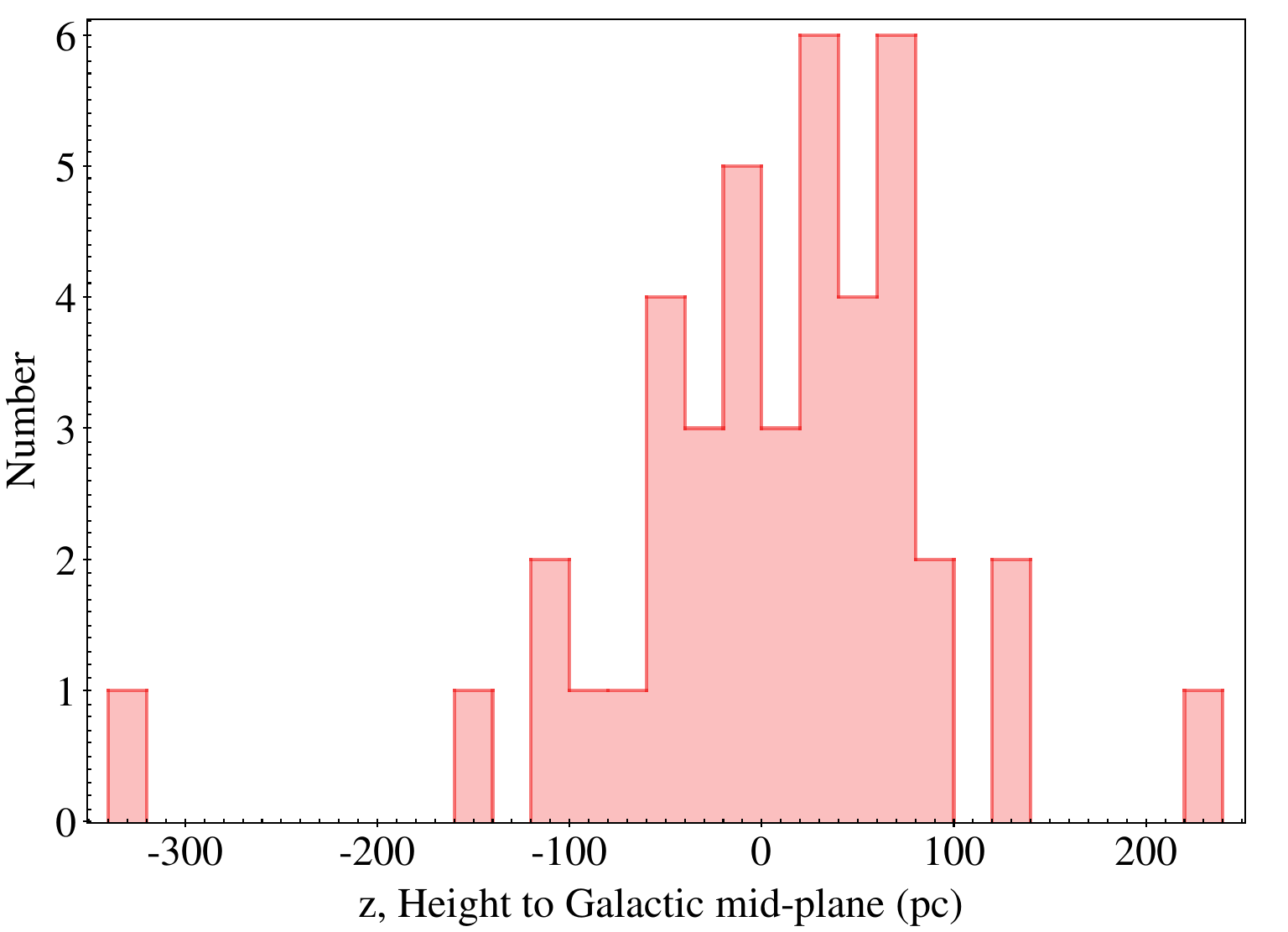}
\includegraphics[width=.33\textwidth]{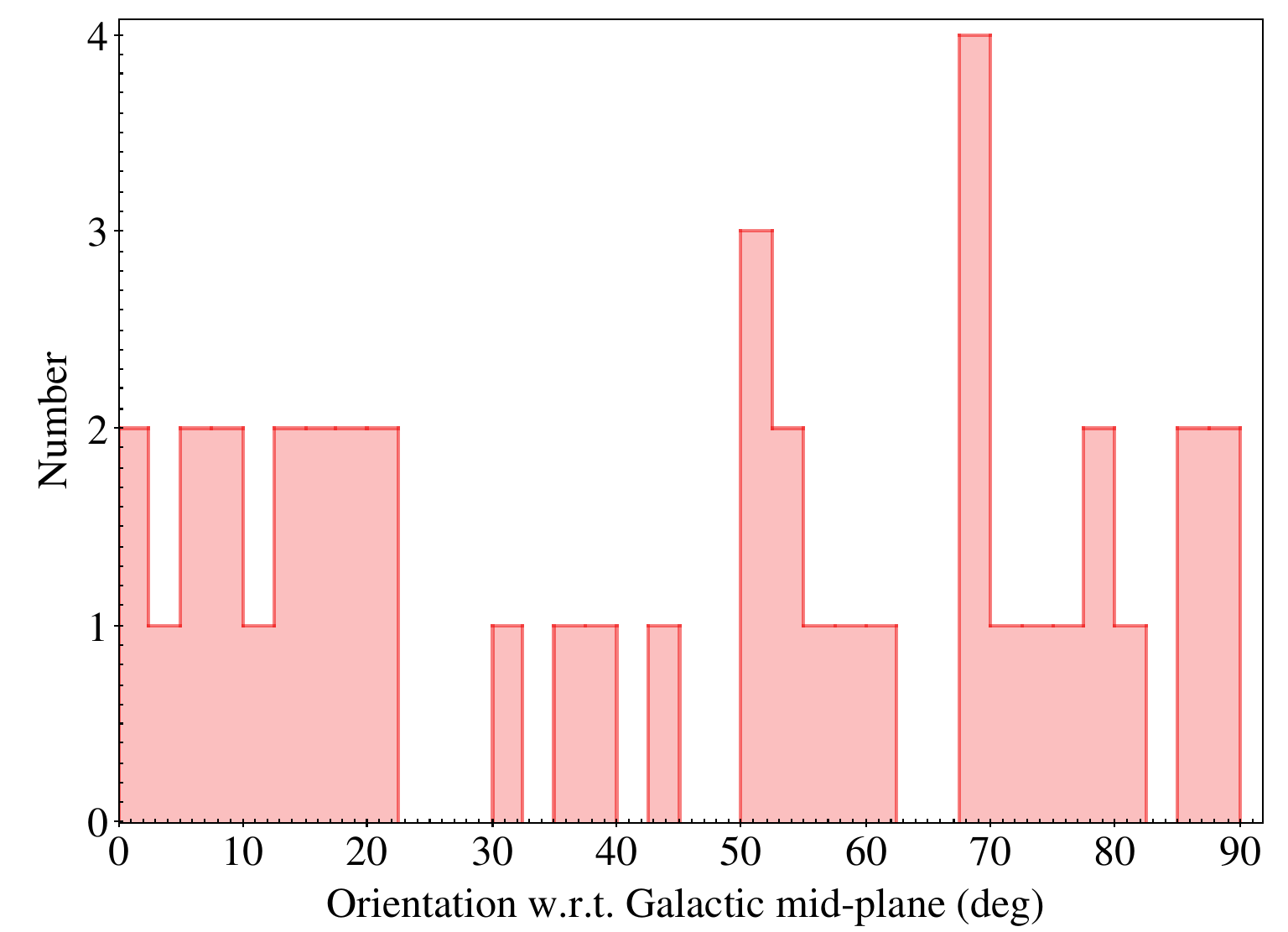}\\
\includegraphics[width=.33\textwidth]{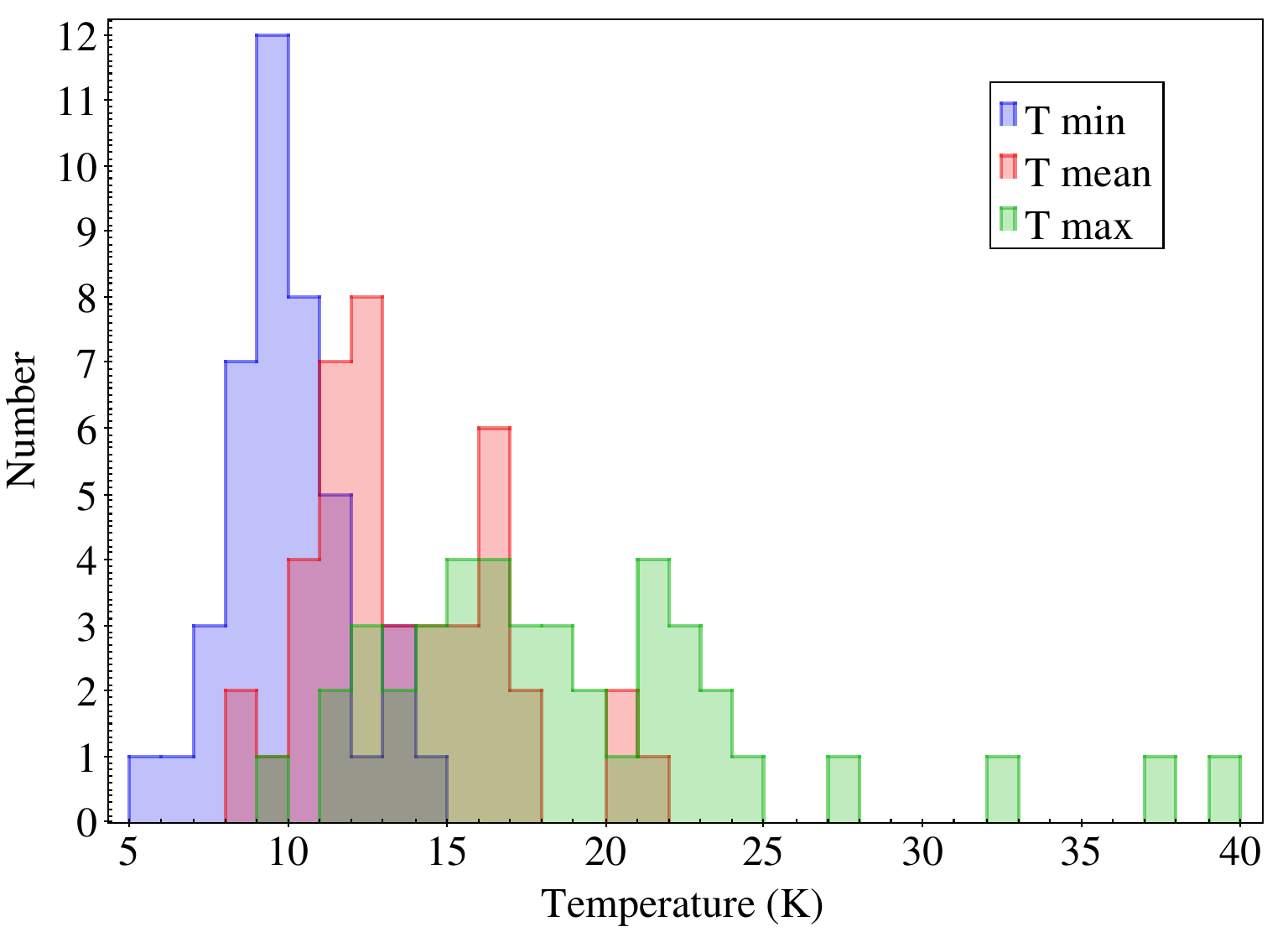}
\includegraphics[width=.33\textwidth]{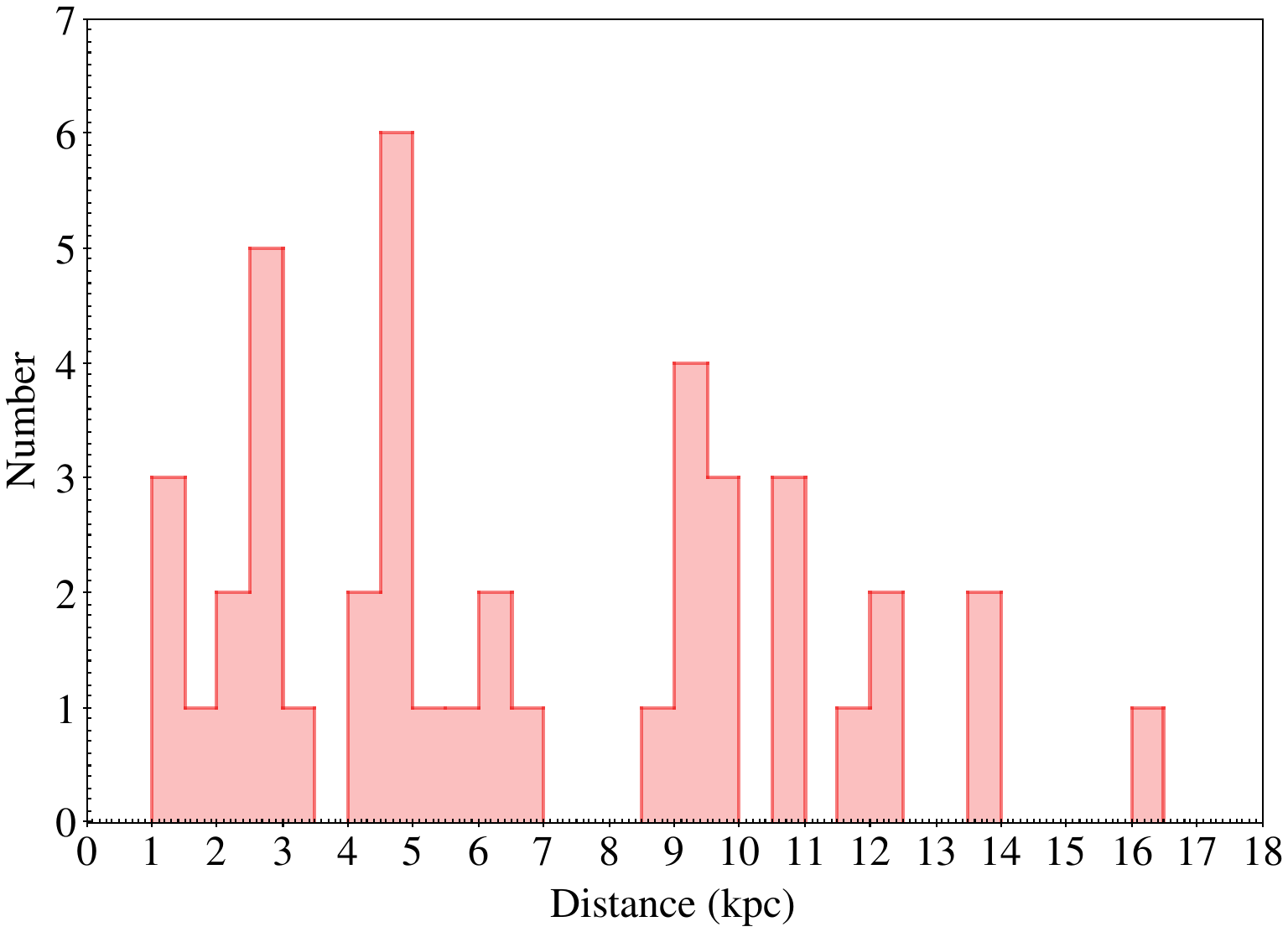}
\caption{Properties of the 42 linear filaments. Note that some parameters are plotted in log scale for better illustration at the long tails in distribution. The temperature histogram shows the minimum, maximum, and mean temperature of the clumps in filaments. Mass, mass/length ratio, and luminosity are lower limits (\S \ref{sec:GlobalProperties}).
}
\label{fig2:hist}
\end{figure*}

\autoref{tab:fl42} lists physical properties of the 42 filaments, and \autoref{fig2:hist} shows distribution of some parameters.
\autoref{fig:mst_a} presents the MST, and
\autoref{fig:rgb_a} illustrate the filaments in two-color far-IR views.

\setlength\tabcolsep{4pt}
\clearpage
\begin{landscape}
\capstartfalse
\begin{table}
\centering
\caption{Physical Properties of the Filaments \label{tab:fl42}}
\tiny
\begin{tabular}{crrrr rrrrr rrrrr rrrrr rrrrrr}
\hline
\hline\\
{ID} &
{$l_1$} &
{$b_1$} &
{$l_2$} &
{$b_2$} &
{$v_{\rm lsr}$} &
{$d$} &
{Len.} &
{Le2e} &
{Lmst} &
{$w$} &
{Aspect} &
{$\Delta v$} &
{Slope} &
{Mass} &
{Lum.} &
{$L/M$} &
{$T$} &
{$N_{70}$} &
{$N_{\rm cl}$} &
{Linearity} &
{$\theta$} &
{$z$} &
{Arm} &
{$P_{\rm v}$} &
{ $\frac{P_{\rm v}}{P_{\rm cl}}$ }  \\
{} &
{$^{\circ}$} &
{$^{\circ}$} &
{$^{\circ}$} &
{$^{\circ}$} &
{$\rm \frac{km}{s}$} &
{kpc} &
{$^{\circ}$} &
{pc} &
{pc} &
{pc} &
{ratio} &
{$\rm \frac{km}{s}$} &
{$\rm \frac{km/s}{pc}$} &
{\msun} &
{\lsun} &
{$\frac{L_\odot}{M_\odot}$} &
{K} &
{} &
{} &
{} &
{$^{\circ}$} &
{pc} &
{} &
{pc} &
{} \\
{(1)} 
& {(2)} & {(3)} & {(4)} 
& {(5)} & {(6)} 
& {(7)} 
& {(8)} & {(9)} 
& {(10)} & {(11)} 
& {(12)} & {(13)} 
& {(14)} & {(15)} 
& {(16)} & {(17)}
& {(18)} & {(19)}
& {(20)} & {(21)} & {(22)}
& {(23)}
& {(24)} & {(25)} & {(26)} 
\\
\hline
\\
  F1       & 10.17  & -0.36 & 10.18  & -0.35 & 19.5   & 13.6 & 0.135  & 31.9 & 33.0 & 2.3  &14.1  & 3.2   & 0.08   & 3.5E+04  & 8.7E+05  & 24.83 & 15.5  & 3    & 5    & 12.2   & 71    & -100   & Perseus     &        &     \\      
  F2       & 12.0   & -0.42 & 12.05  & -0.42 & 45.3   & 12.0 & 0.187  & 39.2 & 39.5 & 2.4  &16.4  & 4.3   & -0.12  & 6.1E+03  & 5.6E+03  & 0.9   & 14.5  & 3    & 5    & 20.9   & 10    & -87    &             &        &     \\      
  F3       & 14.62  & 0.33  & 14.61  & 0.38  & 26.6   & 2.7  & 0.144  & 6.8  & 7.0  & 0.4  &15.8  & 1.4   & 0.07   & 7.6E+02  & 2.6E+03  & 3.37  & 16.6  & 4    & 5    & 13.5   & 68    & 34     &             &  1.3   & 0.7 \\      
  F4       & 18.07  & 0.55  & 18.07  & 0.57  & 23.3   & 13.7 & 0.09   & 21.5 & 22.0 & 2.1  &10.6  & 1.2   & 0.06   & 2.4E+04  & 3.9E+03  & 0.16  & 12.4  & 0    & 5    & 10.5   & 89    & 125    & Perseus     &  5.5   & 1.0 \\      
  F5       & 21.7   & -0.28 & 21.71  & -0.26 & 85.3   & 10.6 & 0.149  & 27.6 & 28.1 & 2.2  &12.9  & 1.7   & -0.06  & 1.1E+04  & 3.3E+03  & 0.3   & 12.3  & 1    & 5    & 10.1   & 51    & -53    & Scu-Cen     &  7.5   & 1.1 \\     
  F6       & 26.19  & 0.59  & 26.2   & 0.6   & 105.4  & 9.3  & 0.106  & 17.2 & 17.6 & 1.5  &11.6  & 0.4   & -0.01  & 1.4E+03  & 2.4E+03  & 1.73  & 16.2  & 0    & 5    & 12.3   & 74    & 98     &             &  4.5   & 1.0 \\      
  F7       & 26.46  & 0.72  & 26.5   & 0.71  & 48.0   & 3.1  & 0.158  & 8.5  & 8.6  & 0.4  &22.2  & 0.8   & -0.01  & 7.3E+02  & 3.0E+03  & 4.16  & 15.3  & 2    & 5    & 23.8   & 17    & 59     &             &  2.5   & 1.2 \\      
  F8       & 28.83  & -0.33 & 28.83  & -0.31 & 100.8  & 9.1  & 0.121  & 19.2 & 19.9 & 1.7  &12.1  & 0.4   & 0.01   & 1.3E+04  & 2.0E+03  & 0.16  & 10.9  & 1    & 5    & 14.5   & 85    & -49    &             &  8.45  & 1.7 \\      
  F9       & 29.8   & -0.83 & 29.81  & -0.83 & 82.8   & 4.7  & 0.152  & 12.5 & 13.1 & 0.7  &19.3  & 1.9   & 0.06   & 1.9E+03  & 2.8E+02  & 0.15  & 11.9  & 0    & 7    & 11.7   & 18    & -50    & Scu-Cen     &  2.3   & 1.1 \\      
  F10      & 30.74  & -0.89 & 30.73  & -0.86 & 80.3   & 4.5  & 0.134  & 10.6 & 10.8 & 0.7  &14.5  & 1.1   & -0.1   & 1.7E+03  & 2.5E+02  & 0.15  & 11.4  & 1    & 5    & 13.3   & 56    & -54    &             &  2.8   & 1.0 \\      
  F11      & 32.74  & -0.08 & 32.75  & -0.06 & 36.9   & 11.7 & 0.154  & 31.5 & 32.8 & 1.5  &22.2  & 1.8   & -0.03  & 9.2E+03  & 2.0E+05  & 21.6  & 20.7  & 7    & 7    & 17.8   & 82    & -22    & Sag-Car     &  7.0   & 1.3 \\      
  F12      & 36.88  & -0.5  & 36.88  & -0.47 & 59.7   & 9.8  & 0.099  & 16.8 & 17.2 & 1.3  &13.0  & 2.0   & 0.09   & 1.0E+04  & 2.5E+04  & 2.5   & 12.4  & 2    & 5    & 10.1   & 79    & -78    & Sag-Car     &        &     \\      
  F13      & 40.02  & -0.12 & 40.02  & -0.09 & 83.6   & 4.5  & 0.221  & 17.3 & 18.6 & 0.9  &21.1  & 1.6   & 0.06   & 1.7E+03  & 2.9E+03  & 1.76  & 11.7  & 1    & 7    & 10.6   & 87    & 3      &             &  2.5   & 0.8 \\      
  F14      & 48.6   & 0.36  & 48.61  & 0.35  & 35.9   & 8.6  & 0.102  & 15.3 & 15.7 & 1.6  &10.0  & 1.4   & 0.1    & 4.5E+03  & 2.0E+03  & 0.44  & 13.1  & 1    & 5    & 10.6   & 40    & 72     &             &  3.8   & 1.0 \\      
  F15      & 56.23  & 0.05  & 56.24  & 0.05  & -4.2   & 9.6  & 0.113  & 19.0 & 21.3 & 1.2  &17.9  & 1.7   & 0.07   & 7.6E+03  & 3.0E+04  & 3.95  & 14.0  & 4    & 6    & 12.1   & 22    & 27     & Perseus     &        &     \\      
  F16      & 60.38  & 0.18  & 60.37  & 0.23  & 20.3   & 6.4  & 0.128  & 14.3 & 14.4 & 1.5  &9.7   & 1.1   & -0.11  & 5.5E+02  & 7.3E+02  & 1.33  & 16.5  & 1    & 5    & 18.2   & 77    & 40     &             &        &     \\      
  F17      & 66.1   & 0.38  & 66.09  & 0.38  & 16.0   & 4.9  & 0.136  & 11.7 & 11.9 & 0.7  &17.6  & 2.7   & -0.21  & 1.2E+03  & 1.4E+02  & 0.11  & 11.4  & 0    & 5    & 10.9   & 59    & 51     &             &        &     \\      
  F18      & 70.85  & -0.01 & 70.87  & -0.01 & 9.1    & 1.2  & 0.147  & 3.2  & 3.3  & 0.3  &13.0  & 1.4   & -0.01  & 9.0E+01  & 1.1E+01  & 0.12  & 11.1  & 0    & 5    & 10.4   & 54    & 24     &             &        &     \\      
  F19      & 71.83  & 1.22  & 71.85  & 1.22  & 15.2   & 1.9  & 0.103  & 3.4  & 3.5  & 0.3  &10.3  & 0.9   & -0.29  & 2.4E+02  & 4.2E+01  & 0.17  & 11.0  & 2    & 5    & 11.5   & 16    & 66     &             &        &     \\      
  F20      & 80.66  & 1.28  & 80.68  & 1.26  & -64.1  & 9.1  & 0.133  & 21.0 & 21.6 & 1.4  &15.6  & 2.1   & -0.11  & 2.3E+03  & 6.2E+03  & 2.73  & 17.3  & 5    & 6    & 10.6   & 22    & 232    &             &        &     \\      
  F21      & 103.68 & 0.43  & 103.68 & 0.44  & -61.5  & 4.2  & 0.108  & 8.0  & 8.1  & 0.5  &15.7  & 1.9   & 0.24   & 2.8E+02  & 9.5E+02  & 3.38  & 14.5  & 3    & 5    & 11.2   & 51    & 61     &             &        &     \\      
  F22      & 182.03 & -0.23 & 182.02 & -0.22 & -10.4  & 4.4  & 0.146  & 11.1 & 11.7 & 0.7  &17.1  & 0.8   & -0.09  & 2.5E+03  & 2.1E+02  & 0.08  & 9.4   & 4    & 7    & 11.1   & 68    & 21     &             &        &     \\      
  F23      & 189.25 & 0.71  & 189.24 & 0.73  & 10.4   & 2.8  & 0.16   & 7.8  & 8.0  & 0.4  &19.1  & 1.1   & -0.15  & 3.5E+02  & 3.4E+01  & 0.1   & 10.0  & 1    & 6    & 14.8   & 69    & 68     & Perseus     &        &     \\      
  F24      & 224.02 & -1.67 & 224.03 & -1.63 & 17.6   & 1.3  & 0.124  & 2.8  & 2.8  & 0.2  &14.5  & 0.0   & 0.0    & 4.8E+01  & 4.4E+01  & 0.92  & 12.6  & 3    & 5    & 12.2   & 78    & -9     &             &        &     \\      
  F25      & 225.44 & -0.24 & 225.48 & -0.23 & 16.6   & 1.2  & 0.191  & 3.9  & 4.0  & 0.3  &15.1  & 0.4   & -0.04  & 1.4E+02  & 4.5E+00  & 0.03  & 8.6   & 0    & 5    & 19.4   & 15    & 23     &             &        &     \\      
  F26      & 239.68 & -0.8  & 239.68 & -0.81 & 65.0   & 5.2  & 0.094  & 8.6  & 9.1  & 0.6  &15.4  & 0.3   & -0.04  & 2.1E+02  & 9.6E+01  & 0.47  & 13.4  & 2    & 5    & 12.3   & 7     & -37    & Perseus     &        &     \\      
  F27      & 283.78 & -0.38 & 283.78 & -0.37 & -2.6   & 5.5  & 0.091  & 8.8  & 9.2  & 1.3  &7.2   & 0.0   & 0.0    & 7.8E+02  & 2.2E+03  & 2.82  & 20.0  & 1    & 5    & 11.7   & 36    & -13    & Sag-Car     &        &     \\      
  F28      & 290.77 & -1.43 & 290.77 & -1.41 & 9.3    & 6.9  & 0.138  & 16.5 & 16.8 & 1.0  &16.1  & 1.9   & -0.15  & 9.2E+02  & 6.3E+03  & 6.86  & 13.8  & 1    & 5    & 15.9   & 69    & -153   &             &        &     \\      
  F29      & 305.22 & 0.5   & 305.26 & 0.49  & -46.7  & 4.8  & 0.16   & 13.4 & 13.8 & 0.8  &18.1  & 2.2   & -0.14  & 2.6E+03  & 9.2E+02  & 0.35  & 12.8  & 3    & 6    & 10.7   & 14    & 63     &             &  2.6   & 0.9 \\      
  F30      & 305.41 & 0.71  & 305.45 & 0.74  & -46.5  & 4.9  & 0.257  & 22.1 & 23.4 & 0.7  &32.9  & 1.5   & 0.06   & 2.9E+03  & 9.1E+02  & 0.32  & 11.7  & 2    & 9    & 10.4   & 31    & 78     &             &        &     \\      
  F31      & 319.97 & 0.56  & 319.98 & 0.56  & -40.1  & 2.5  & 0.121  & 5.3  & 5.5  & 0.5  &11.8  & 1.2   & 0.03   & 2.8E+02  & 4.3E+01  & 0.15  & 10.9  & 0    & 5    & 11.4   & 10    & 45     &             &        &     \\      
  F32      & 323.69 & 0.63  & 323.72 & 0.63  & -46.9  & 10.6 & 0.134  & 24.8 & 26.6 & 1.9  &14.4  & 1.5   & 0.04   & 5.3E+03  & 4.1E+04  & 7.7   & 16.9  & 3    & 6    & 11.3   & 2     & 124    &             &        &     \\      
  F33      & 324.16 & -0.66 & 324.15 & -0.62 & -33.3  & 9.6  & 0.183  & 30.6 & 30.9 & 2.0  &15.6  & 1.2   & 0.0    & 5.2E+03  & 1.1E+03  & 0.22  & 12.4  & 0    & 5    & 13.4   & 90    & -108   &             &        &     \\      
  F34      & 328.81 & 0.07  & 328.81 & 0.07  & -32.9  & 12.1 & 0.062  & 13.1 & 13.2 & 1.7  &7.9   & 2.6   & 0.26   & 6.1E+03  & 2.6E+04  & 4.22  & 16.0  & 5    & 5    & 15.4   & 53    & 15     &             &  3.3   & 1.0 \\      
  F35      & 338.56 & 0.22  & 338.58 & 0.21  & -36.5  & 2.8  & 0.12   & 5.8  & 5.8  & 0.4  &14.7  & 2.7   & -0.11  & 7.7E+02  & 4.9E+03  & 6.38  & 17.8  & 4    & 5    & 17.6   & 3     & 30     & Nor-Out     &  1.85  & 1.3 \\      
  F36      & 338.82 & -0.47 & 338.87 & -0.48 & -38.8  & 2.9  & 0.367  & 18.3 & 20.1 & 0.4  &48.0  & 3.5   & 0.01   & 3.5E+03  & 4.2E+02  & 0.12  & 10.4  & 7    &11    & 10.2   & 8     & -4     & Scu-Cen     &  2.0   & 1.0 \\      
  F37      & 340.79 & 0.57  & 340.79 & 0.59  & -124.9 & 9.4  & 0.144  & 23.5 & 24.2 & 1.6  &14.9  & 0.8   & -0.03  & 5.3E+03  & 1.6E+03  & 0.31  & 12.8  & 0    & 5    & 11.0   & 62    & 94     &             &        &     \\      
  F38      & 342.93 & -0.77 & 342.95 & -0.75 & -26.7  & 2.0  & 0.118  & 4.1  & 4.1  & 0.4  &9.7   & 0.7   & -0.14  & 1.4E+03  & 6.2E+01  & 0.04  & 8.9   & 1    & 5    & 15.8   & 44    & -6     & Nor-Out     &  1.14  & 1.1 \\      
  F39      & 346.02 & -0.02 & 346.06 & -0.03 & -80.3  & 10.8 & 0.189  & 35.7 & 36.8 & 1.4  &25.8  & 2.8   & -0.08  & 1.2E+04  & 6.4E+04  & 5.52  & 16.0  & 5    & 6    & 10.2   & 7     & -1     &             &  4.0   & 0.5 \\      
  F40      & 348.04 & -0.44 & 348.05 & -0.43 & -94.7  & 6.0  & 0.104  & 10.9 & 11.1 & 1.1  &10.3  & 2.1   & -0.08  & 6.4E+03  & 4.3E+04  & 6.71  & 15.6  & 4    & 5    & 12.2   & 2     & -33    &             &  3.0   & 1.1 \\      
  F41      & 348.73 & -0.29 & 348.77 & -0.26 & -19.3  & 2.2  & 0.184  & 7.1  & 7.1  & 0.4  &16.1  & 0.7   & -0.11  & 2.9E+02  & 1.3E+02  & 0.44  & 12.1  & 1    & 5    & 12.9   & 19    & 9      &             &  1.6   & 0.9 \\      
  F42      & 352.62 & -1.08 & 352.63 & -1.07 & -0.6   & 16.3 & 0.069  & 19.6 & 20.0 & 1.1  &18.5  & 1.7   & -0.09  & 5.1E+04  & 1.4E+06  & 27.86 & 21.9  & 5    & 5    & 12.7   & 51    & -321   & Sag-Car     &        &     \\
  \hline
\end{tabular}
\end{table}

\capstarttrue
\end{landscape}
\clearpage

\setlength\tabcolsep{4pt}
\setcounter{table}{0}
\clearpage
\begin{landscape}
\capstartfalse
\begin{table}
\centering
\caption{Continued.}
\tiny
\begin{tabular}{crrrr rrrrr rrrrr rrrrr rrrrrr}
\hline
\hline\\
{ID} &
{$l_1$} &
{$b_1$} &
{$l_2$} &
{$b_2$} &
{$v_{\rm lsr}$} &
{$d$} &
{Len.} &
{Le2e} &
{Lmst} &
{$w$} &
{Aspect} &
{$\Delta v$} &
{Slope} &
{Mass} &
{Lum.} &
{$L/M$} &
{$T$} &
{$N_{70}$} &
{$N_{\rm cl}$} &
{Linearity} &
{$\theta$} &
{$z$} &
{Arm} &
{$P_{\rm v}$} &
{ $\frac{P_{\rm v}}{P_{\rm cl}}$ }  \\
{} &
{$^{\circ}$} &
{$^{\circ}$} &
{$^{\circ}$} &
{$^{\circ}$} &
{$\rm \frac{km}{s}$} &
{kpc} &
{$^{\circ}$} &
{pc} &
{pc} &
{pc} &
{ratio} &
{$\rm \frac{km}{s}$} &
{$\rm \frac{km/s}{pc}$} &
{\msun} &
{\lsun} &
{$\frac{L_\odot}{M_\odot}$} &
{K} &
{} &
{} &
{} &
{$^{\circ}$} &
{pc} &
{} &
{pc} &
{} \\
{(1)} 
& {(2)} & {(3)} & {(4)} 
& {(5)} & {(6)} 
& {(7)} 
& {(8)} & {(9)} 
& {(10)} & {(11)} 
& {(12)} & {(13)} 
& {(14)} & {(15)} 
& {(16)} & {(17)}
& {(18)} & {(19)}
& {(20)} & {(21)} & {(22)}
& {(23)}
& {(24)} & {(25)} & {(26)} 
\\
  \hline \\
  Min      &-177.97 & -1.67 &        &       & -124.9 & 1.2  & 0.062  & 2.8  & 2.8  & 0.2  &7.2   & 0.0   & 0.00   & 4.8E+01  & 4.5E+00  & 0.03  & 8.6   & 0.0  & 5.0  & 10.1   & 1.9   & -320.5 &             &  1.1   & 0.5 \\      
  Max      & 103.68 & 1.28  &        &       & 105.4  & 16.3 & 0.367  & 39.2 & 39.5 & 2.4  &48.0  & 4.3   & 0.29   & 5.1E+04  & 1.4E+06  & 27.86 & 21.9  & 7.0  & 11.0 & 23.8   & 90.0  & 232.2  &             &  8.5   & 1.7 \\      
  Avg      &  -8.92 & -0.05 &        &       & 4.8    & 6.7  & 0.142  & 15.5 & 16.0 & 1.1  &16.2  & 1.5   & 0.08   & 5.8E+03  & 6.6E+04  & 3.47  & 13.7  & 2.2  & 5.6  & 13.1   & 44.5  & 6.8    &             &  3.6   & 1.0 \\      
  Med      &   1.40 & -0.05 &        &       & 9.9    & 5.4  & 0.135  & 13.9 & 14.1 & 1.0  &15.3  & 1.5   & 0.07   & 2.1E+03  & 2.0E+03  & 0.69  & 12.8  & 2.0  & 5.0  & 12.2   & 51.2  & 18.0   &             &  2.8   & 1.0 \\      
  Std      &  66.76 & 0.66  &        &       & 54.0   & 4.0  & 0.053  & 9.4  & 9.7  & 0.6  &7.0   & 0.9   & 0.07   & 9.9E+03  & 2.6E+05  & 6.41  & 3.1   & 2.0  & 1.2  & 3.2    & 29.3  & 88.2   &             &  2.1   & 0.2 \\      
  Skewness &   -0.9 & -0.2  &        &       & -0.1   & 0.5  & 2.1    & 0.7  & 0.7  & 0.5  &2.7   & 0.7   & 1.2    & 3.3      & 4.7      & 2.9   & 0.7   & 0.8  & 3.0  & 1.6    & 0.0   & -1.0   &             &  1.1   & 0.8 \\      
  Kurtosis &    0.5 & -0.1  &        &       & -0.3   & -0.8 & 7.5    & -0.1 & -0.3 & -0.9 &10.4  & 0.8   & 1.4    & 12.1     & 22.5     & 8.1   & 0.3   & -0.2 & 10.2 & 2.3    & -1.5  & 4.1    &             &  0.4   & 3.2 \\ 
\hline \\
\end{tabular}
\tablefoot{
Col. (1) Assigned filament ID.
Col. (2-5) Galactic coordinates of the two end-to-end clumps. Statistics (last rows) are computed by converting $l$ to [-180, 180] deg.
Col. (6) Mean LSR velocity of the clumps in a filament.
Col. (7) Distance (\S \ref{sec:GlobalProperties}).
\\
Col. (8-9) End-to-end length, in angular and physical units.
Col. (10) Curved length following the MST, i.e., connecting the clumps.
Col. (11) Width, the diameter of the largest clump in the filament.
Col. (12) Aspect ratio, length divided by width.
Col. (13) Difference between the min. and max. velocity of the clumps in a filament. 
Col. (14) Slope by fitting clump velocity along the filament (\S \ref{sec:GlobalProperties}). Statistics (last rows) are given for $|\rm Slope|$.
\\
Col. (15-16) Sum of clump mass/luminosity, which gives lower limit for filament mass/luminosity (\S \ref{sec:GlobalProperties}).
Col. (17) Luminosity/mass ratio.
Col. (18) Mean dust temperature of the clumps.
Col. (19-20) Number of 70\um\ sources, and number of HiGAL clumps.
Col. (21) Linearity, as defined in Paper \citetalias{me16} and refined in Paper \citetalias{Ge22} (\S \ref{sec:DataMethod}).
\\
Col. (22) Angle between filament and Galactic mid-plane.
Col. (23) Height from the Galactic mid-plane.
Col. (24) Associated spiral arm, if any.
Col. (25) Velocity oscillation period determined from PV analysis (\S \ref{sec:VeloOscillation}).
Col. (26) Ratio between velocity period and density period (mean separation between clumps, \S \ref{sec:VeloOscillation}).
\\
\\
The last rows list statistics of the parameters.
Skewness ($S$) measures symmetry of a distribution.
$S = 0$ means symmetric distribution, and negative/positive $S$ mean asymmetric tails with lower/higher values around the mean, respectively.
Kurtosis ($K$) measures how spread the distribution is compared to a normal distribution. A normal distribution has $K=0$. $K>0$ means the distribution is more centrally peaked than normal distribution, and $K<0$ is flatter.
\\
\\
Cross match:
six filaments (F1, 3, 7, 35, 36, 40) (partially) overlap with previously known filaments.
F1 is a small part of an X-shape filament (F5 in Paper \citetalias{me16} and F7 in Paper \citetalias{Ge22}).
F7 is part of the X-shaped \her\ cold filament CFG26 \citep{me15}.
F36 is the central and linear part of the well kown S-shaped ``Nessie'' filament \citep{Jackson2010,me15}.
F40 is partly overlapped with F90 in Paper \citetalias{Ge22}.
F3, F35, F36 are part of G014.478+0.736, G338.528+0.214, G338.680-0.455 in \cite{LiGX16-FLcat}, respectively.
}
\end{table}

\capstarttrue
\end{landscape}
\clearpage

\begin{figure*}
\centering
\includegraphics[width=.28\textwidth]{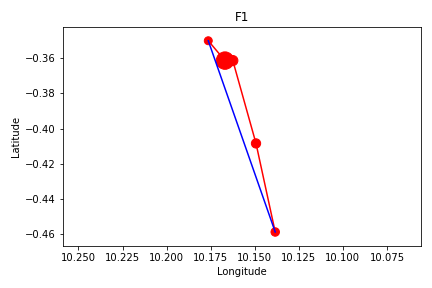}
\includegraphics[width=.28\textwidth]{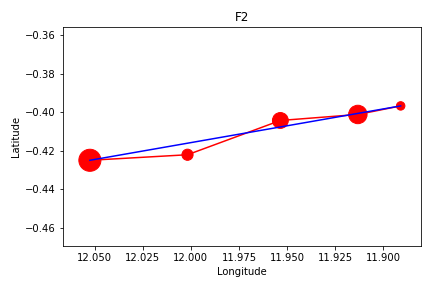}
\includegraphics[width=.28\textwidth]{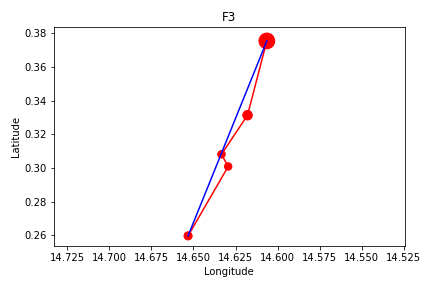}
\includegraphics[width=.28\textwidth]{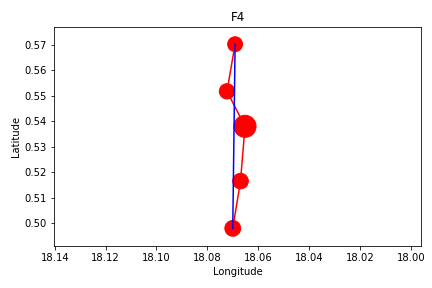}
\includegraphics[width=.28\textwidth]{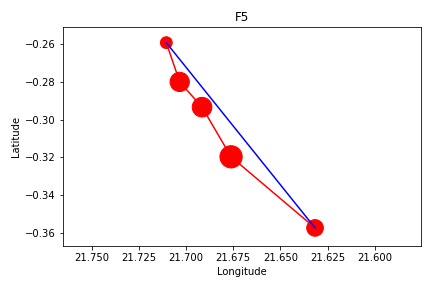}
\includegraphics[width=.28\textwidth]{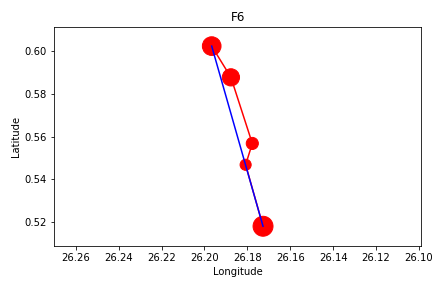}
\includegraphics[width=.28\textwidth]{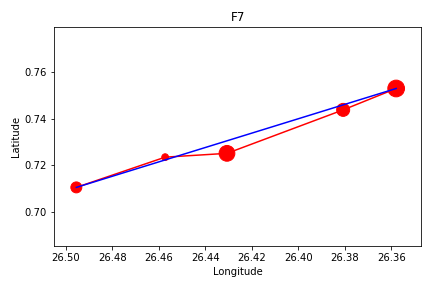}
\includegraphics[width=.28\textwidth]{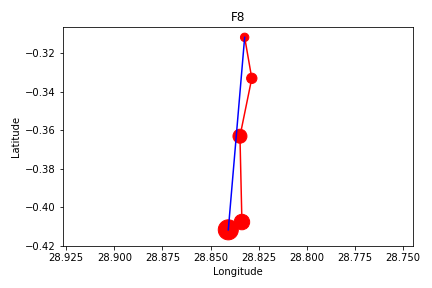}
\includegraphics[width=.28\textwidth]{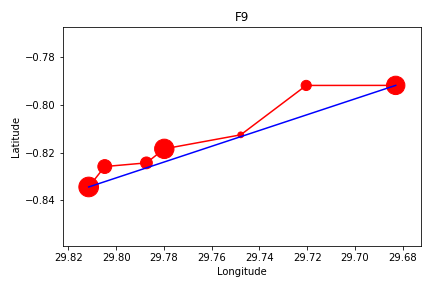}
\includegraphics[width=.28\textwidth]{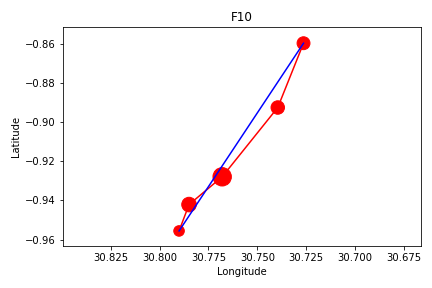}
\includegraphics[width=.28\textwidth]{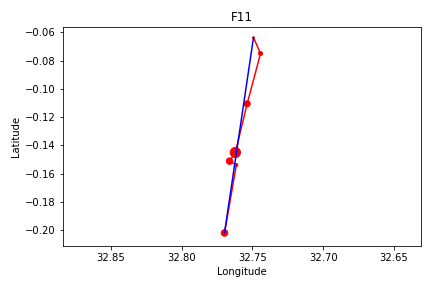}
\includegraphics[width=.28\textwidth]{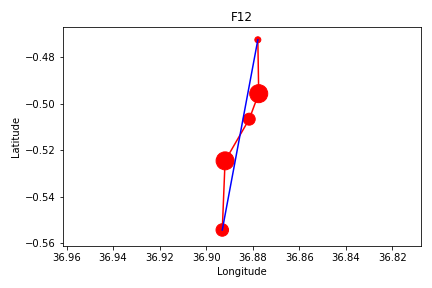}
\includegraphics[width=.28\textwidth]{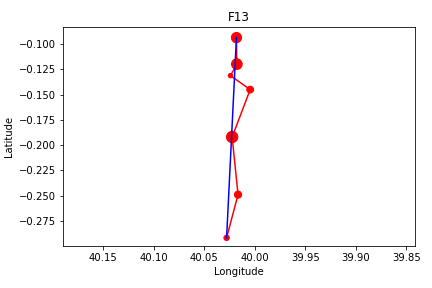}
\includegraphics[width=.28\textwidth]{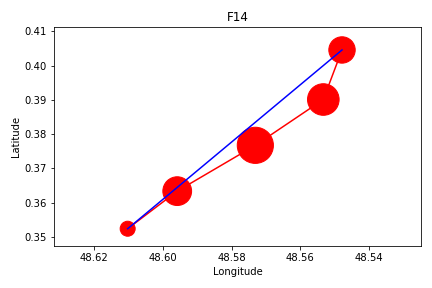}
\includegraphics[width=.28\textwidth]{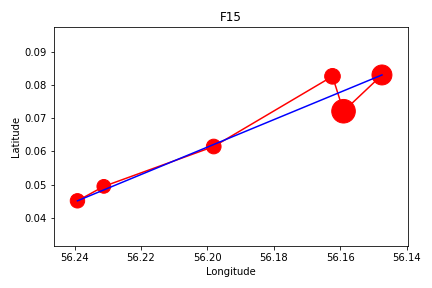}
\includegraphics[width=.28\textwidth]{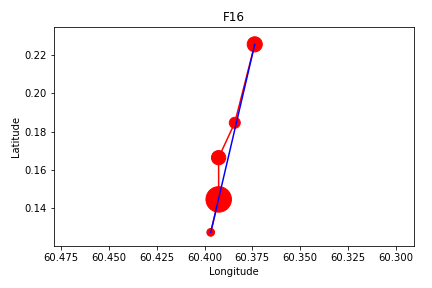}
\includegraphics[width=.28\textwidth]{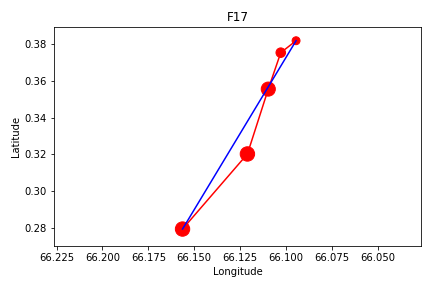}
\includegraphics[width=.28\textwidth]{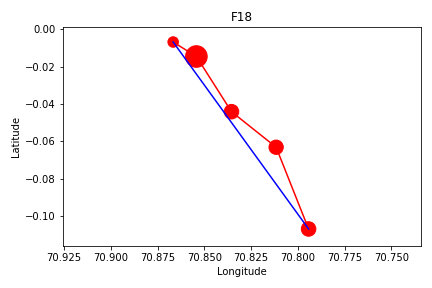}
\includegraphics[width=.28\textwidth]{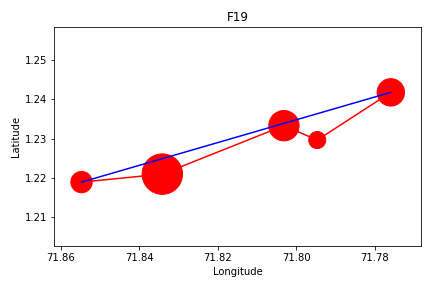}
\includegraphics[width=.28\textwidth]{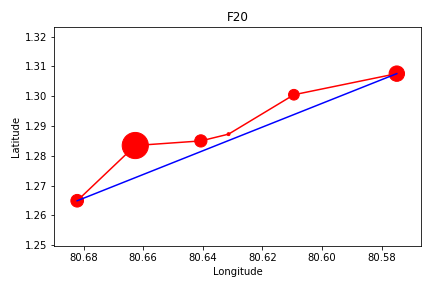}
\includegraphics[width=.28\textwidth]{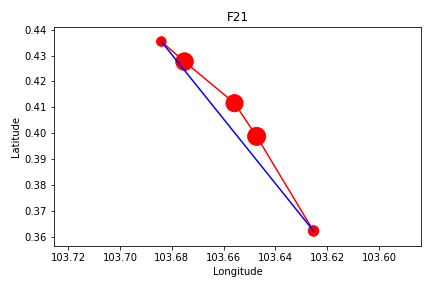}
\vskip -.2cm
\caption{MST for F1 to F21. The red circles represent the position and size of the \her/HiGAL clumps, and the red segments are the MST edges connecting the clumps. The blue line marks the filament end-to-end.
}
\label{fig:mst_a}
\end{figure*}

\setcounter{figure}{1}
\begin{figure*}
\centering
\includegraphics[width=.28\textwidth]{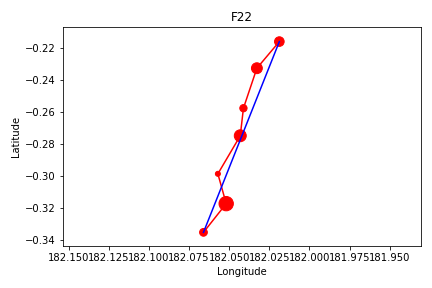}
\includegraphics[width=.28\textwidth]{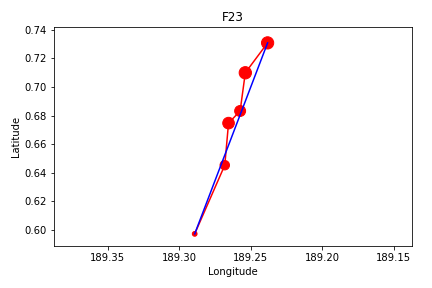}
\includegraphics[width=.28\textwidth]{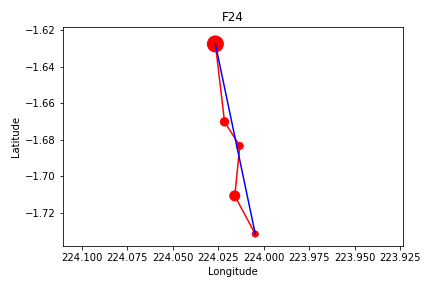}
\includegraphics[width=.28\textwidth]{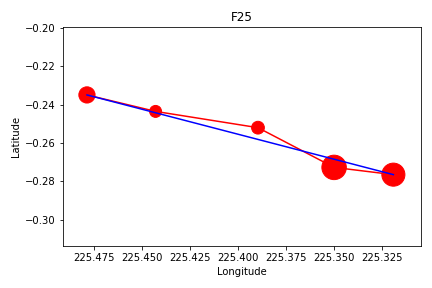}
\includegraphics[width=.28\textwidth]{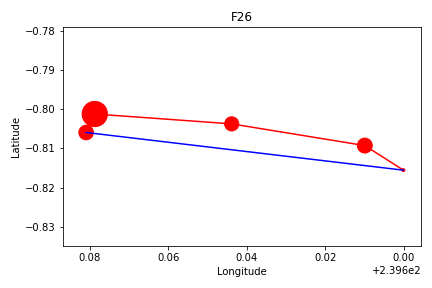}
\includegraphics[width=.28\textwidth]{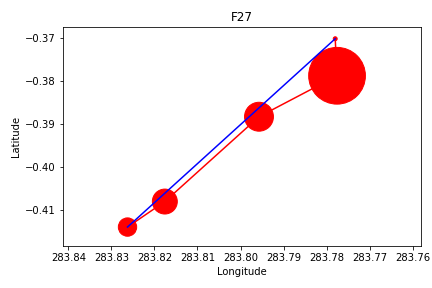}
\includegraphics[width=.28\textwidth]{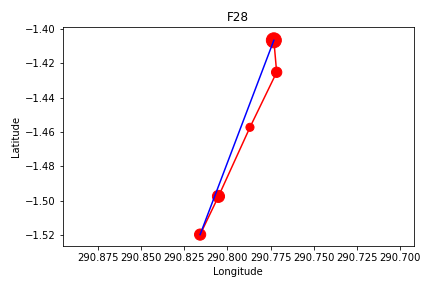}
\includegraphics[width=.28\textwidth]{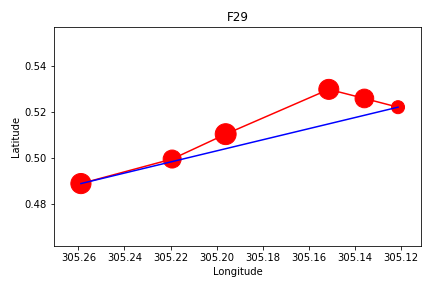}
\includegraphics[width=.28\textwidth]{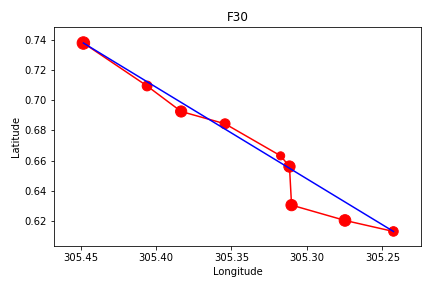}
\includegraphics[width=.28\textwidth]{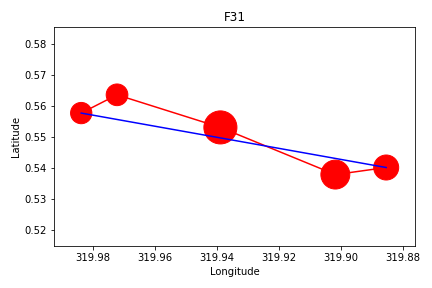}
\includegraphics[width=.28\textwidth]{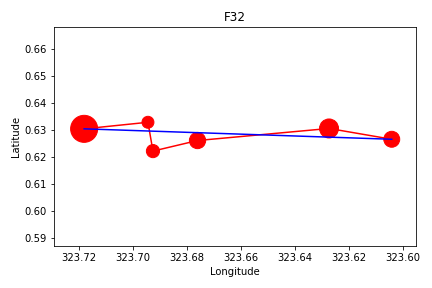}
\includegraphics[width=.28\textwidth]{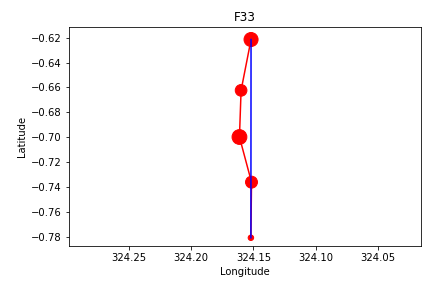}
\includegraphics[width=.28\textwidth]{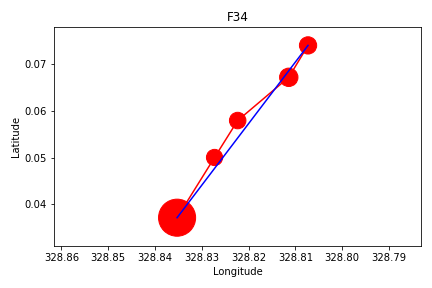}
\includegraphics[width=.28\textwidth]{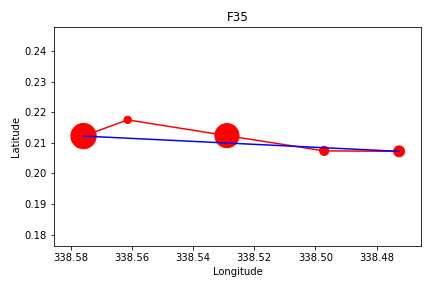}
\includegraphics[width=.28\textwidth]{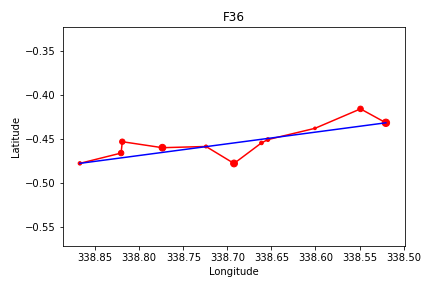}
\includegraphics[width=.28\textwidth]{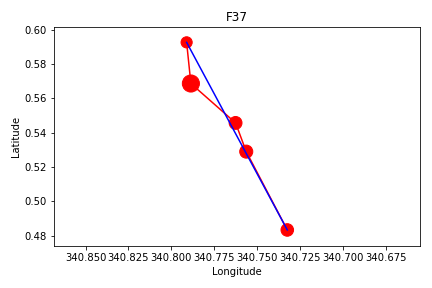}
\includegraphics[width=.28\textwidth]{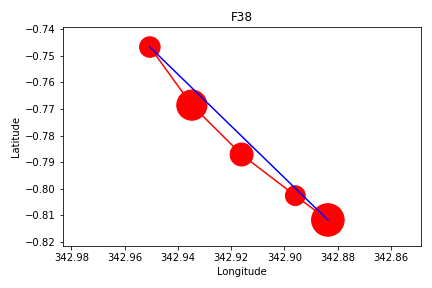}
\includegraphics[width=.28\textwidth]{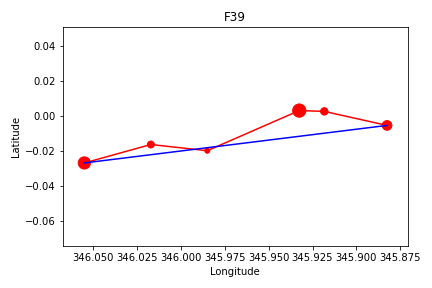}
\includegraphics[width=.28\textwidth]{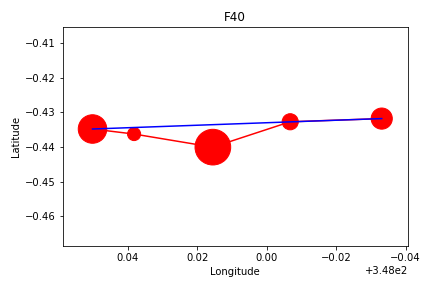}
\includegraphics[width=.28\textwidth]{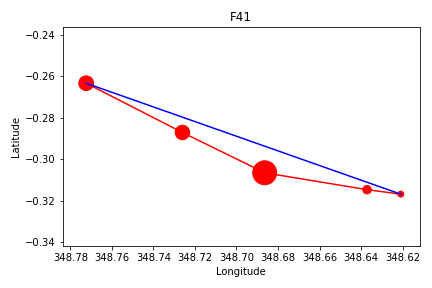}
\includegraphics[width=.28\textwidth]{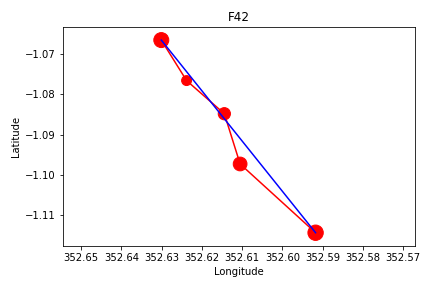}
\caption{Continued. MST for F22 to F42.
}
\label{fig:mst_b}
\end{figure*}

\begin{figure*}
\centering
\includegraphics[width=.28\textwidth]{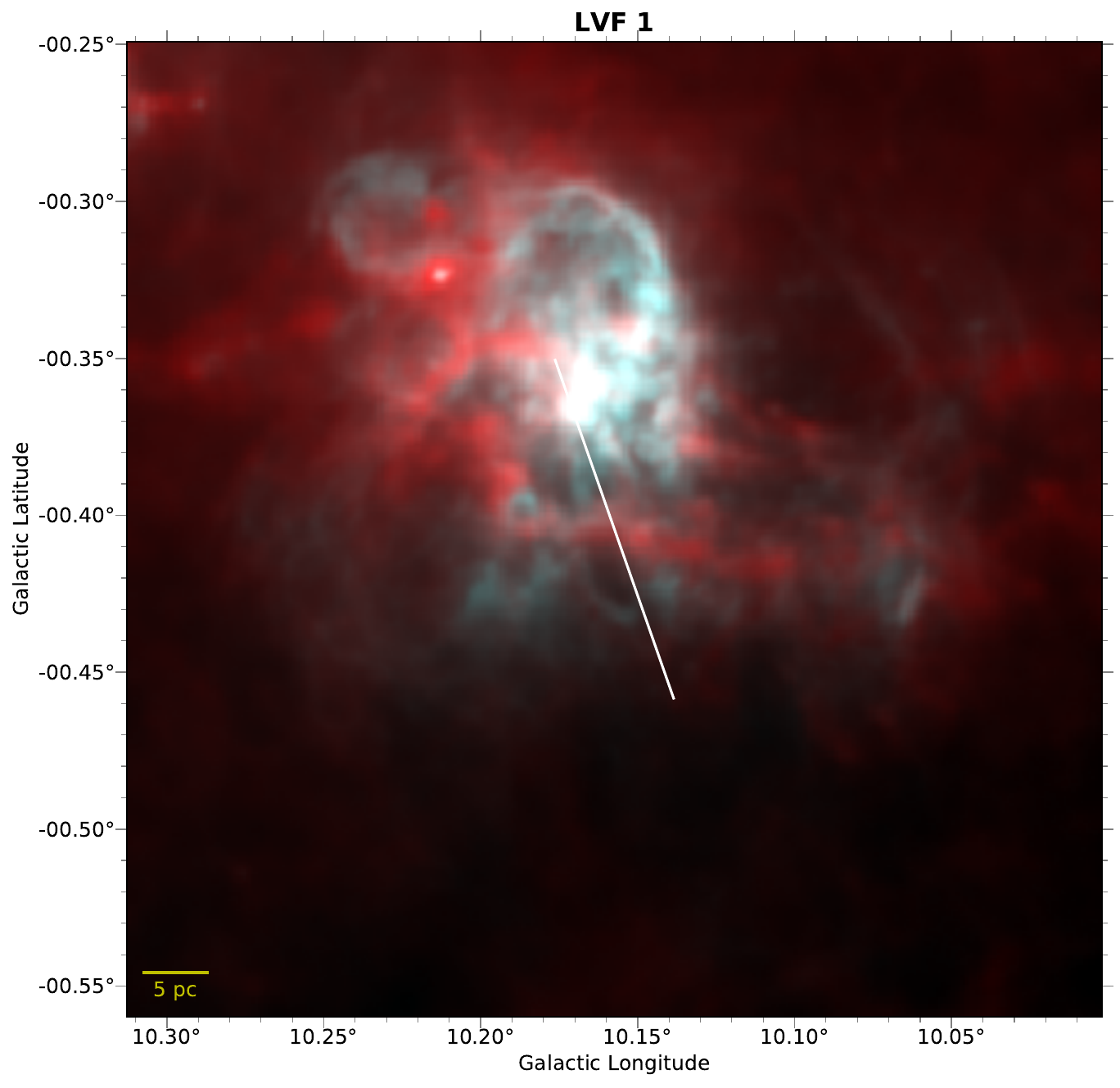}
\includegraphics[width=.28\textwidth]{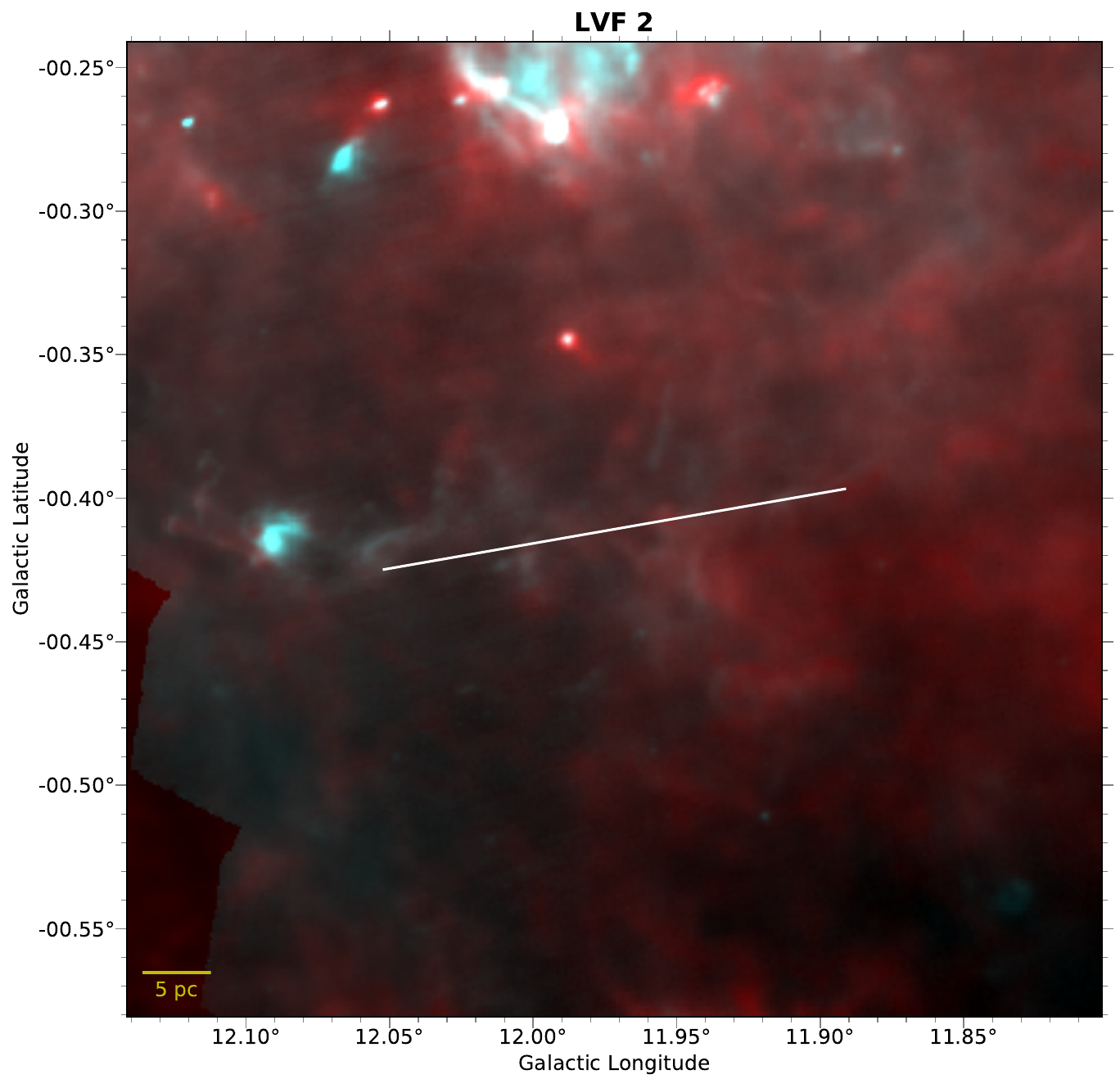}
\includegraphics[width=.28\textwidth]{rgb/LVF3.pdf}
\includegraphics[width=.28\textwidth]{rgb/LVF4.pdf}
\includegraphics[width=.28\textwidth]{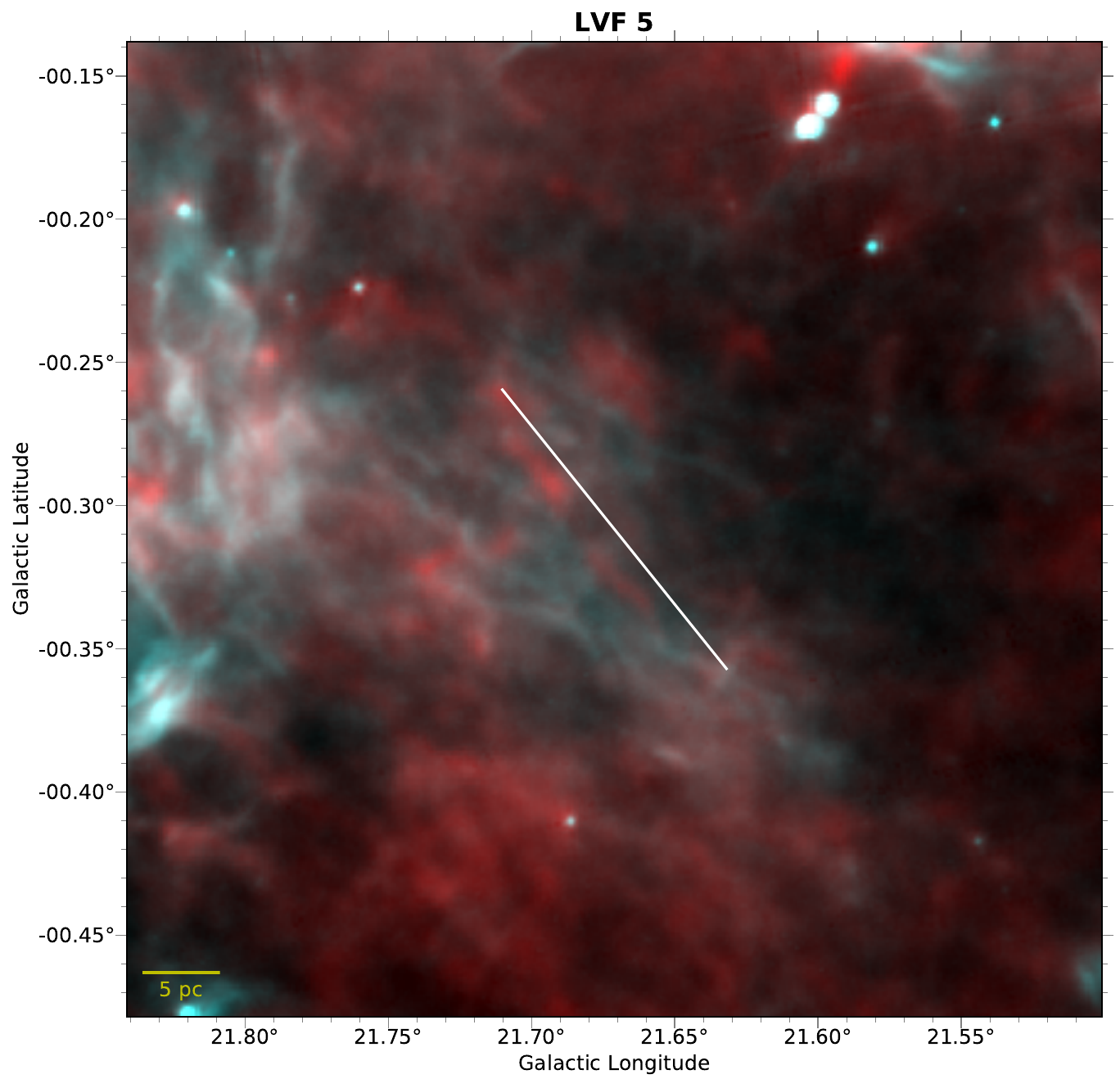}
\includegraphics[width=.28\textwidth]{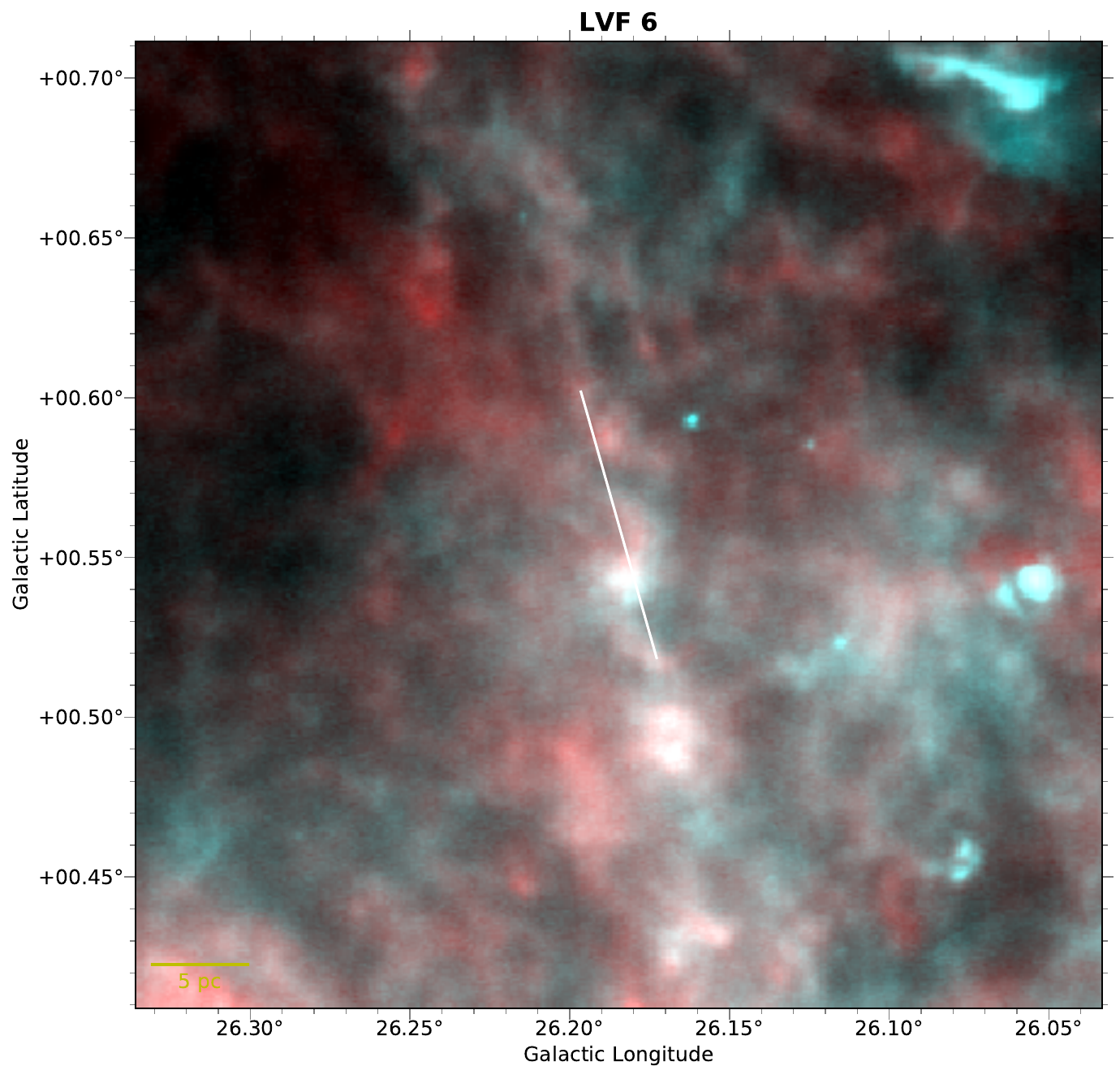}
\includegraphics[width=.28\textwidth]{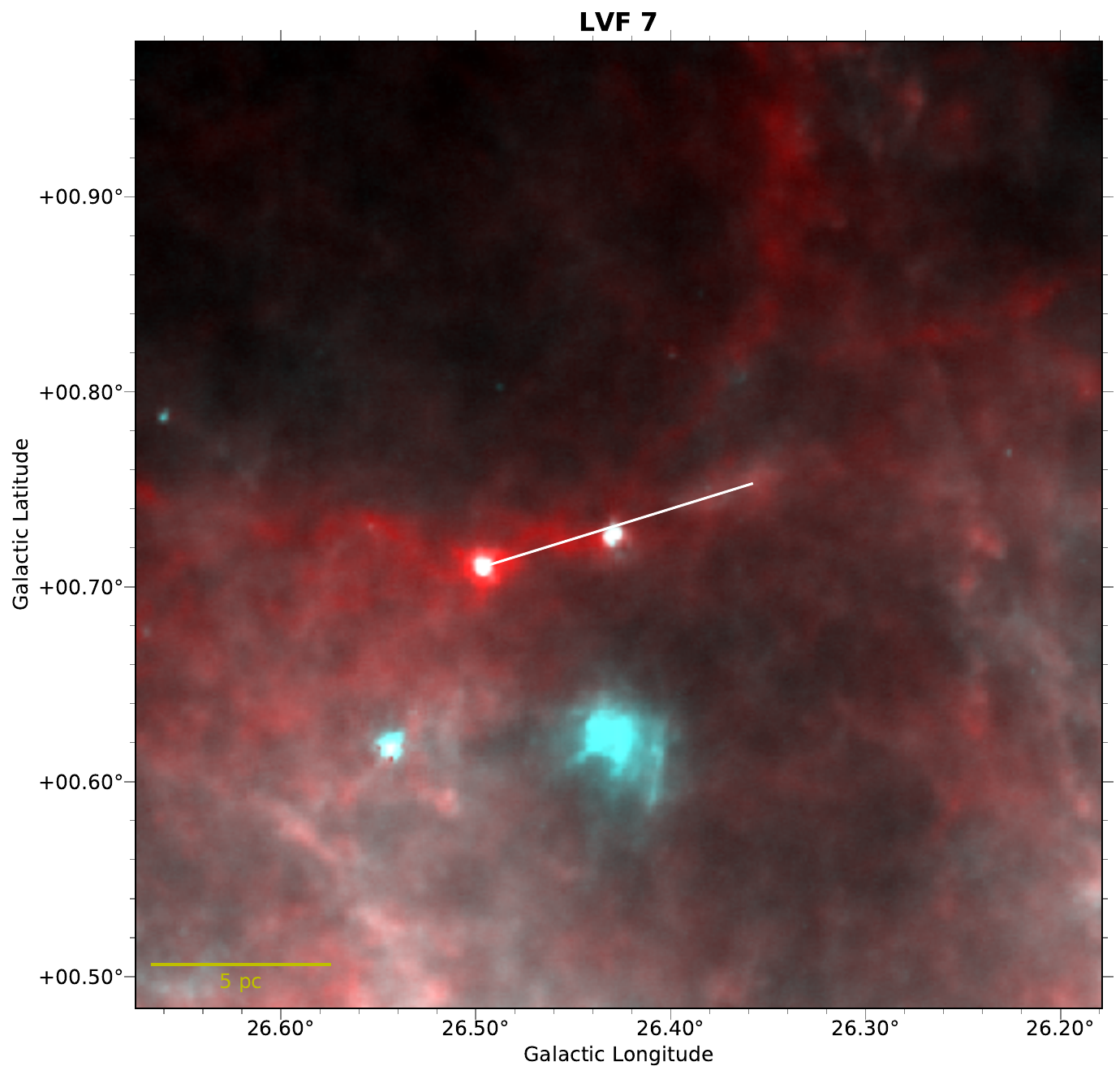}
\includegraphics[width=.28\textwidth]{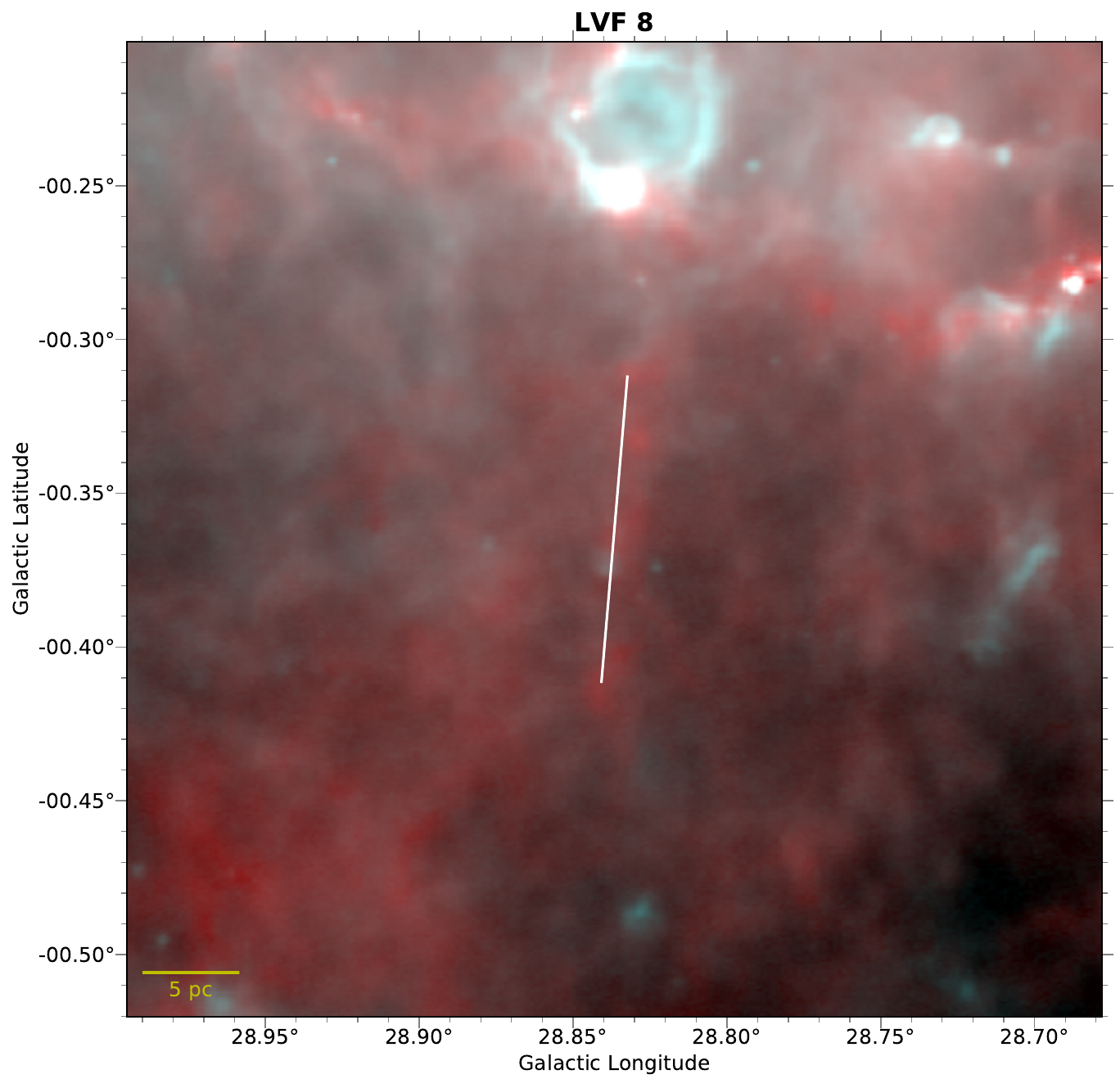}
\includegraphics[width=.28\textwidth]{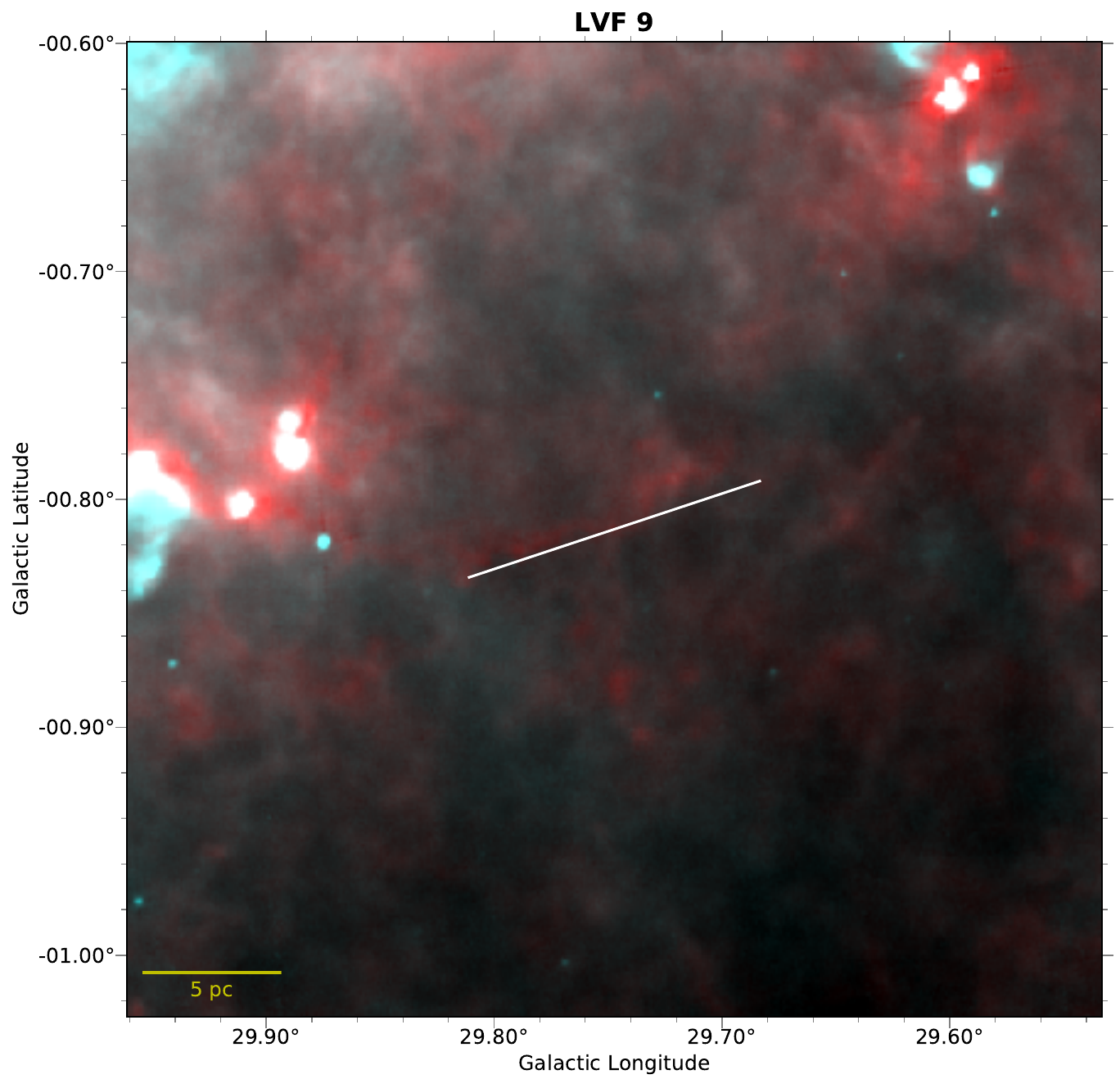}
\includegraphics[width=.28\textwidth]{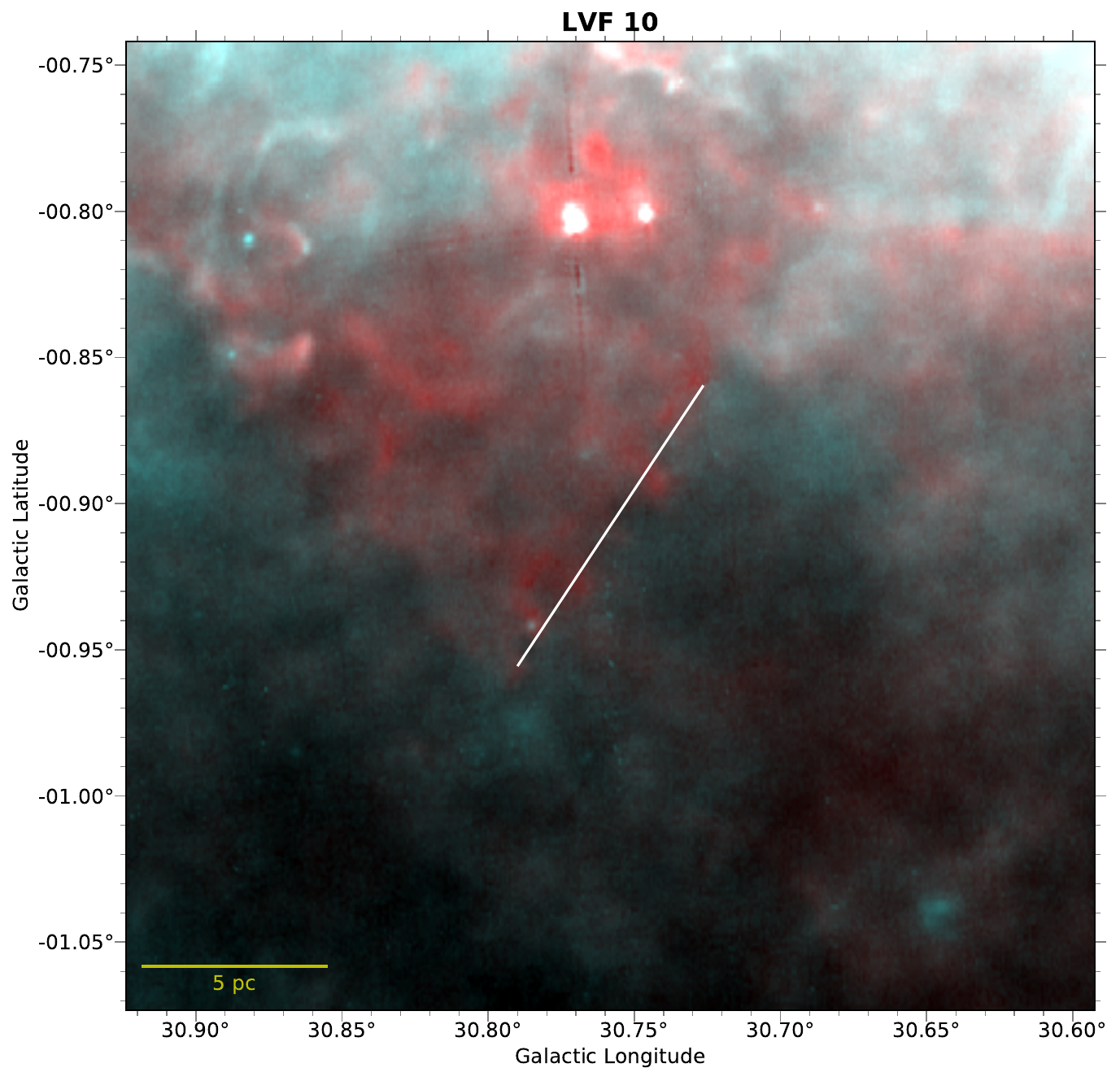}
\includegraphics[width=.28\textwidth]{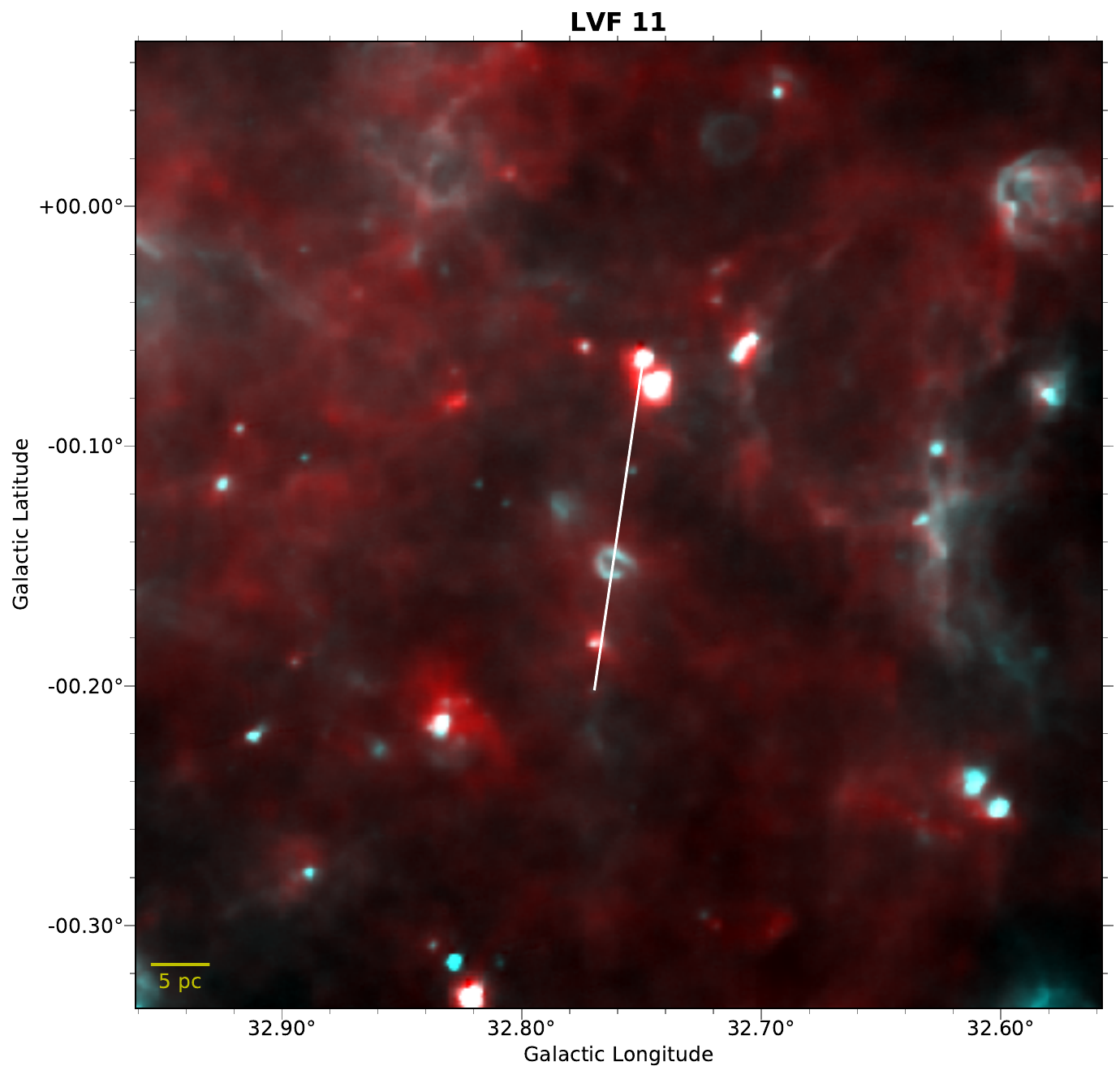}
\includegraphics[width=.28\textwidth]{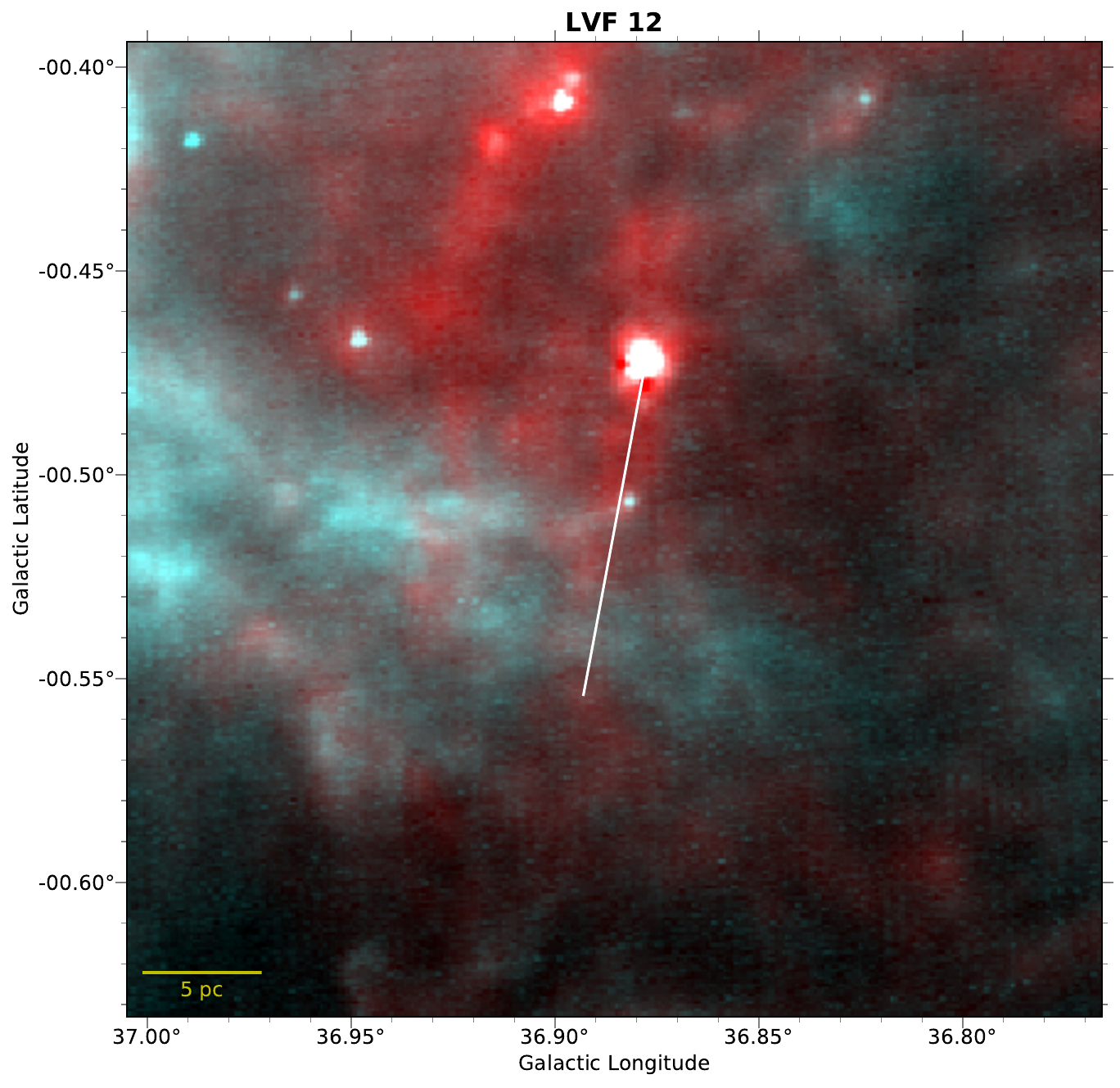}
\includegraphics[width=.28\textwidth]{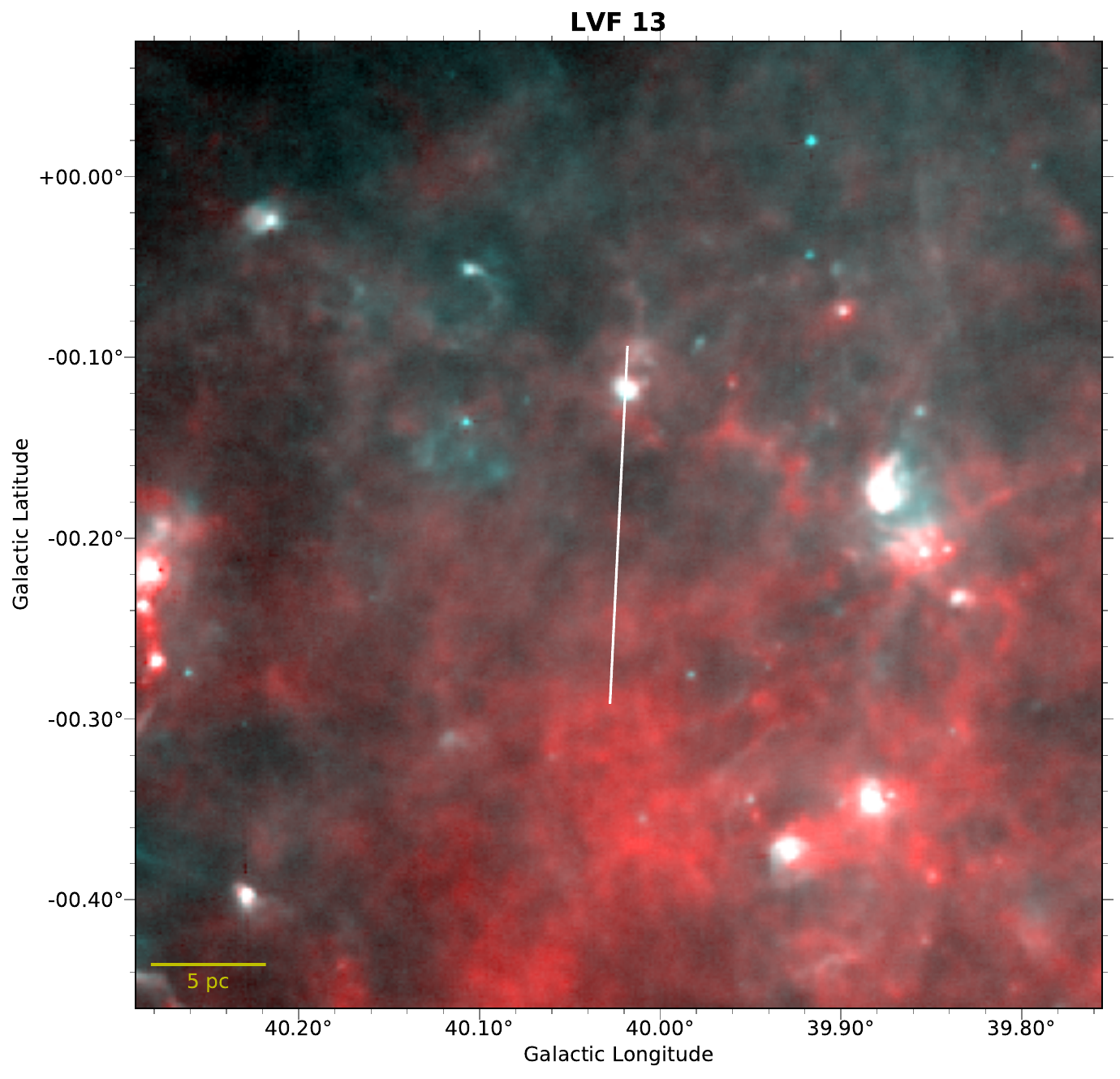}
\includegraphics[width=.28\textwidth]{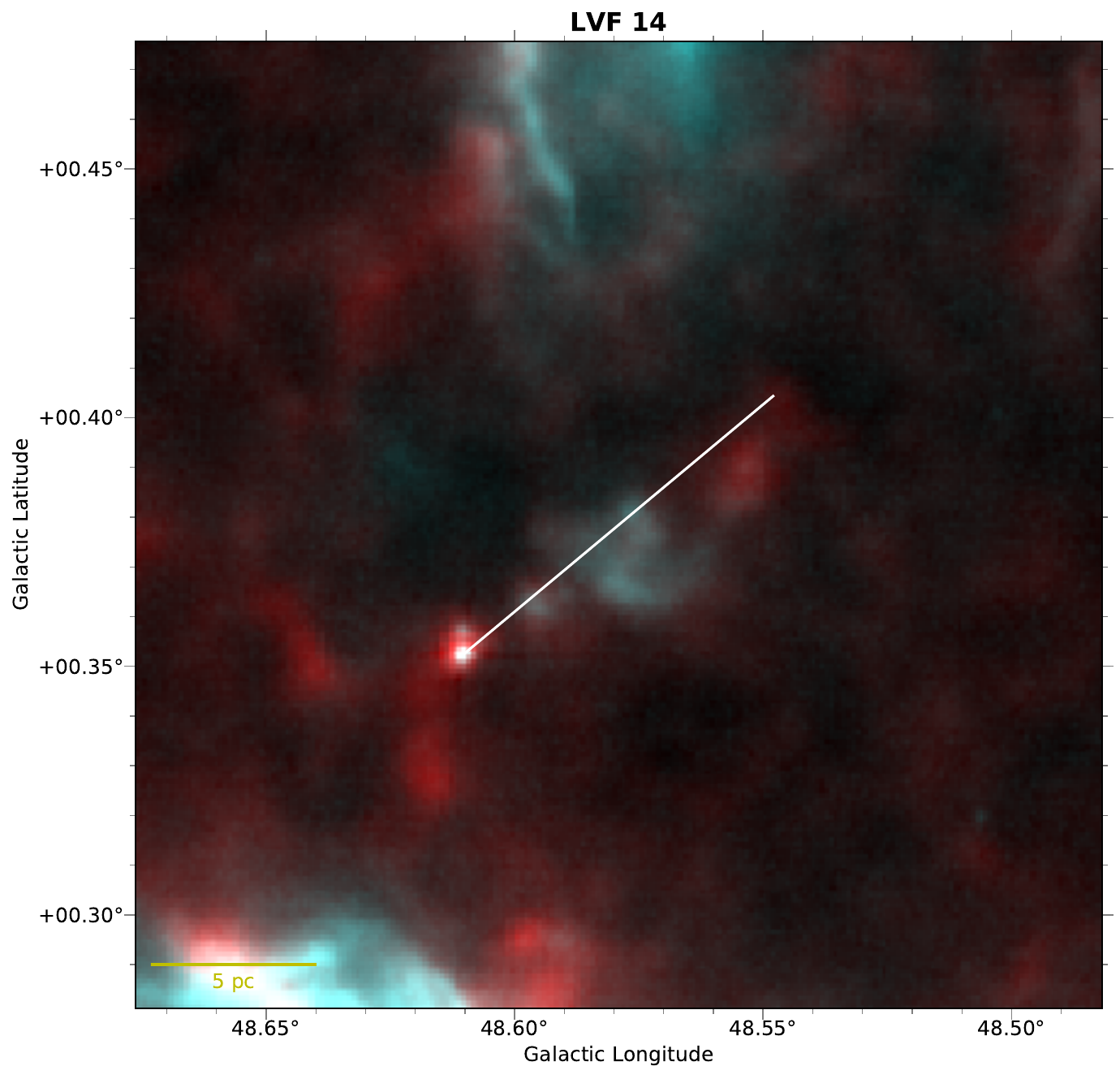}
\includegraphics[width=.28\textwidth]{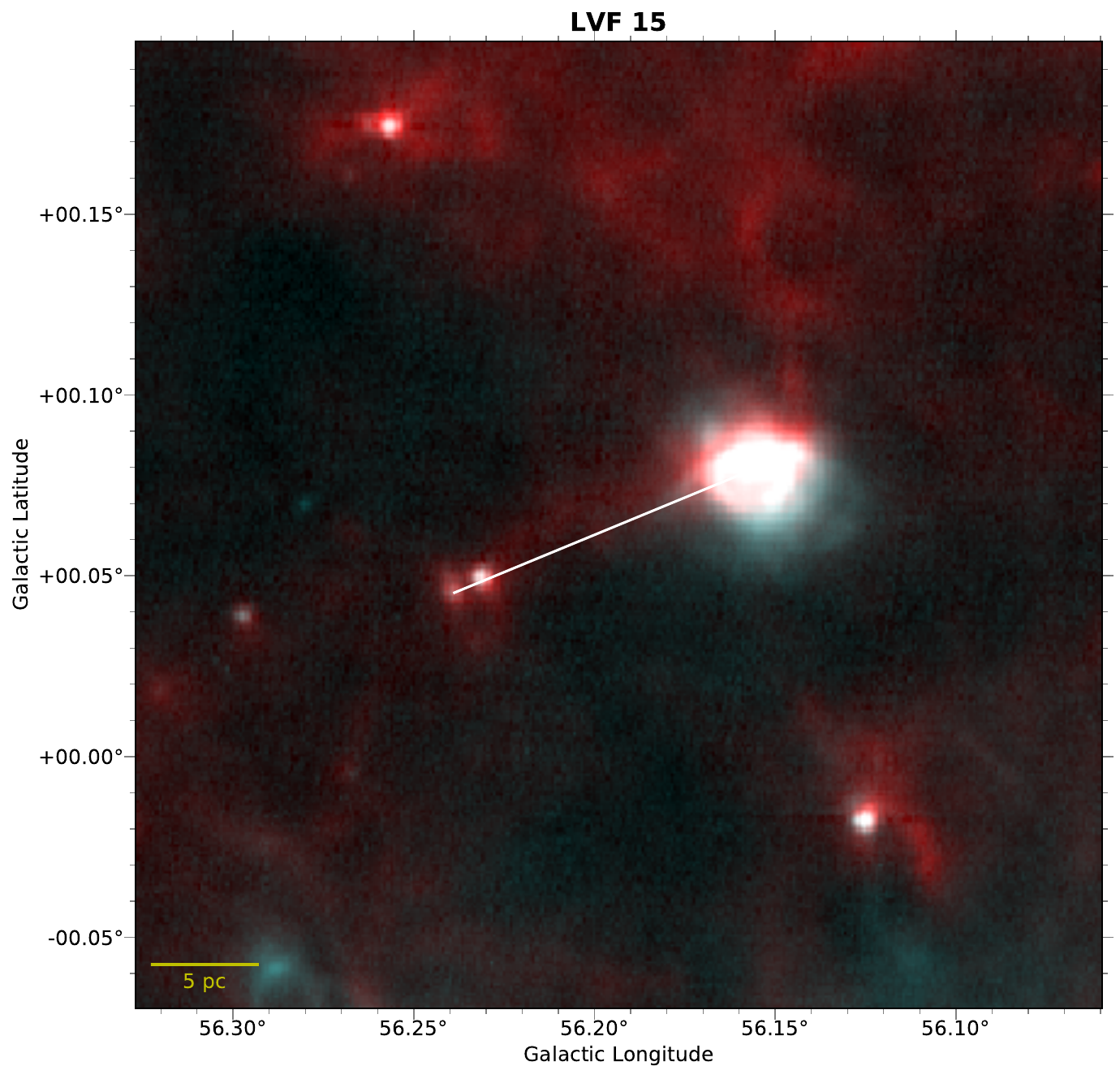}
\vskip .2cm
\caption{Two-color view of the linear filaments F1 to F15. \her\ 250\um/70\um\ are shown in red/cyan. The filament is marked by end-to-end line in white.
}
\label{fig:rgb_a}
\end{figure*}

\setcounter{figure}{2}
\begin{figure*}
\centering
\includegraphics[width=.28\textwidth]{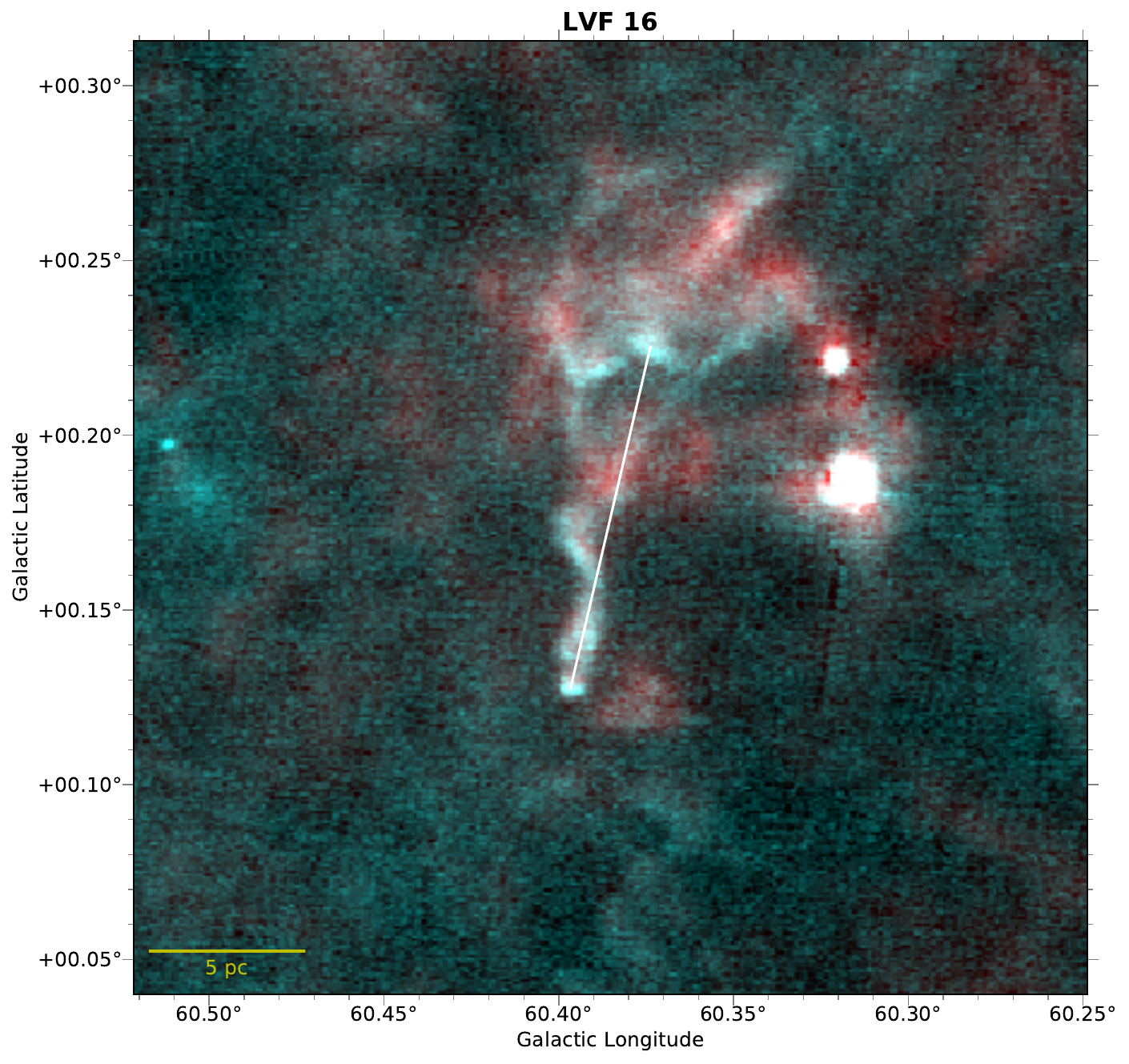}
\includegraphics[width=.28\textwidth]{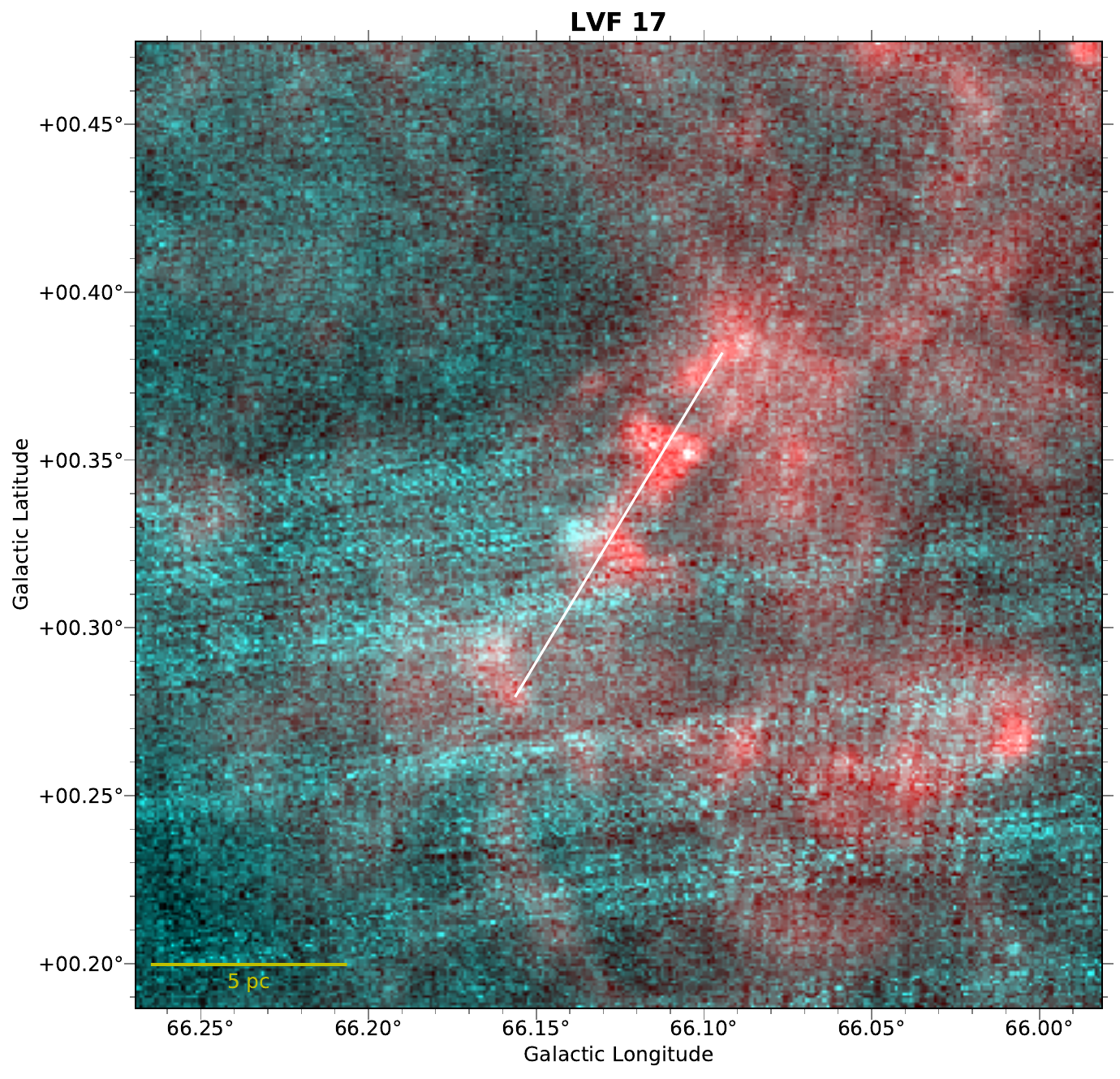}
\includegraphics[width=.28\textwidth]{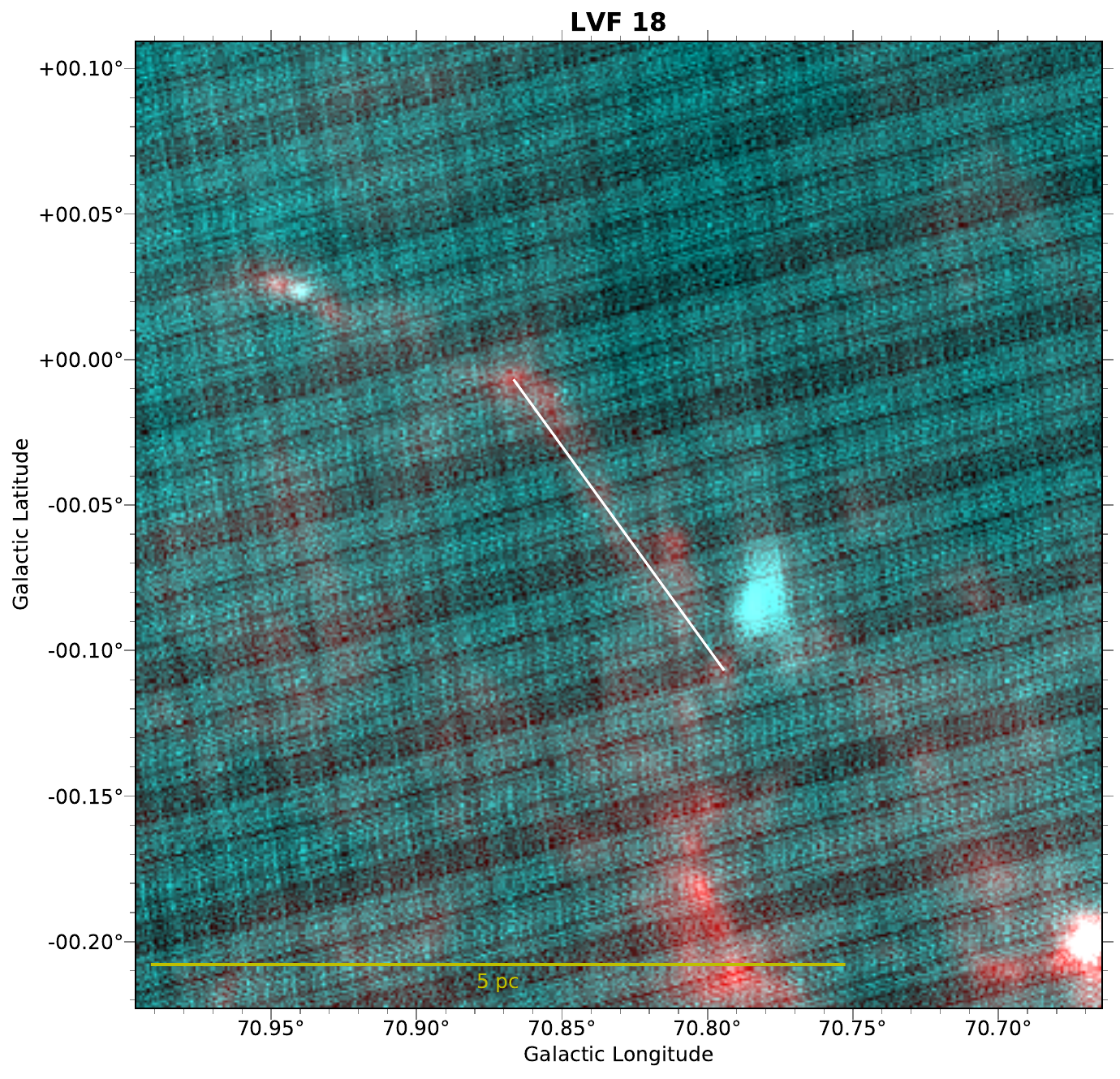}
\includegraphics[width=.28\textwidth]{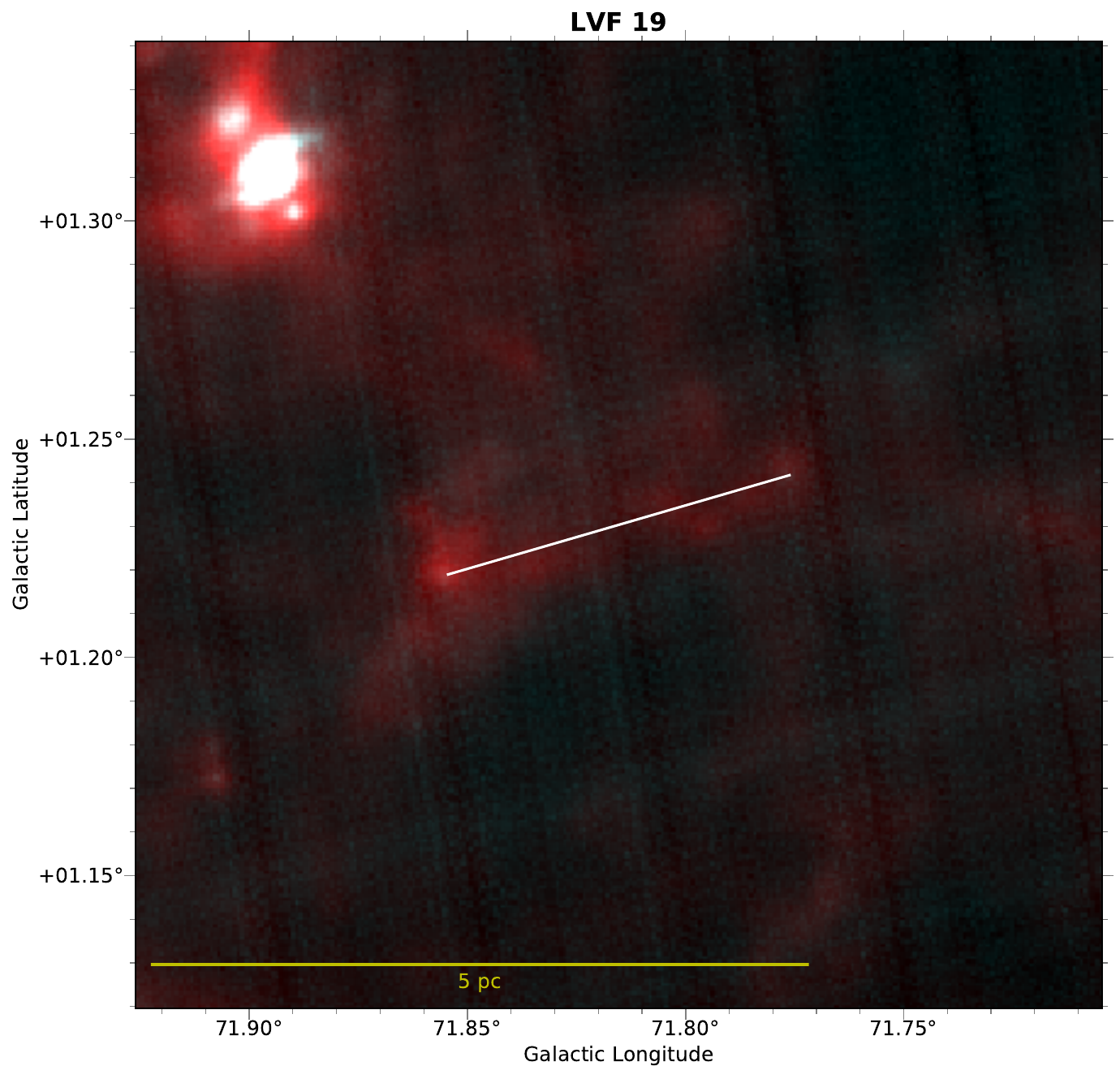}
\includegraphics[width=.28\textwidth]{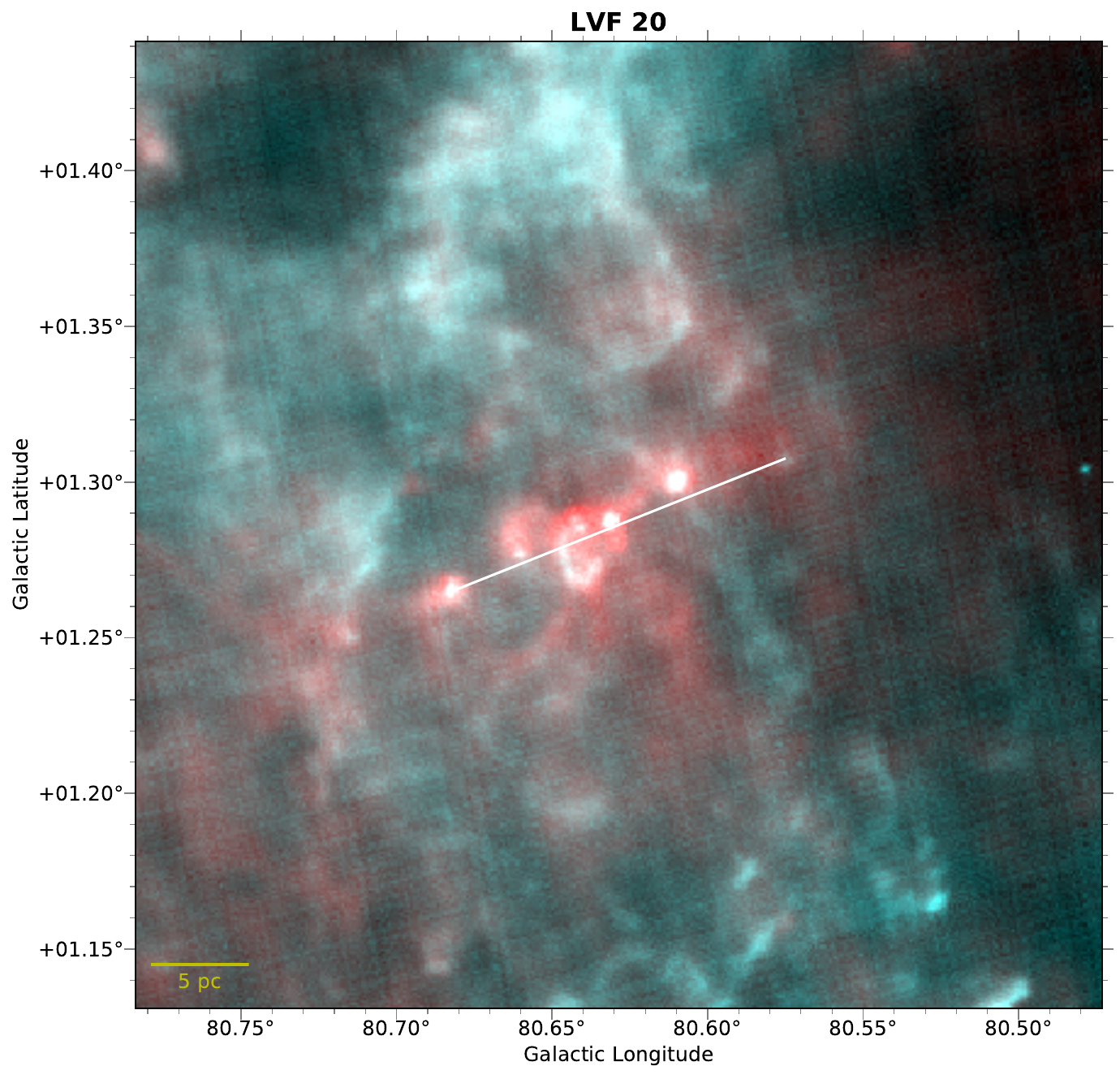}
\includegraphics[width=.28\textwidth]{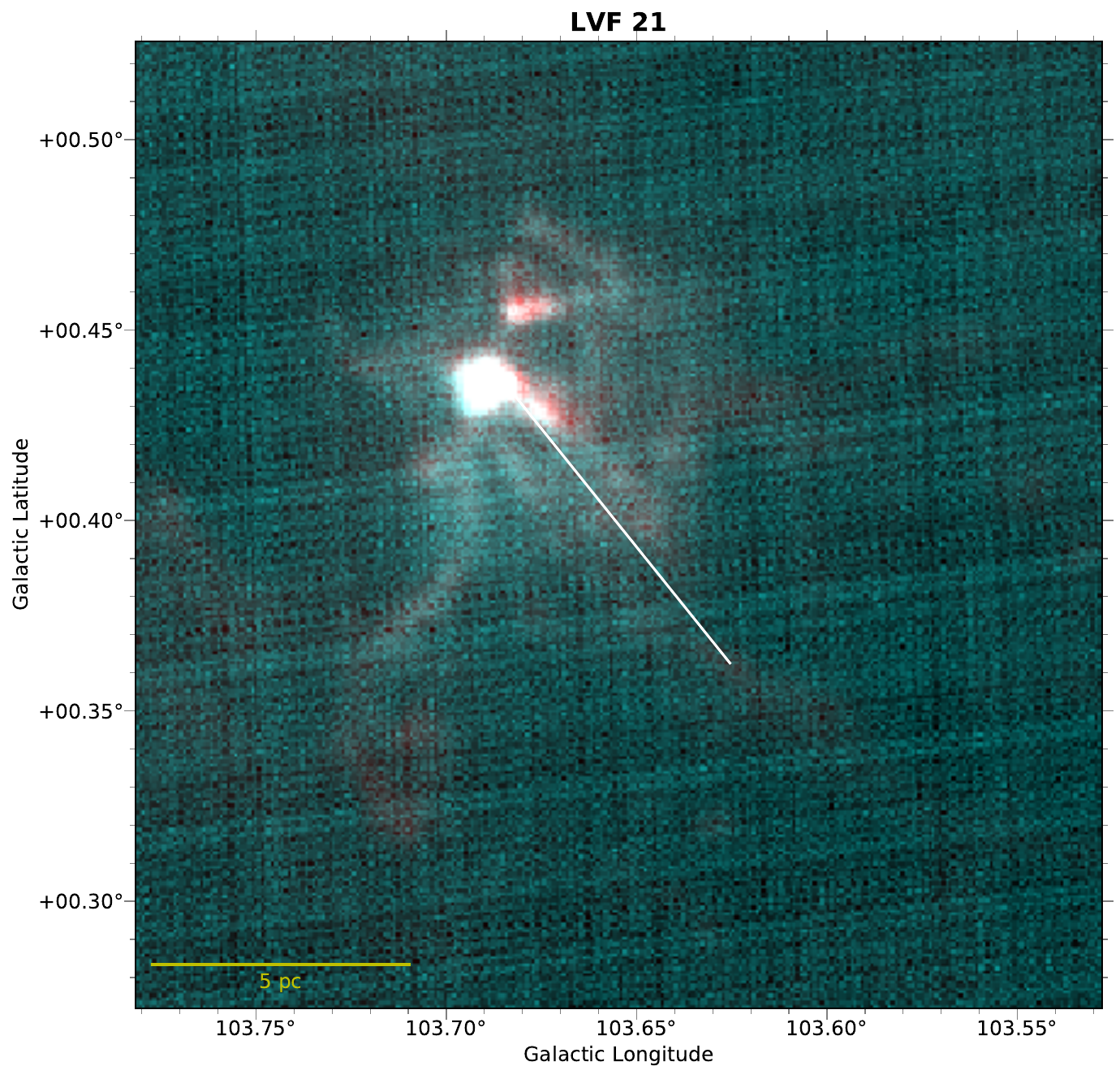}
\includegraphics[width=.28\textwidth]{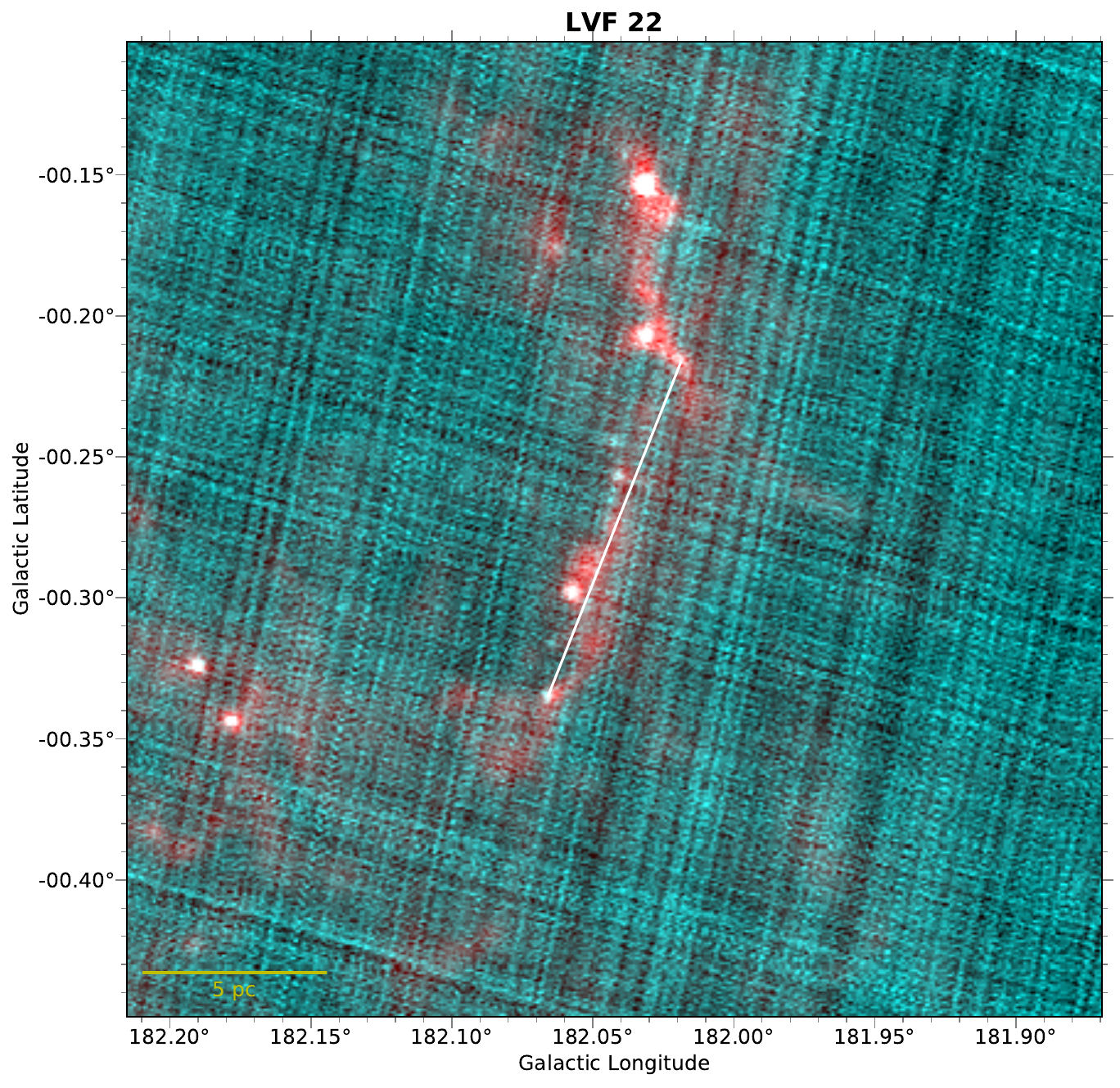}
\includegraphics[width=.28\textwidth]{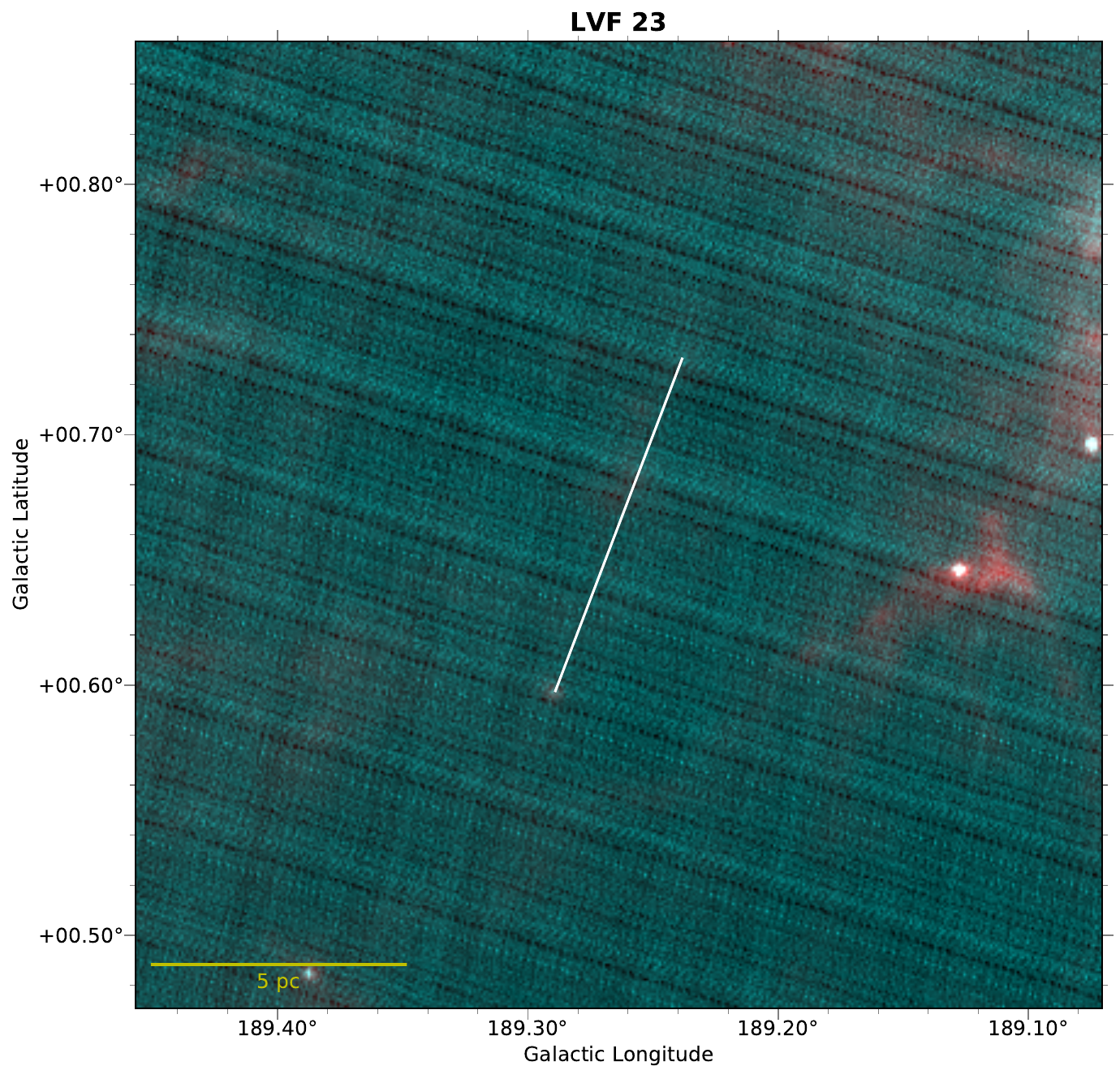}
\includegraphics[width=.28\textwidth]{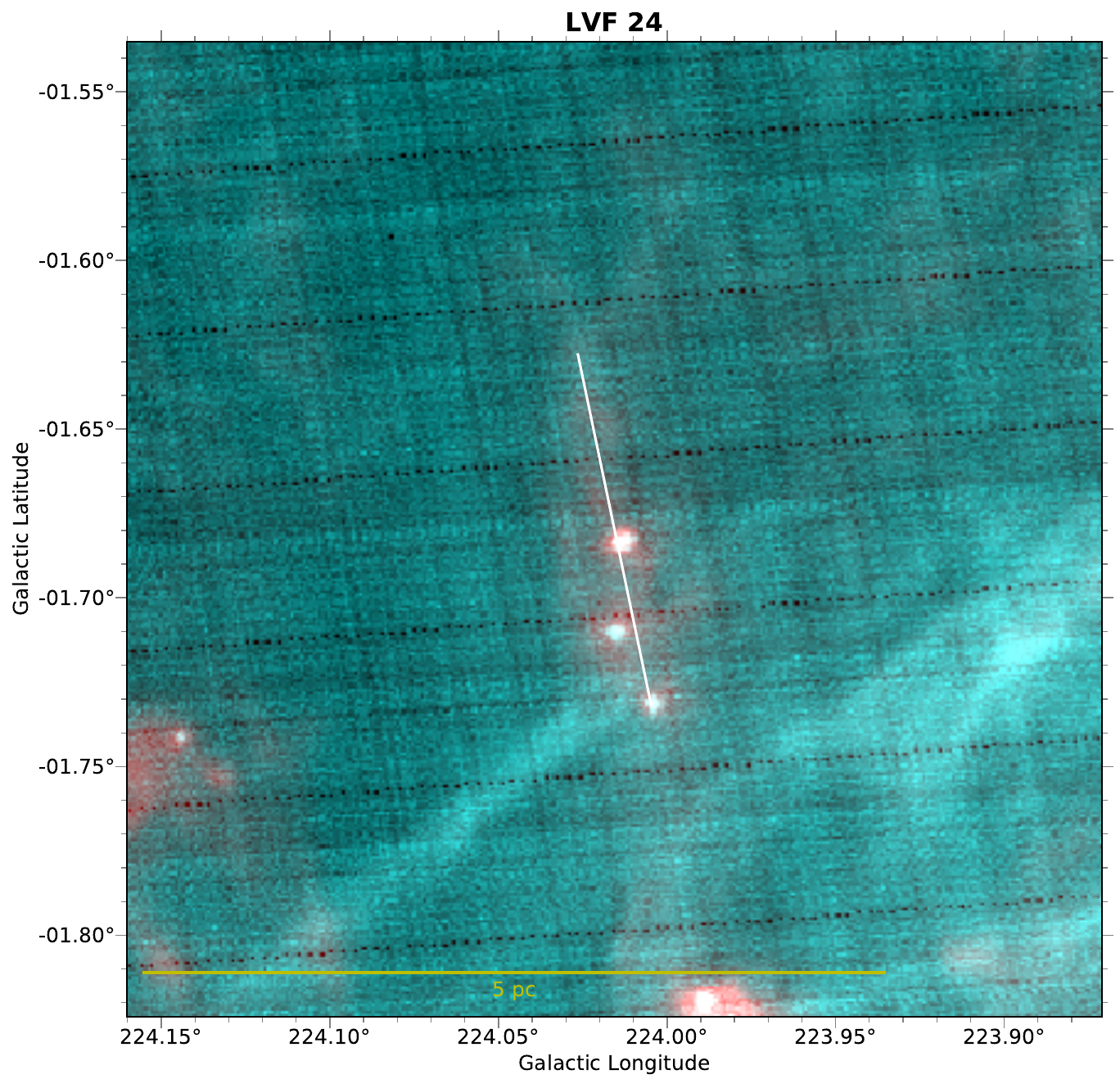}
\includegraphics[width=.28\textwidth]{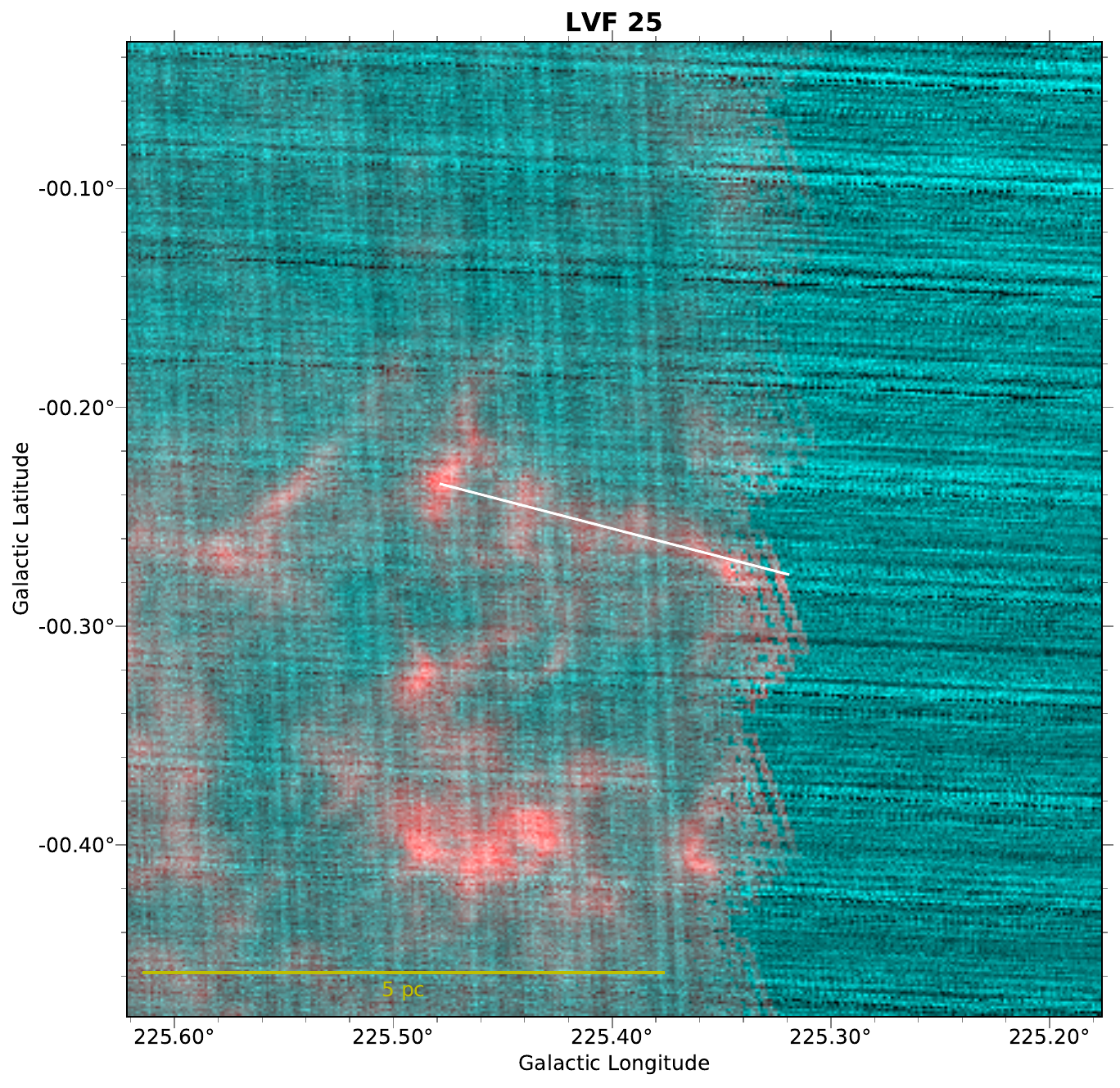}
\includegraphics[width=.28\textwidth]{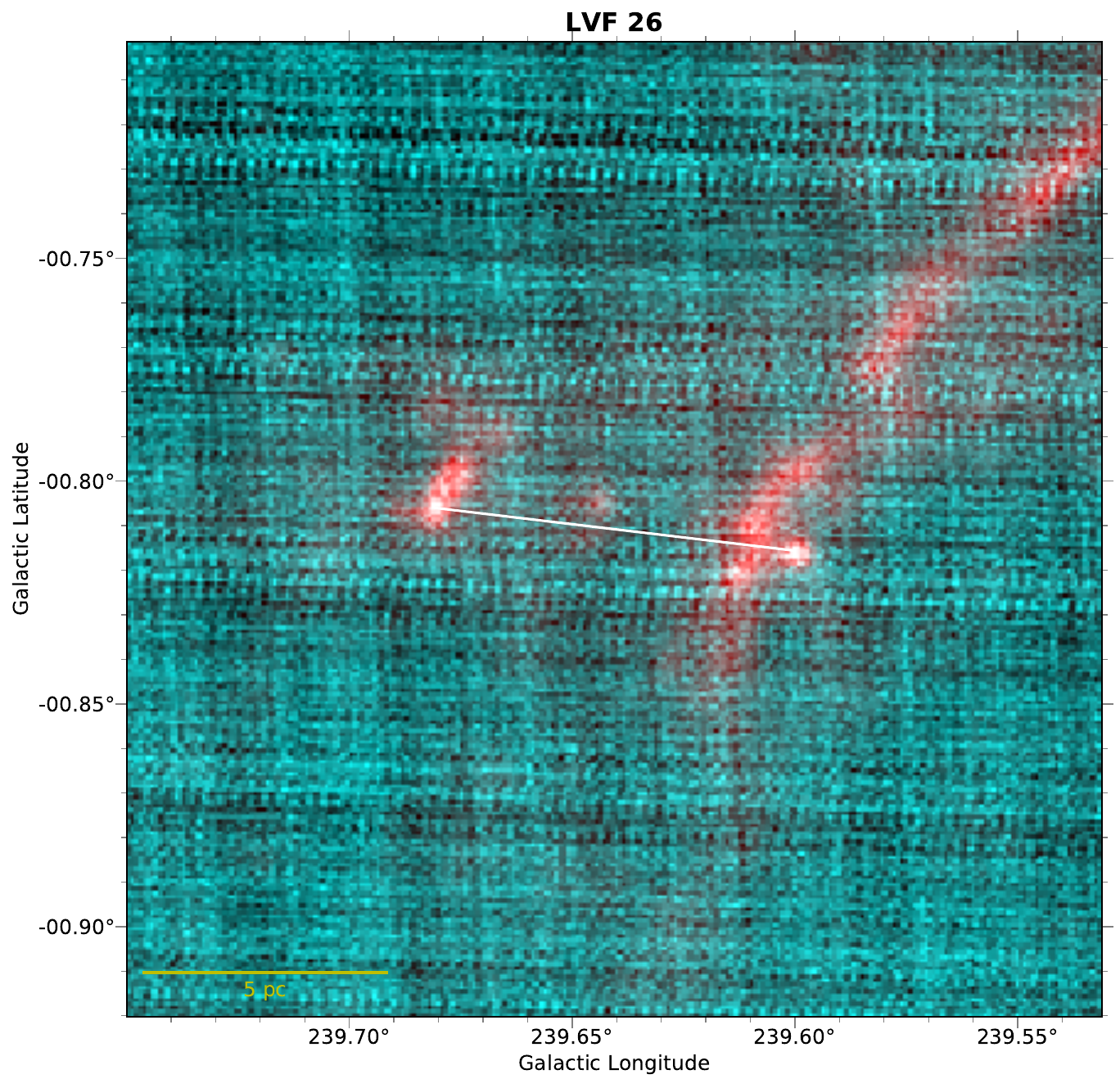}
\includegraphics[width=.28\textwidth]{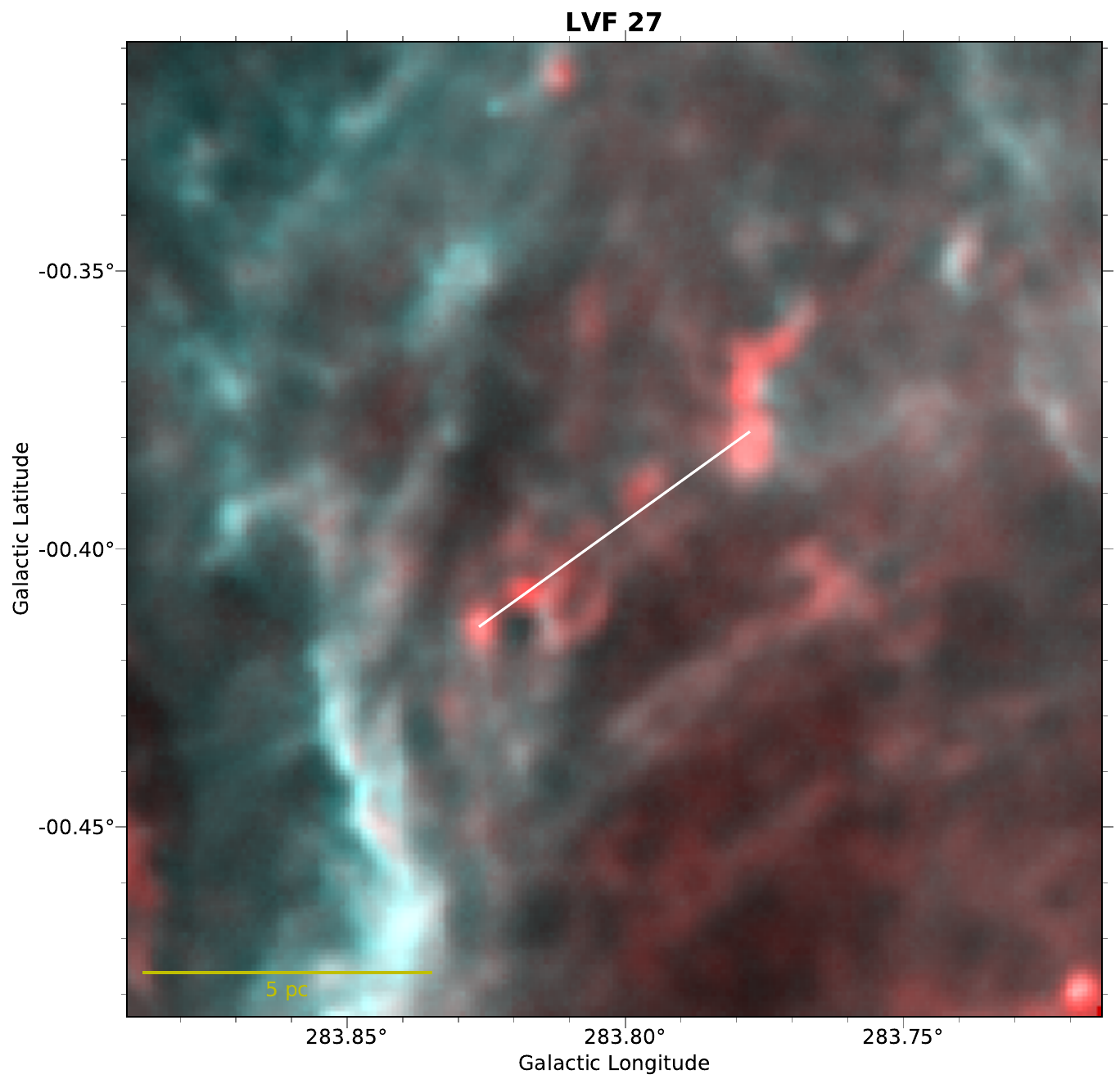}
\includegraphics[width=.28\textwidth]{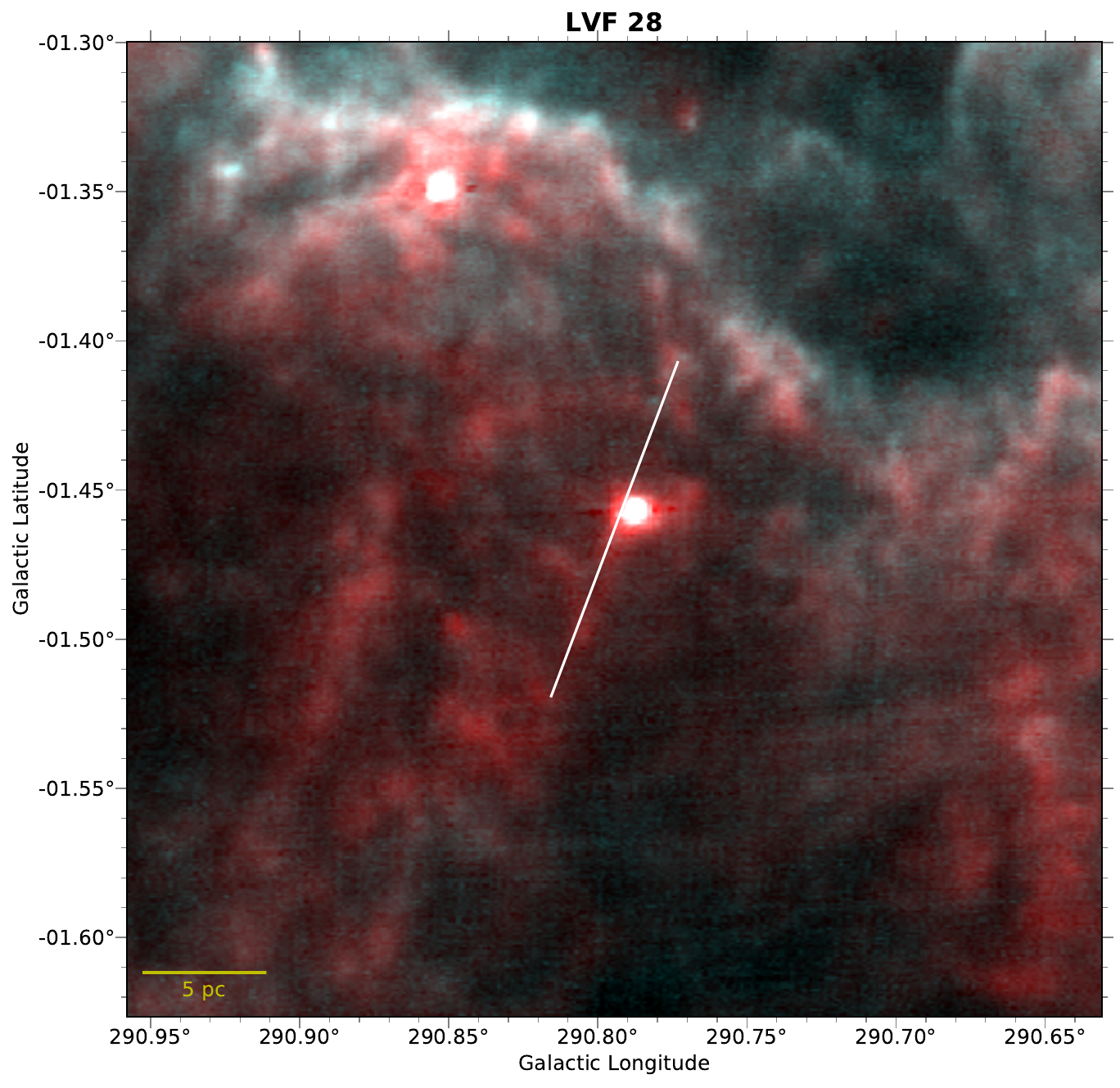}
\includegraphics[width=.28\textwidth]{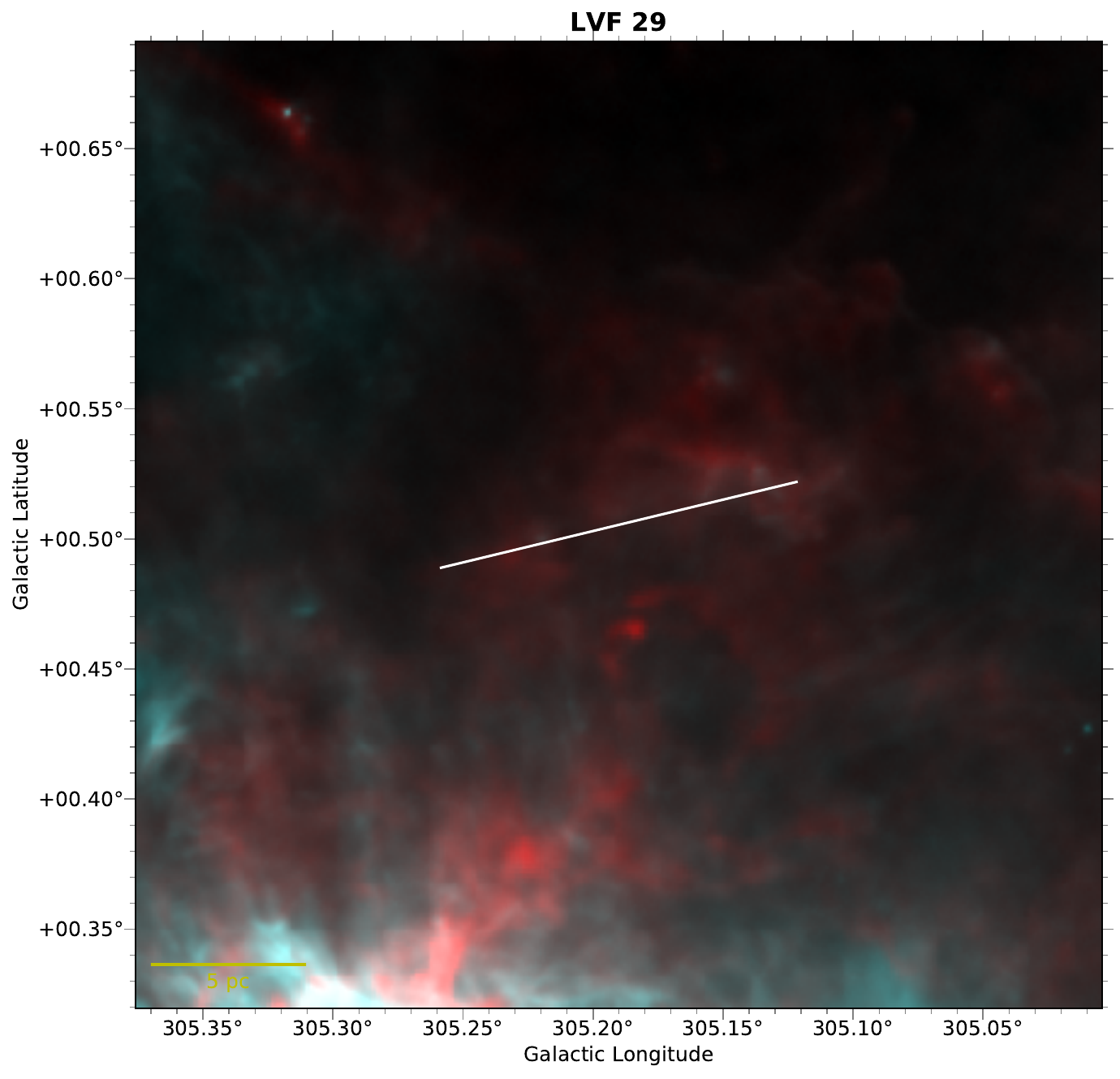}
\includegraphics[width=.28\textwidth]{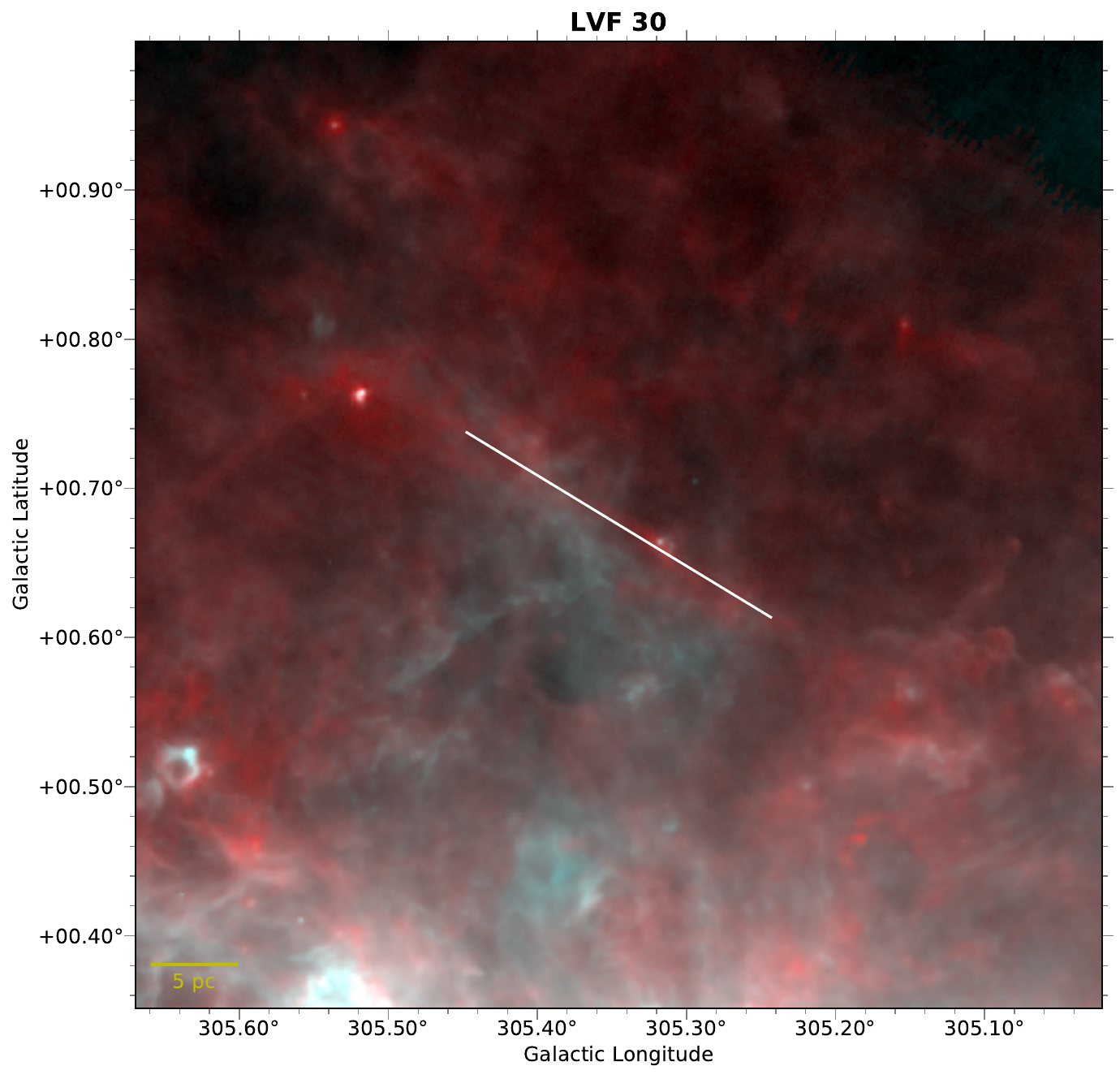}
\caption{Continued for F16 to F30.
}
\label{fig:rgb_b}
\end{figure*}

\setcounter{figure}{2}
\begin{figure*}
\centering
\includegraphics[width=.28\textwidth]{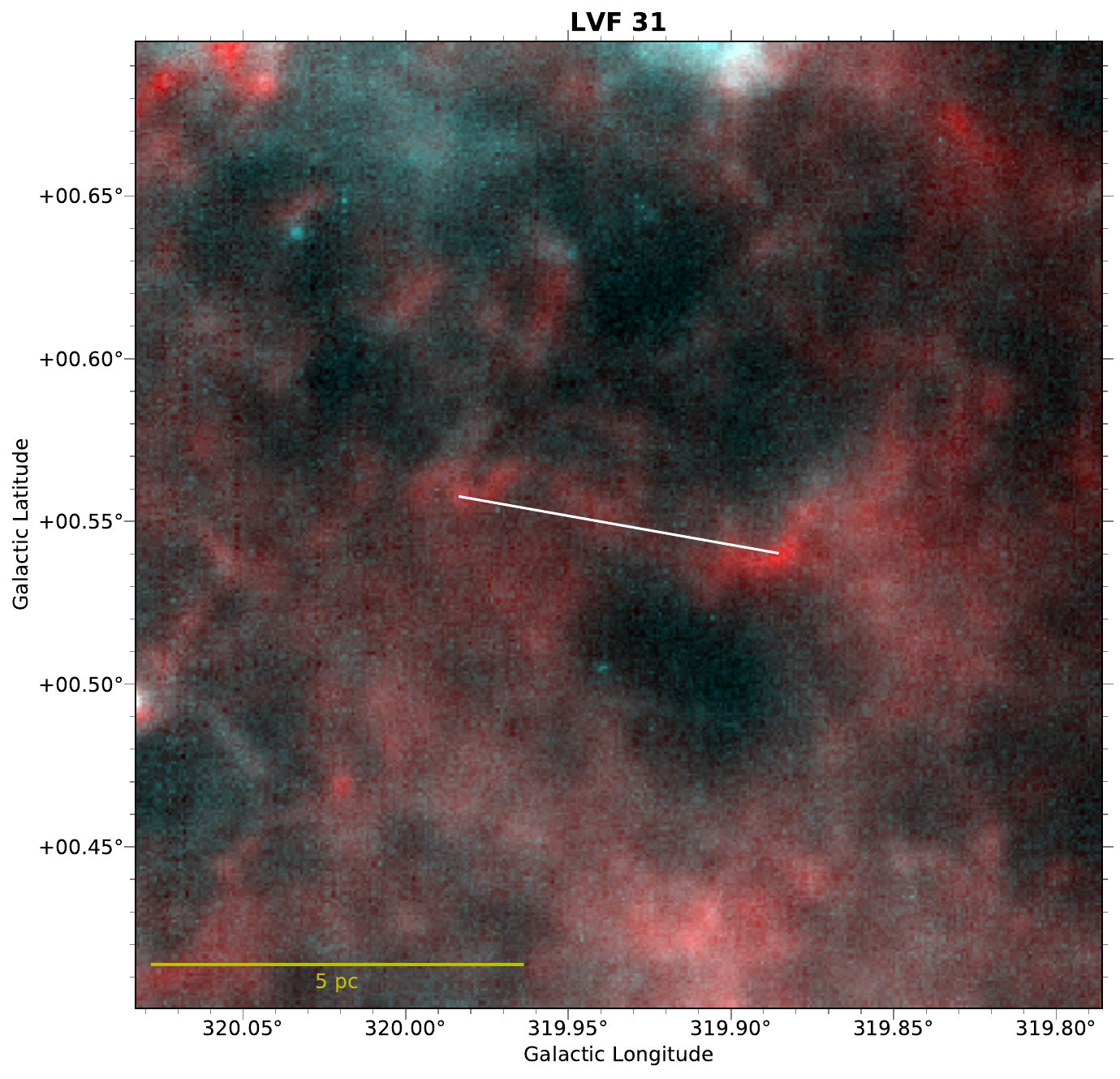}
\includegraphics[width=.28\textwidth]{rgb/LVF32.pdf}
\includegraphics[width=.28\textwidth]{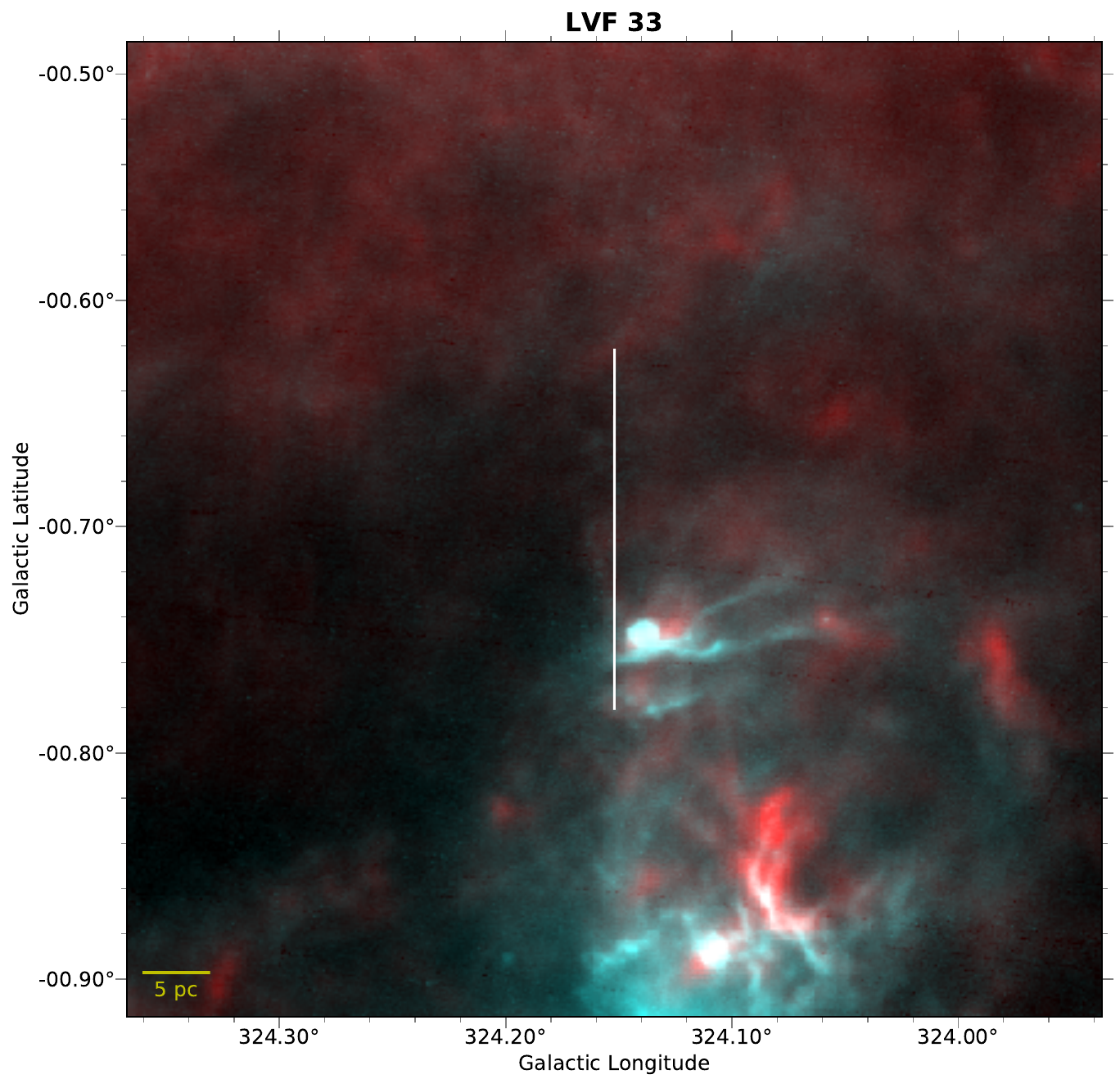}
\includegraphics[width=.28\textwidth]{rgb/LVF34.pdf}
\includegraphics[width=.28\textwidth]{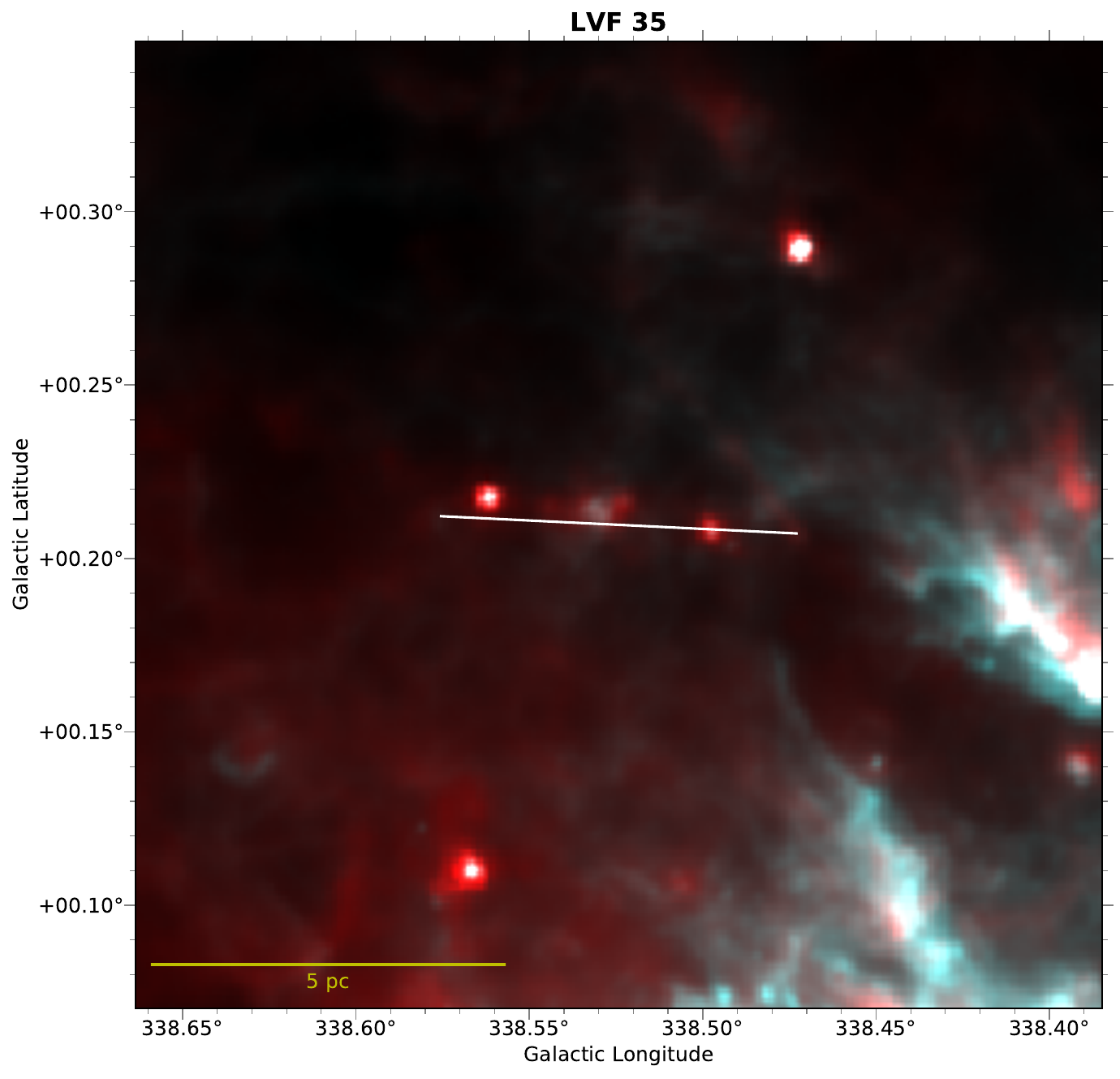}
\includegraphics[width=.28\textwidth]{rgb/LVF36.pdf}
\includegraphics[width=.28\textwidth]{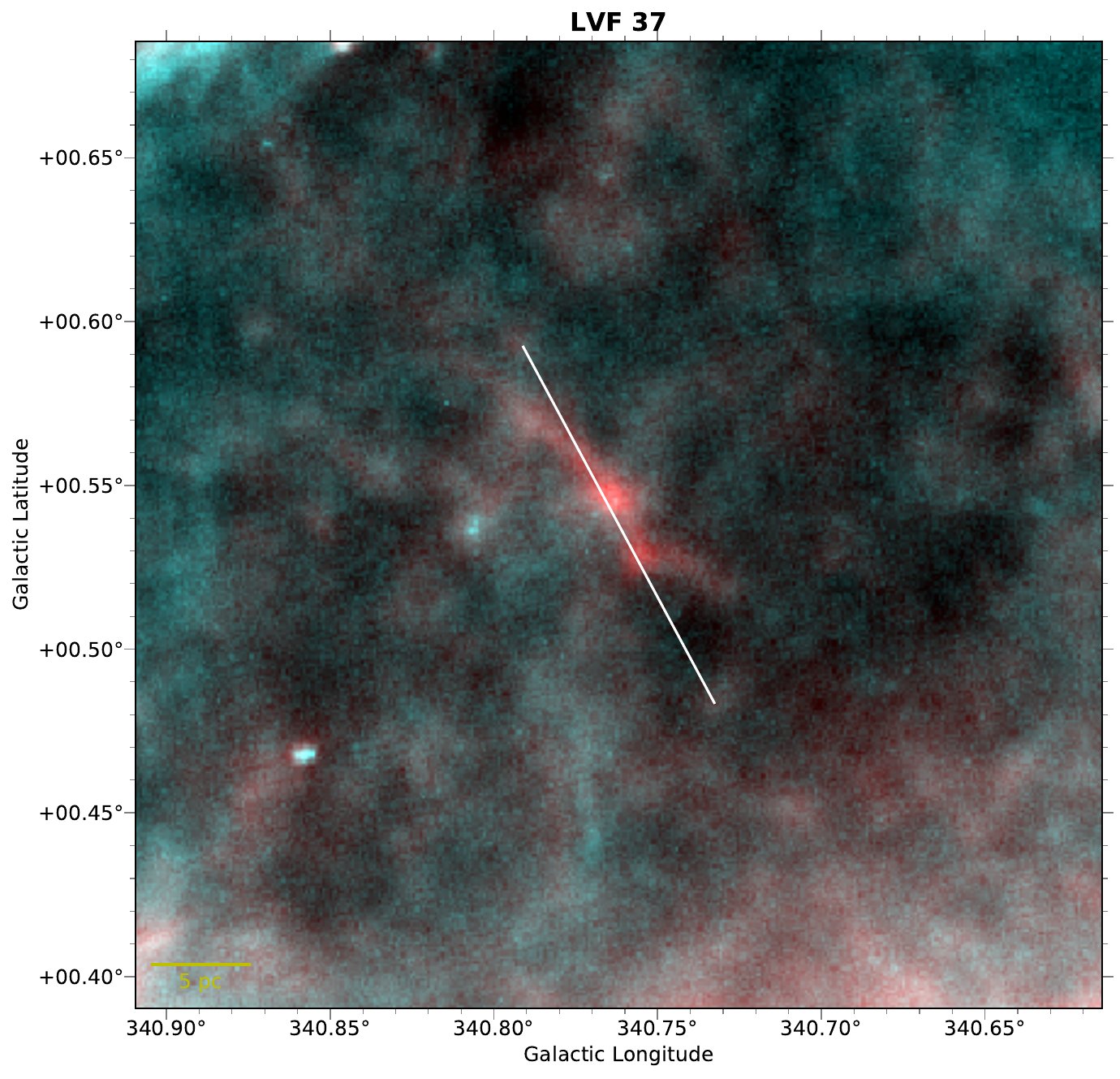}
\includegraphics[width=.28\textwidth]{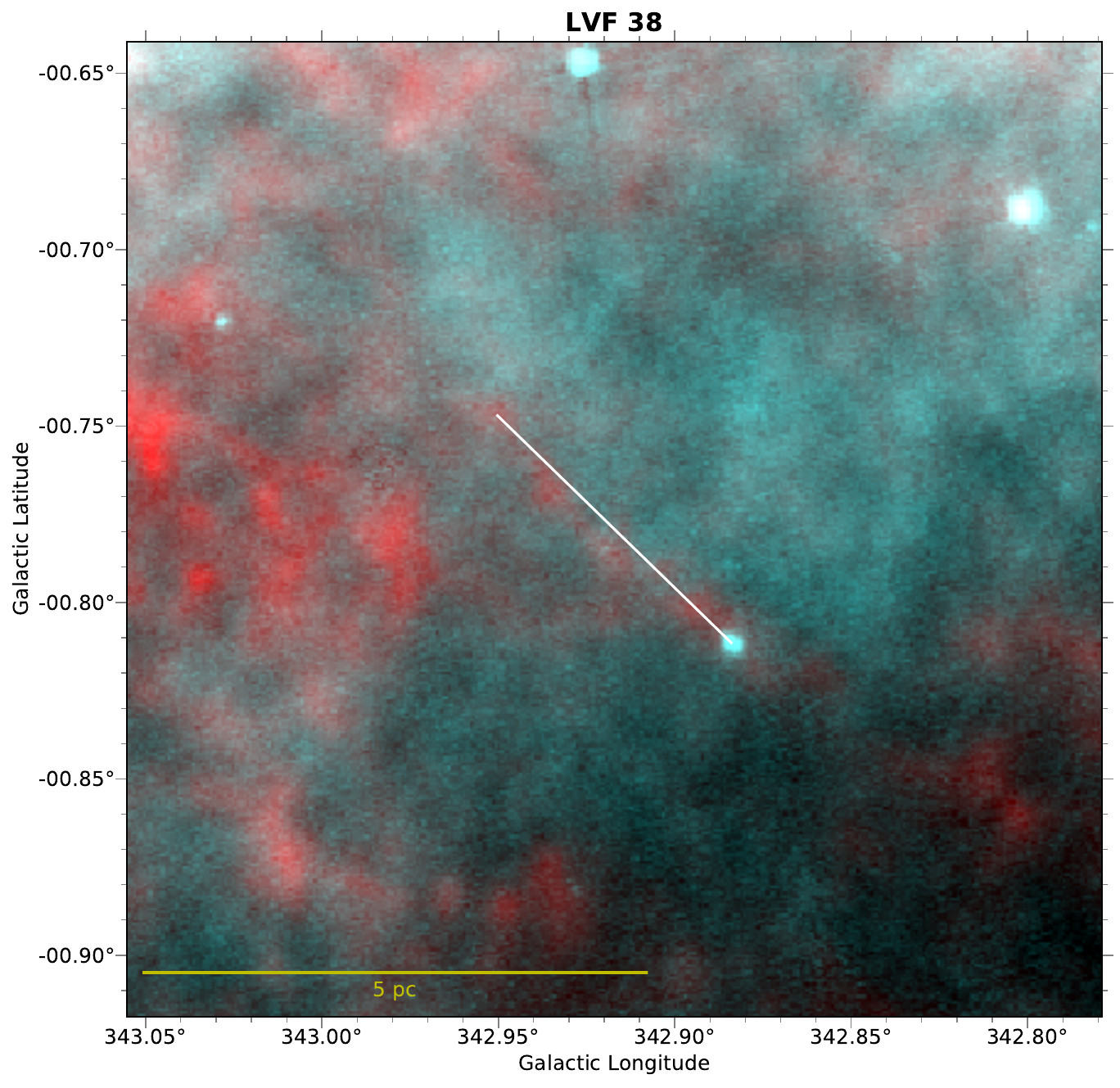}
\includegraphics[width=.28\textwidth]{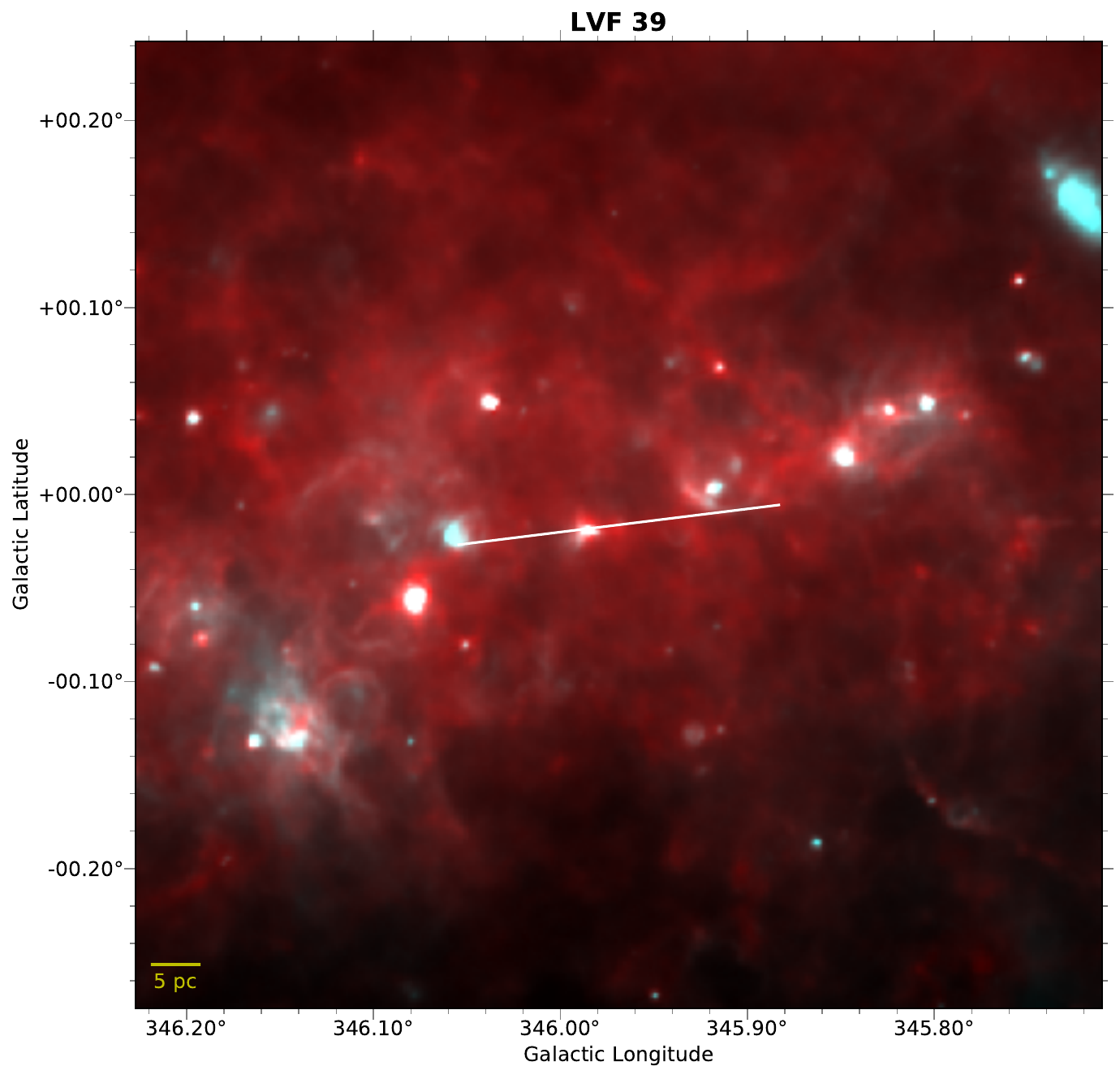}
\includegraphics[width=.28\textwidth]{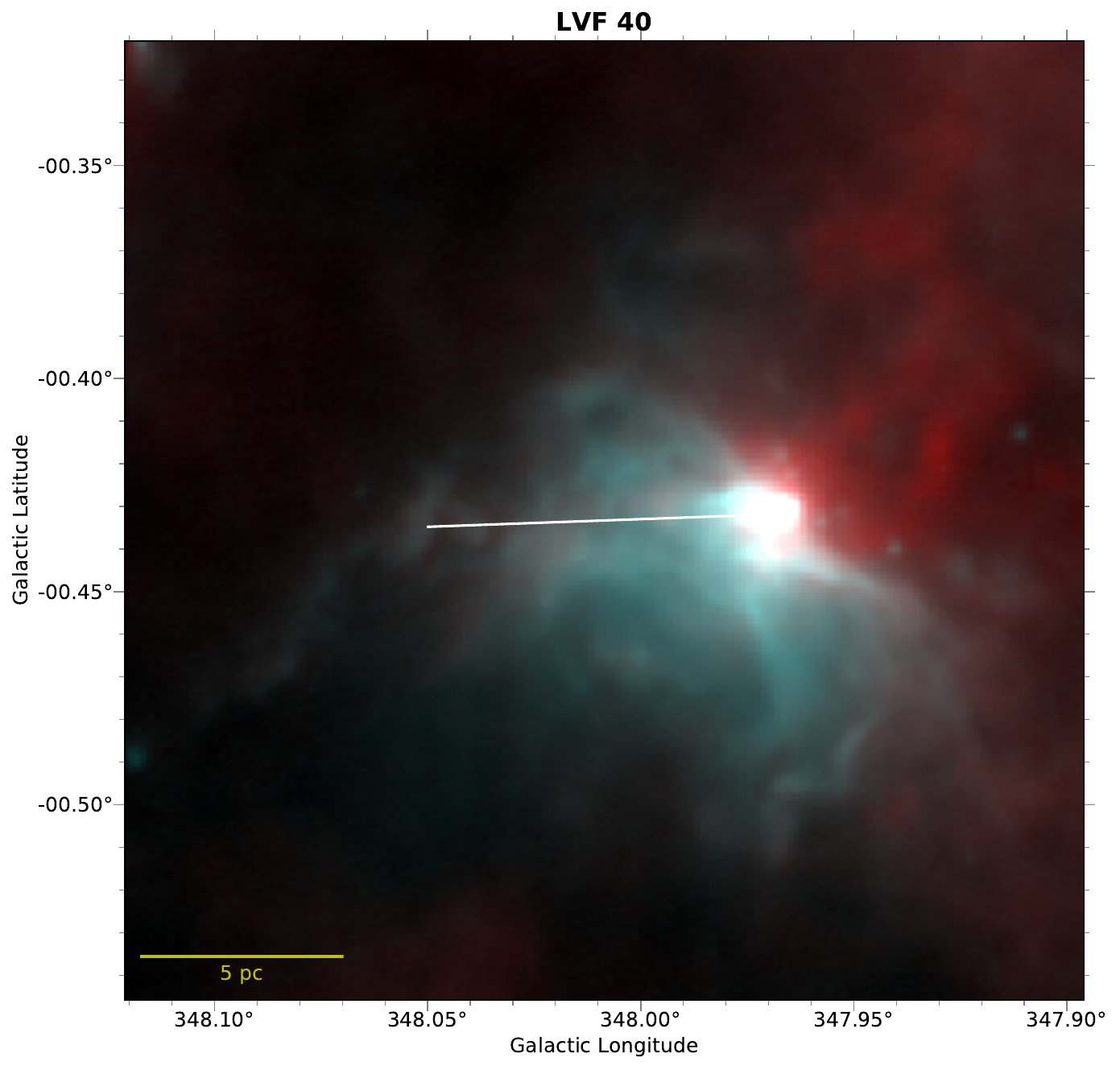}
\includegraphics[width=.28\textwidth]{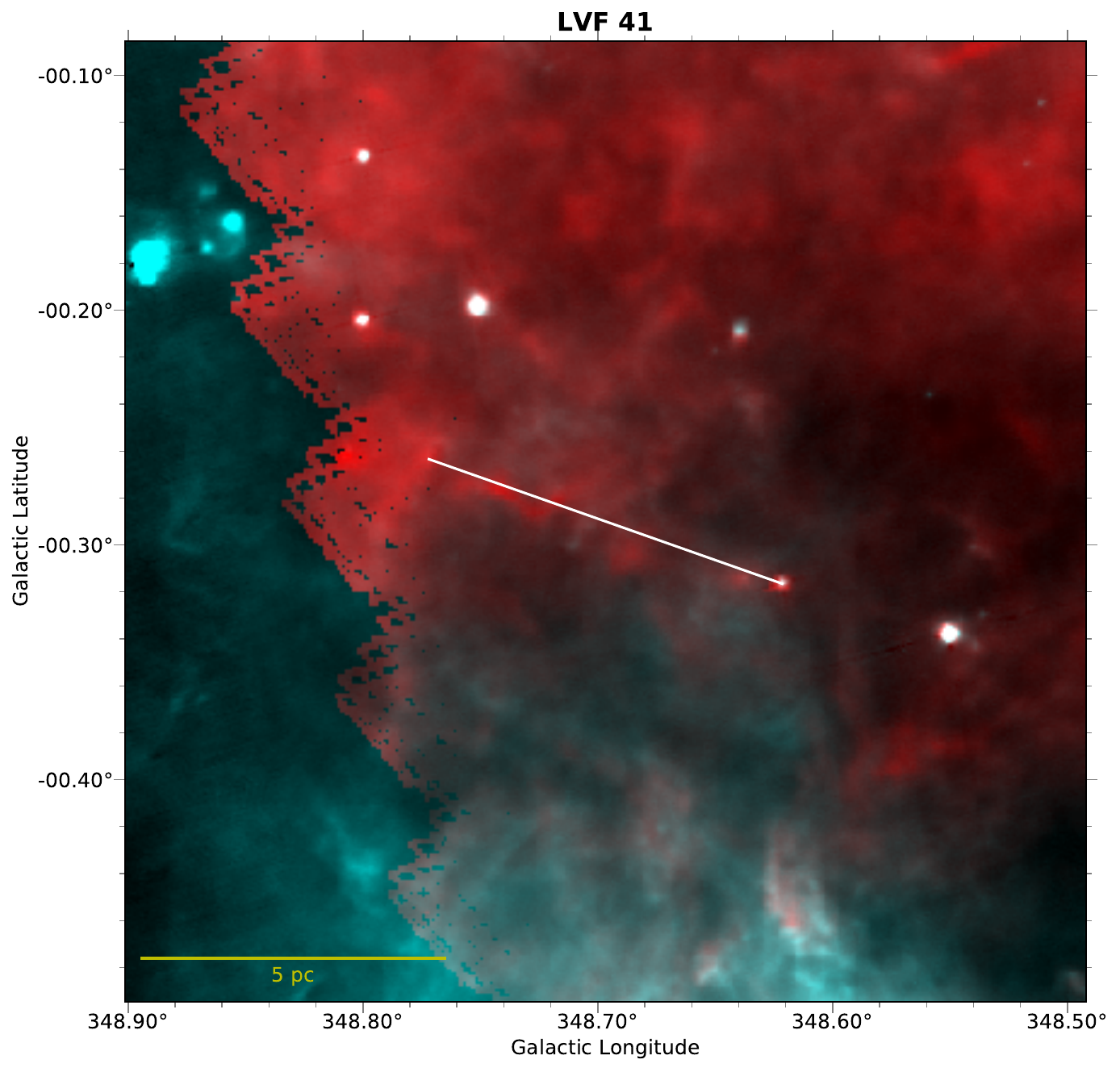}
\includegraphics[width=.28\textwidth]{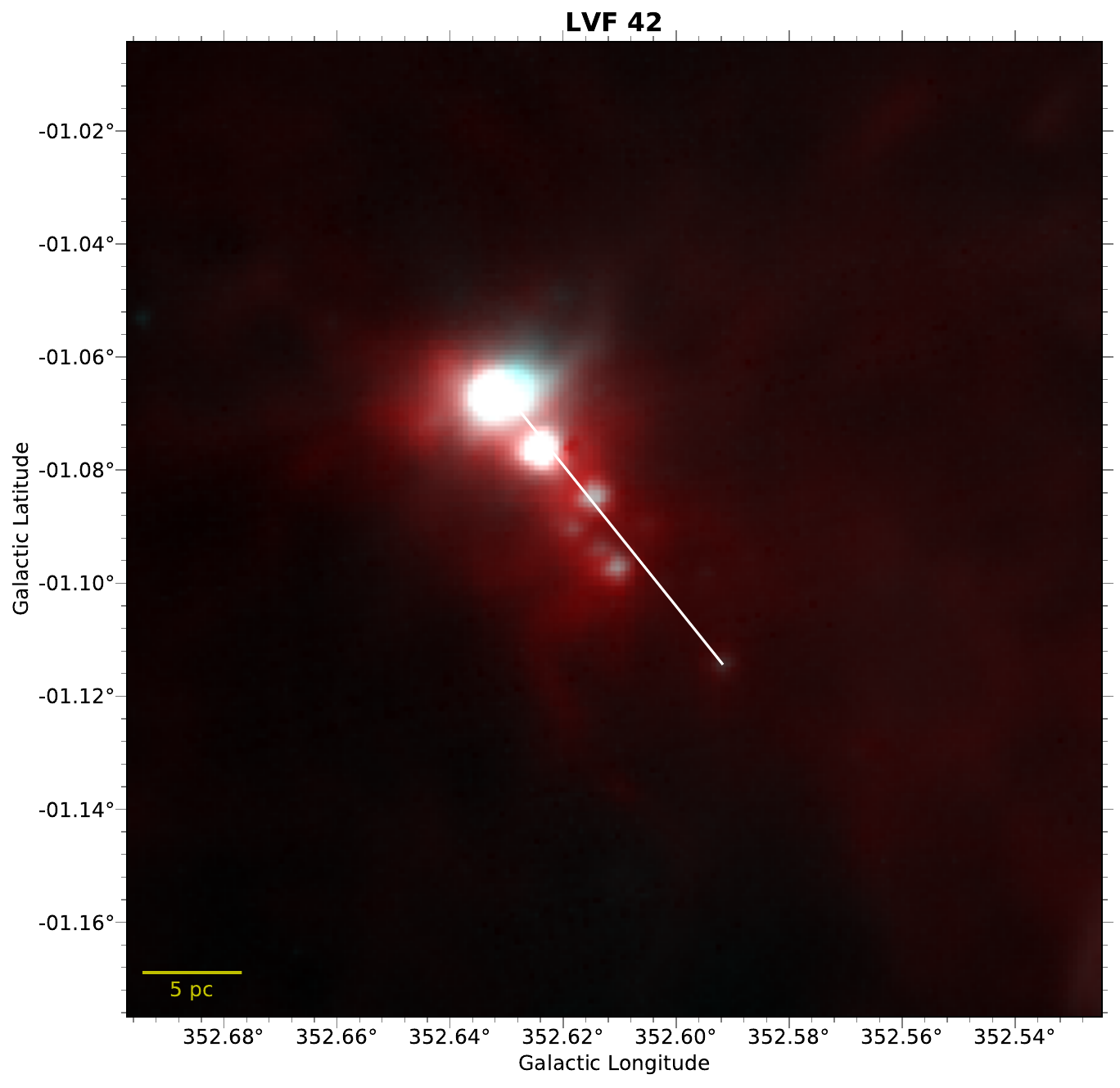}
\caption{Continued for F31 to F42.
}
\label{fig:rgb_c}
\end{figure*}

\end{appendix} 
\end{document}